\documentclass[12pt]{amsart}
\usepackage{amssymb,multicol,amscd,amsmath}

\def\vstavka[#1]{{\bf #1}}

\usepackage{hyperref}

\makeatletter \@addtoreset{equation}{section} \makeatother

\newcommand{\Add}[3]{\setcounter{#1}{#2}\addtocounter{#1}{#3}}
\newcommand{\MAdd}[6]{\Add{#1}{#2}{#3}\addtocounter{#1}{#4}\addtocounter{#1}{#5}\addtocounter{#1}{#6}}

\newcommand{\Inc}[1]{\addtocounter{#1}{1}}
\newcommand{\Dec}[1]{\addtocounter{#1}{-1}}

\newcommand{\MulFive}[2]{\MAdd{#1}{#2}{#2}{#2}{#2}{#2}}
\newcommand{\MulTen}[2]{\MulFive{#1}{#2}\addtocounter{#1}{\value{#1}}}

\newcounter{PictureWidth}     \newcounter{PictureHeight}  

\newcounter{YoungWidth}\newcounter{YoungHeight}  
\newcounter{ULYoungX}\newcounter{ULYoungY}  
\newcounter{URYoungX}\newcounter{URYoungY}  
\newcounter{DLYoungX}\newcounter{DLYoungY}  

\newcounter{TempYoungWidth}\newcounter{TempYoungHeight}  
\newcounter{TempURYoungX}\newcounter{TempDLYoungY}       

\newcounter{CurrentYoungWidth}\newcounter{CurrentYoungHeight}
\newcounter{CurrentURYoungX}\newcounter{CurrentURYoungY}
\newcounter{CurrentDLYoungX}\newcounter{CurrentDLYoungY}

\newcounter{Temp}\newcounter{TempA}\newcounter{TempB}

\newenvironment{Young}[3]{#3\InitializeYoung{#1}{#2}\SetToRow{1}\SetToCol{1}%
\begin{picture}(\value{PictureWidth},\value{PictureHeight})}{\end{picture}}

\newcommand{\InitializeYoung}[2]{{%
\MulTen{Temp}{#1}\MAdd{PictureWidth}{\LSkipValue}{\value{Temp}}{\RSkipValue}{0}{0}%
\MulTen{Temp}{#2}\MAdd{PictureHeight}{\USkipValue}{\value{Temp}}{\DSkipValue}{0}{0}%
\setcounter{YoungWidth}{0}\setcounter{YoungHeight}{0}%
\setcounter{ULYoungX}{\LSkipValue}\Add{ULYoungY}{\value{PictureHeight}}{-\USkipValue}%
\setcounter{URYoungX}{\value{ULYoungX}}\setcounter{URYoungY}{\value{ULYoungY}}%
\setcounter{DLYoungX}{\value{ULYoungX}}\setcounter{DLYoungY}{\value{ULYoungY}}%
\setcounter{CurrentYoungWidth}{0}\setcounter{CurrentYoungHeight}{0}%
\setcounter{CurrentURYoungX}{\value{ULYoungX}}\setcounter{CurrentURYoungY}{\value{ULYoungY}}%
\setcounter{CurrentDLYoungX}{\value{ULYoungX}}\setcounter{CurrentDLYoungY}{\value{ULYoungY}}%
\ZeroTempCounters}}

\newcommand{\ZeroTempCounters}{\setcounter{TempYoungWidth}{0}\setcounter{TempYoungHeight}{0}%
\setcounter{TempURYoungX}{0}\setcounter{TempDLYoungY}{0}}

\newcommand{\Put}[1]{{\put(\value{URYoungX},\value{DLYoungY}){{#1}}%
\addtocounter{YoungWidth}{\value{TempYoungWidth}}\addtocounter{YoungHeight}{\value{TempYoungHeight}}%
\setcounter{CurrentDLYoungX}{\value{URYoungX}}\setcounter{CurrentURYoungY}{\value{DLYoungY}}%
\addtocounter{URYoungX}{\value{TempURYoungX}}\addtocounter{DLYoungY}{\value{TempDLYoungY}}%
\setcounter{CurrentURYoungX}{\value{URYoungX}}\setcounter{CurrentDLYoungY}{\value{DLYoungY}}%
\setcounter{CurrentYoungWidth}{\value{TempYoungWidth}}\setcounter{CurrentYoungHeight}{\value{TempYoungHeight}}%
\ZeroTempCounters}}

\newcommand{\PutToRight}[1]{{\put(\value{URYoungX},\value{URYoungY}){#1}%
\addtocounter{YoungWidth}{\value{TempYoungWidth}}%
\setcounter{CurrentDLYoungX}{\value{URYoungX}}\setcounter{CurrentURYoungY}{\value{URYoungY}}%
\addtocounter{URYoungX}{\value{TempURYoungX}}%
\setcounter{CurrentURYoungX}{\value{URYoungX}}\Add{CurrentDLYoungY}{\value{ULYoungY}}{\value{TempDLYoungY}}%
\setcounter{CurrentYoungWidth}{\value{TempYoungWidth}}\setcounter{CurrentYoungHeight}{\value{TempYoungHeight}}%
\ZeroTempCounters}}

\newcommand{\PutToDown}[1]{{\put(\value{DLYoungX},\value{DLYoungY}){#1}%
\addtocounter{YoungHeight}{\value{TempYoungHeight}}%
\setcounter{CurrentDLYoungX}{\value{DLYoungX}}\setcounter{CurrentURYoungY}{\value{DLYoungY}}%
\addtocounter{DLYoungY}{\value{TempDLYoungY}}%
\Add{CurrentURYoungX}{\value{DLYoungX}}{\value{TempURYoungX}}\setcounter{CurrentDLYoungY}{\value{DLYoungY}}%
\setcounter{CurrentYoungWidth}{\value{TempYoungWidth}}\setcounter{CurrentYoungHeight}{\value{TempYoungHeight}}%
\ZeroTempCounters}}

\newcounter{HLines}\newcounter{VLines}

\newcommand{\Block}[2]{{\begin{picture}(0,0)
\Add{HLines}{#2}{1}\Add{VLines}{#1}{1}%
\BlockRectLines{#1}{#2}%
\multiput(0,0)(0,-10){\value{HLines}}{\usebox{\HLineBox}}%
\multiput(0,0)(10,0){\value{VLines}}{\usebox{\VLineBox}}%
\multiput(5,-5)(0,-10){#2}{\multiput(0,0)(10,0){#1}%
{\raisebox{-0.5\height}{\makebox[0pt][c]{\ensuremath{\Mark}}}\usebox{\HatchBox}}}%
\setcounter{TempYoungWidth}{#1}\addtocounter{TempYoungHeight}{#2}%
\setcounter{TempURYoungX}{\value{HLineLength}}\addtocounter{TempDLYoungY}{-\value{VLineLength}}%
\end{picture}}}

\newcommand{\BlockII}[4]{{\begin{picture}(0,0)
\put(0,0){\Block{#1}{#2}}\put(0,\value{TempDLYoungY}){\Block{#3}{#4}}
\setcounter{TempYoungWidth}{#1}\MulTen{Temp}{#1}\setcounter{TempURYoungX}{\value{Temp}}%
\end{picture}}}

\newcommand{\BlockIII}[6]{{\begin{picture}(0,0)
\put(0,0){\BlockII{#1}{#2}{#3}{#4}}\put(0,\value{TempDLYoungY}){\Block{#5}{#6}}
\setcounter{TempYoungWidth}{#1}\MulTen{Temp}{#1}\setcounter{TempURYoungX}{\value{Temp}}%
\end{picture}}}

\newcommand{\BlockIV}[8]{{\begin{picture}(0,0)
\put(0,0){\BlockIII{#1}{#2}{#3}{#4}{#5}{#6}}\put(0,\value{TempDLYoungY}){\Block{#7}{#8}}
\setcounter{TempYoungWidth}{#1}\MulTen{Temp}{#1}\setcounter{TempURYoungX}{\value{Temp}}%
\end{picture}}}

\newcommand{\LSkipValue}{0}\newcommand{\RSkipValue}{0}  
\newcommand{\USkipValue}{0}\newcommand{\DSkipValue}{0}  

\newcommand{\LSkip}{\renewcommand{\LSkipValue}{10}}

\newcommand{\USkip}{\renewcommand{\USkipValue}{10}}
\newcommand{\DSkip}{\renewcommand{\DSkipValue}{10}}

\newcommand{\LabelStyle}{\scriptstyle}

\newcounter{MiddleX}\newcounter{MiddleY}\newcounter{TempMiddleA}\newcounter{TempMiddleB}
\newcommand{\MiddleX}[3]{\Add{TempMiddleA}{#3}{-#2}\MulFive{TempMiddleB}{\value{TempMiddleA}}%
\Add{MiddleX}{#1}{\value{TempMiddleB}}}
\newcommand{\MiddleY}[3]{\Add{TempMiddleA}{#3}{-#2}\MulFive{TempMiddleB}{\value{TempMiddleA}}%
\Add{MiddleY}{#1}{-\value{TempMiddleB}}}

\newcounter{CurrentRow}\newcounter{CurrentCol}
\newcounter{CurrentRowY}\newcounter{CurrentColX}

\newcommand{\SetToRow}[1]{{\setcounter{CurrentRow}{#1}\MulTen{Temp}{#1}%
\Add{CurrentRowY}{\value{ULYoungY}}{-\value{Temp}}}}
\newcommand{\NextRow}{\Inc{CurrentRow}\addtocounter{CurrentRowY}{-10}}
\newcommand{\PrevRow}{\Dec{CurrentRow}\addtocounter{CurrentRowY}{10}}

\newcommand{\SetToCol}[1]{\setcounter{CurrentCol}{#1}\MulTen{Temp}{#1}%
\MAdd{CurrentColX}{\value{ULYoungX}}{\value{Temp}}{-10}{0}{0}}
\newcommand{\NextCol}{\Inc{CurrentCol}\addtocounter{CurrentColX}{10}}

\newcommand{\LLabelVShift}{2}
\newcommand{\ULabelHShift}{2}\newcommand{\ULabelVShift}{3}
\newcommand{\ULabelMidHShift}{0}\newcommand{\ULabelMidVShift}{2}

\newcommand{\DLabelMidHShift}{0}\newcommand{\DLabelMidVShift}{7}

\newcommand{\LEnumSymb}{}\newcommand{\UEnumSymb}{}

\newcommand{\SetLEnumSymb}[1]{\renewcommand{\LEnumSymb}{#1}}
\newcommand{\SetUEnumSymb}[1]{\renewcommand{\UEnumSymb}{#1}}

\newcounter{LabelX}\newcounter{LabelY}
\newcounter{Index}

\newcommand{\LlapLabel}[1]{\Add{LabelY}{\value{CurrentRowY}}{\LLabelVShift}%
\put(\value{ULYoungX},\value{LabelY}){\llap{\ensuremath{\LabelStyle#1}}}%
\NextRow}

\newcommand{\LlapEnum}[2]{\setcounter{Index}{#1}\multiput(0,0)(0,0){#2}{\LlapLabel{\LEnumSymb_{\theIndex}}\Inc{Index}}}

\newcommand{\ULabel}[1]{\Add{LabelX}{\value{CurrentColX}}{\ULabelHShift}%
\Add{LabelY}{\value{ULYoungY}}{\ULabelVShift}%
\put(\value{LabelX},\value{LabelY}){\ensuremath{\LabelStyle#1}}%
\NextCol}

\newcommand{\ULabelMid}[1]{\MiddleX{\value{CurrentDLYoungX}}{0}{\value{CurrentYoungWidth}}%
\Add{LabelX}{\value{MiddleX}}{\ULabelMidHShift}%
\Add{LabelY}{\value{ULYoungY}}{\ULabelMidVShift}%
\put(\value{LabelX},\value{LabelY}){\makebox[0pt][c]{\ensuremath{\LabelStyle#1}}}}

\newcommand{\UEnum}[2]{\setcounter{Index}{#1}\multiput(0,0)(0,0){#2}{\ULabel{\UEnumSymb_{\theIndex}}\Inc{Index}}}

\newcommand{\DLabelMid}[1]{\MiddleX{\value{CurrentDLYoungX}}{0}{\value{CurrentYoungWidth}}%
\Add{LabelX}{\value{MiddleX}}{\DLabelMidHShift}%
\Add{LabelY}{\value{CurrentDLYoungY}}{-\DLabelMidVShift}%
\put(\value{LabelX},\value{LabelY}){\makebox[0pt][c]{\ensuremath{\LabelStyle#1}}}}

\newcommand{\DotsDist}{5}\newcommand{\DotsNumb}{3}
\newcommand{\LDotsHShift}{6}
\newcommand{\UDotsHShift}{3}\newcommand{\UDotsVShift}{4}

\newcommand{\LDotsBetw}[1]{\PrevRow\Add{Temp}{#1}{-1}\MiddleY{\value{CurrentRowY}}{\value{CurrentRow}}{\value{Temp}}%
\Add{LabelX}{\value{ULYoungX}}{-\LDotsHShift}\MAdd{LabelY}{\value{MiddleY}}{\DotsDist}{-1}{0}{0}%
\multiput(\value{LabelX},\value{LabelY})(0,-\DotsDist){\DotsNumb}{\ensuremath{.}}%
\SetToRow{#1}}

\newcommand{\UDots}{\Add{LabelX}{\value{CurrentColX}}{\UDotsHShift}%
\Add{LabelY}{\value{ULYoungY}}{\UDotsVShift}%
\multiput(\value{LabelX},\value{LabelY})(\DotsDist,0){\DotsNumb}{\ensuremath{.}}%
\NextCol\NextCol}

\newcommand{\UDotsBetw}[1]{\MiddleX{\value{CurrentColX}}{\value{CurrentCol}}{#1}%
\MAdd{LabelX}{\value{MiddleX}}{-\DotsDist}{-1}{0}{0}\Add{LabelY}{\value{ULYoungY}}{\UDotsVShift}%
\multiput(\value{LabelX},\value{LabelY})(\DotsDist,0){\DotsNumb}{\ensuremath{.}}%
\SetToCol{#1}}

\newcounter{HLineLength}\newcounter{VLineLength}
\newcounter{HLineRef}\newcounter{VLineRef}
\newcounter{HLineDots}\newcounter{VLineDots}

\newsavebox{\HLineBox}\newsavebox{\VLineBox}
\newsavebox{\HatchBox}

\newcommand{\BlockRectLines}[2]{%
\setcounter{HLineDots}{0}\multiput(0,0)(0,0){#1}{\addtocounter{HLineDots}{\value{DotsPerBox}}}%
\setcounter{VLineDots}{0}\multiput(0,0)(0,0){#2}{\addtocounter{VLineDots}{\value{DotsPerBox}}}%
\MulTen{HLineLength}{#1}\MulTen{VLineLength}{#2}%
\MulFive{HLineRef}{#1}\MulFive{VLineRef}{#2}%
\sbox{\HLineBox}{\HLine}\sbox{\VLineBox}{\VLine}\sbox{\HatchBox}{\Hatch}}

\newcommand{\HLine}{\line(1,0){\value{HLineLength}}}
\newcommand{\VLine}{\line(0,-1){\value{VLineLength}}}

\newcounter{DotsPerBox}

\newcommand{\Mark}{}

\newcommand{\Hatch}{}

\newcounter{BraceX}\newcounter{BraceY}\newcounter{BraceSize}

\newcounter{MBlockHeight}   
\newcommand{\MBlockHeight}[4]{{\MAdd{MBlockHeight}{#1}{#2}{#3}{#4}{0}}}

\newcommand{\LEnumNumb}{2}\newcommand{\UEnumNumb}{2}

\newcommand{\UEnumDotsBlockD}[8]{\MBlockHeight{#2}{#4}{#6}{#8}\begin{Young}{#1}{\value{MBlockHeight}}{\USkip}%
\Put{\Block{#1}{#2}}\PutToDown{\Block{#3}{#4}}\PutToDown{\Block{#5}{#6}}\PutToDown{\Block{#7}{#8}}
\SetToCol{1}\UEnum{1}{\UEnumNumb}\UDots
\end{Young}}

\newcommand{\LUEnumMDotsLabelBlockC}[7]{\MBlockHeight{#2}{#4}{#6}{0}\begin{Young}{#1}{\value{MBlockHeight}}{\LSkip\USkip}%
\Put{\Block{#1}{#2}}\PutToDown{\Block{#3}{#4}}\PutToDown{\Block{#5}{#6}}
\SetToRow{1}\LlapEnum{1}{\LEnumNumb}\LDotsBetw{\value{MBlockHeight}}\LlapLabel{\LEnumSymb_{#7}}
\SetToCol{1}\UEnum{1}{\UEnumNumb}\UDotsBetw{#1}\ULabel{\UEnumSymb_{\LEnumSymb_1}}
\end{Young}}

\newcommand{\LUEnumMDotsLabelBlockD}[9]{\MBlockHeight{#2}{#4}{#6}{#8}\begin{Young}{#1}{\value{MBlockHeight}}{\LSkip\USkip}%
\Put{\Block{#1}{#2}}\PutToDown{\Block{#3}{#4}}\PutToDown{\Block{#5}{#6}}\PutToDown{\Block{#7}{#8}}
\SetToRow{1}\LlapEnum{1}{\LEnumNumb}\LDotsBetw{\value{MBlockHeight}}\LlapLabel{\LEnumSymb_{#9}}
\SetToCol{1}\UEnum{1}{\UEnumNumb}\UDotsBetw{#1}\ULabel{\UEnumSymb_{\LEnumSymb_1}}
\end{Young}}

\def\Distance{12}\def\HalfDistance{6}
\newcounter{LengthRow} \newcounter{XRow} \newcounter{YRow} \newcounter{XEnd} \newcounter{YEnd}
\newcounter{XDelta}  \addtocounter{XDelta}{\Distance}
\newsavebox{\Elem}\sbox{\Elem}{\put(0,0){\circle*{3}}}%
\newsavebox{\ElemArrows}\sbox{\ElemArrows}{\put(0,0){\vector(-1,1){\Distance}}\put(0,0){\vector(1,1){\Distance}}\put(0,0){\usebox{\Elem}}}%
\newsavebox{\ArrowsElem}\sbox{\ArrowsElem}{\put(-\Distance,-\Distance){\vector(1,1){\Distance}}\put(\Distance,-\Distance)%
{\vector(-1,1){\Distance}}\put(0,0){\usebox{\Elem}}}%
\newcommand{\adModuleStruct}[1]{\setcounter{LengthRow}{0}\setcounter{XRow}{0}\setcounter{YRow}{0}%
\begin{picture}(0,0)%
\multiput(0,0)(0,0){#1}{\addtocounter{YRow}{-\Distance}}%
\multiput(0,\value{YRow})(0,\Distance){#1}{\addtocounter{LengthRow}{1}\addtocounter{YRow}{\Distance}%
\multiput(\value{XRow},0)(\value{XDelta},0){\value{LengthRow}}{\usebox{\ElemArrows}}%
\addtocounter{XRow}{-\Distance}}%
\addtocounter{LengthRow}{1}%
\multiput(\value{XRow},\value{YRow})(\value{XDelta},0){\value{LengthRow}}{\usebox{\Elem}}%
\addtocounter{YRow}{\Distance}%
\multiput(0,\value{YRow})(0,\Distance){#1}{\addtocounter{LengthRow}{-1}\addtocounter{XRow}{\Distance}\addtocounter{YRow}{\Distance}%
\multiput(\value{XRow},0)(\value{XDelta},0){\value{LengthRow}}{\usebox{\ArrowsElem}}}%
\end{picture}}
\newsavebox{\Etc}\sbox{\Etc}{\linethickness{0.8pt}\qbezier[3](0,0)(\HalfDistance,\HalfDistance)(\Distance,\Distance)}%
\newsavebox{\ArrowElem}\sbox{\ArrowElem}{\put(0,0){\usebox{\Elem}}\put(0,0){\vector(1,1){\Distance}}\put(\Distance,\Distance){\usebox{\Elem}}} %
\newsavebox{\ArrowElemArrow}\sbox{\ArrowElemArrow}{\put(0,0){\usebox{\ArrowElem}}\put(\Distance,\Distance){\vector(-1,1){\Distance}}}%
\newcommand{\twModuleLine}[1]{\setcounter{XEnd}{0}\setcounter{YEnd}{0}%
\multiput(0,0)(\Distance,\Distance){#1}{\usebox{\ArrowElemArrow}\addtocounter{XEnd}{\Distance}\addtocounter{YEnd}{\Distance}}%
\put(\value{XEnd},\value{YEnd}){\usebox{\Etc}}}%
\newcommand{\twModuleLastLine}[1]{\setcounter{XEnd}{0}\setcounter{YEnd}{0}%
\multiput(0,0)(\Distance,\Distance){#1}{\usebox{\ArrowElem}\addtocounter{XEnd}{\Distance}\addtocounter{YEnd}{\Distance}}%
\put(\value{XEnd},\value{YEnd}){\usebox{\Etc}}}%
\newcommand{\twModuleStruct}[2]{\setcounter{LengthRow}{#1}\setcounter{YRow}{0}%
\multiput(0,0)(0,\value{XDelta}){#2}{\twModuleLine{\value{LengthRow}}\addtocounter{LengthRow}{-1}\addtocounter{YRow}{\value{XDelta}}}%
\put(0,\value{YRow}){\twModuleLastLine{\value{LengthRow}}}}%

\newcommand{\half}{\frac{1}{2}}
\def\nn{\nonumber}

\def\be{\begin{equation}\begin{aligned}}
\def\ee{\end{aligned}\end{equation}}

\def\bal{\begin{align}}
\def\eal{\end{align}}

\def\ptl{\partial}
\newcommand{\dpd}[1]{\frac{\ptl}{\ptl #1}}
\newcommand{\ddpd}[2]{\frac{\ptl^2}{\ptl #1 \ptl #2}}
\newcommand{\ddpddot}[2]{\frac{\ptl^2}{\ptl #1\cdot \ptl #2}}

 \def\cB{\mathcal{B}} \def\cC{\mathcal{C}}
\def\cD{\mathcal{D}} \def\cE{\mathcal{E}} \def\cF{\mathcal{F}}
 \def\cH{\mathcal{H}} 
 \def\cK{\mathcal{K}} \def\cL{\mathcal{L}}
\def\cM{\mathcal{M}}  
\def\cP{\mathcal{P}}  
\def\cS{\mathcal{S}}  
  \def\cX{\mathcal{X}}
\def\cY{\mathcal{Y}} \def\cZ{\mathcal{Z}}

 \newcommand{\oC}{\mathbb{C}}
 \newcommand{\oZ}{\mathbb{Z}}

\newcommand{\agen}{\mathfrak{g}}
\newcommand{\asubgen}{\mathfrak{h}}
\newcommand{\aso}{\mathfrak{so}}
\newcommand{\asu}{\mathfrak{su}}
\newcommand{\au}{\mathfrak{u}}

\newcommand{\agl}{\mathfrak{gl}}
\newcommand{\asl}{\mathfrak{sl}}
\newcommand{\aiu}{\mathfrak{iu}}
\newcommand{\aisu}{\mathfrak{isu}}

\newcommand{\aisun}[1]{\mathfrak{isu}^{#1}}
\newcommand{\ahsc}{\mathfrak{hsc}(4)}

\newcommand{\acu}{\mathfrak{cu}(1,0|8)}
\newcommand{\ahuzero}{\mathfrak{hu}_0(1,0|8)}

\def\ideal{\mathfrak{I}}

\def\uL{\cL} \def\uD{\cD} \def\uP{\cP} \def\uK{\cK} 
\def\buL{\bar{\uL}} \def\uX{\cX_{\au(k,k)}{}}

\def\adinfL{({\rm ad}^\infty_{\uL}){}} \def\adinfbL{({\rm ad}^\infty_{\buL}{})}
\def\adinfP{({\rm ad}^\infty_{\uP}){}} \def\adinfK{({\rm ad}^\infty_{\uK}){}} \def\adinfD{{\rm ad}^\infty_{\uD}{}} \def\adinfPNO{{\rm ad}^\infty_{\PNO}{}}
\def\adX{({\rm ad}_{\uX}){}}

\def\adL{({\rm ad}_{\uL}){}} \def\adbL{({\rm ad}_{\buL}{})}
\def\adP{({\rm ad}_{\uP}){}} \def\adK{({\rm ad}_{\uK}){}} \def\adD{{\rm ad}_{\uD}{}} 

\def\trProj{\Pi^\bot}
\def\ttrProj{\tilde{\Pi}^\bot}

\def\na{n_a} \def\nb{n_b} \def\nba{n_\ba} \def\nbb{n_\bb}
\def\bn{\bar n}  \def\tbn{\bar{\tilde{n}}}

\def\tb{\tilde{b}} 
\def\tbb{\bar{\tilde{b}}} \def\ntbb{n_{\tbb}}

\def\twinfL{({\rm tw}^\infty_{\uL}){}} \def\twinfbL{({\rm tw}^\infty_{\buL}{})}
\def\twinfP{({\rm tw}^\infty_{\uP}){}} \def\twinfK{({\rm tw}^\infty_{\uK}){}} \def\twinfD{{\rm tw}^\infty_{\uD}{}} \def\twinfPNO{{\rm tw}^\infty_{\PNO}{}}

\def\twL{({\rm tw}_{\uL}){}} \def\twbL{({\rm tw}_{\buL}{})}
\def\twP{({\rm tw}_{\uP}){}} \def\twK{({\rm tw}_{\uK}){}} \def\twD{{\rm tw}_{\uD}{}} \def\twPNO{{\rm tw}_{\PNO}{}}

  \def\gP{P} \def\gK{K} \def\gD{D} \def\gPNO{Z} \def\gModule{M}

\def\btwP{(\bar{{\rm tw}}_{\uP}){}}

\def\tPNO{\tilde{\PNO}}
\def\tD{\tilde{\uD}}

\def\adbasis{\cB{}}
\def\adbasisanscoef{d}
\def\admonomial{m{}}
\def\twbasis{\tilde{\cB}{}}

\def\twbasisterm{\tilde{\cB}^{\rm t\,}{}}
\def\twbasisanscoef{\tilde{d}}
\def\twmonomial{\tilde{m}{}}
\def\twmonomialterm{\tilde{m}^{\rm t}{}}
\def\tf{\tilde{f}}

\def\adbasefunc{g}
\def\twbasefunc{\tilde{g}}

\def\confweight{\Delta}
\def\twconfweight{\tilde{\confweight}}

\def\xiuL{\stackrel{\xi}{\cL}} \def\xibuL{\stackrel{\xi}{\bar{\uL}}}
\def\xiCasimir{\stackrel{\xi}{\Casimir}}

\def\admodule{\cM}
\def\admodules[#1]{\admodule_{#1}}
\def\admoduleinf{\admodule^\infty}
\def\admoduleinfs[#1]{\admoduleinf_{#1}}

\def\twmodule{\tilde{\cM}}
\def\twmodules[#1]{\twmodule_{#1}}
\def\twmodulen[#1]{\twmodule^{#1}}
\def\twmodulens[#1][#2]{\twmodule^{#1}_{#2}}
\def\twmoduleinf{\twmodule^\infty}
\def\twmoduleinfs[#1]{\twmoduleinf_{#1}}

\def\btwmodule{\bar{\tilde{\cM}}}
\def\btwmodules[#1]{\btwmodule_{#1}}
\def\btwmodulen[#1]{\btwmodule^{#1}}
\def\btwmodulens[#1][#2]{\btwmodule^{#1}_{#2}}
\def\btwmoduleinf{\btwmodule^\infty}
\def\btwmoduleinfs[#1]{\btwmoduleinf_{#1}}

\def\twsubmodulen[#1]{\tilde{\ideal}^{#1}}
\def\twsubmodulens[#1][#2]{\tilde{\ideal}^{#1}_{#2}}

\def\opsinfo{{s}_1}
\def\opsinft{{s}_2}
\def\opsinf{{s}}

\def\optsinfo{{\tilde{s}}_1}
\def\optsinft{{\tilde{s}}_2}
\def\opbtsinfo{\bar{{\tilde{s}}}_1}
\def\opbtsinft{\bar{{\tilde{s}}}_2}

\def\gaugepar{\varepsilon}
\def\physf{\varphi}
\def\adeq{E}
\def\badeq{\bar{\adeq}}
\def\adsyzo{S}
\def\adsyzt{S}

\def\Weyltens{C}
\def\tweq{\tilde{E}}
\def\twsyz{\tilde{S}}

\def\func{\varphi}

\def\lcws{l}

\def\deg{{\rm deg}}
\newcommand{\formrank}[1]{p^{#1}}

\def\extdiff{{\rm d}}

\def\Gd{\Delta}

\def\Go{\Omega}

\def\ga{\alpha}
\def\gb{\beta}
\def\gga{\gamma}
\def\gd{\delta}
\def\gs{\sigma}
\def\gt{\theta}
\def\gl{\lambda}
\def\gep{\epsilon}
\def\go{\omega}

\def\da{{\dot \ga}}
\def\db{{\dot \gb}}
\def\dg{{\dot \gga}}
\def\dd{{\dot \gd}}

\def\ba{{\bar a}}
\def\bb{{\bar b}}

\def\comp{\mathfrak{C}}
\def\csp{\cC}

\def\twcomp{\tilde{\comp}}
\def\twcsp{\tilde{\csp}}

\newcommand{\twcoord}[1]{(#1)}

\def\vp{v^{\prime}}
\def\qp{q^{\prime}}
\def\tp{t^{\prime}}
\def\nap{\na^{\prime}}
\def\nbp{\nb^{\prime}}
\def\nbap{\nba^{\prime}}
\def\ntbbp{\ntbb^{\prime}}
\def\monp{\twmonomial^{\prime}}

\def\comp{\mathfrak{C}}
\def\csp{\cC}
\def\diffg{{\ptl}}
\def\anticomm{\Theta}
\def\cohom{\cH}

\def\rewcompmap[#1]{\left.#1\right|_{\stackrel{\psi=\xi}{c=b}}}

\def\Ch[#1]{\compel^{p,(#1)}_{s_1,s_2,N}}

\def\eh[#1]{\epsilon^{p-1,(#1)}_{s_1,s_2,N}}
\def\tCh[#1]{\tilde{\compel}^{p,(#1)}_{s_1,s_2,N}}

\def\bt{\begin{tabular}}
\def\et{\end{tabular}}

\def\mcol{h}

\newcommand{\uurl}{\gl}\newcommand{\uucl}{\mu}
\newcommand{\lurl}{\tilde{\gl}}\newcommand{\lucl}{\tilde{\mu}}
\newcommand{\udrl}{\gl'}\newcommand{\udcl}{\mu'}
\newcommand{\ldrl}{\tilde{\gl}'}\newcommand{\ldcl}{\tilde{\mu}'}

\newcommand{\UUCase}{\SetLEnumSymb{\uurl}\SetUEnumSymb{\uucl}}
\newcommand{\LUCase}{\SetLEnumSymb{\lurl}\SetUEnumSymb{\lucl}}
\newcommand{\UDCase}{\SetLEnumSymb{\udrl}\SetUEnumSymb{\udcl}}
\newcommand{\LDCase}{\SetLEnumSymb{\ldrl}\SetUEnumSymb{\ldcl}}

\newcommand{\Cas}{\cC}
\newcommand{\Casimir}{\Cas^2}
\newcommand{\UUTotal}{\Sigma}\newcommand{\LUTotal}{\tilde{\Sigma}}

\newcommand{\Ydiag}{\cY}
\newcommand{\limone}[1]{\Big|_{#1}}
\newcommand{\limtwo}[2]{\Big|_{\stackrel{\scriptstyle #1}{\scriptstyle #2}}}

\newcommand{\transpose}{{\rm T}}
\newcommand{\symm}[1]{(#1)}

\newcommand{\compel}{C}

\newcommand{\compa}[1]{\stackrel{a}{(#1)}}
\newcommand{\compb}[1]{\stackrel{b}{(#1)}}
\newcommand{\compba}[1]{\stackrel{\ba}{(#1)}}
\newcommand{\compbb}[1]{\stackrel{\bb}{(#1)}}
\newcommand{\compxi}[1]{\stackrel{\xi}{#1}}

\def\tcohom{\tilde{\cohom}}
\def\btcohom{\bar{\tcohom}}
\def\hcohom{\hat{\cohom}}

\def\tgen{\tilde{\cF}}
\def\tgenc{\tilde{F}}

\def\tmcohomo{\tilde{h}^{-;1}}

\newcommand{\Utens}{\otimes^U}
\newcommand{\Ttens}{\otimes^\bot}

\def\PNO{\cZ}
\def\compconj{\zeta}

\def\Curv{R}
\def\GaugeF{\omega}
\def\WeylF{C}
\def\bWeylF{\bar{C}}
\def\adsm{\gs_-}
\def\twsm{\tilde{\gs}_-}
\def\btwsm{\bar{\tilde{\gs}}_-}
\def\ChEsm{\gs{}}
\def\bChEsm{\bar{\gs}{}}
\def\ChEsminf{\ChEsm^{\infty}}
\def\bChEsminf{\bChEsm^{\infty}}
\def\oppsi{\psi}
\def\boppsi{\bar{\psi}}

\def\Curvinf{\Curv^\infty}
\def\GaugeFinf{\GaugeF^\infty}
\def\WeylFinf{\WeylF^\infty}
\def\bWeylFinf{\bWeylF^\infty}
\def\adsminf{\adsm^\infty}
\def\twsminf{\twsm^\infty}
\def\btwsminf{\btwsm^\infty}

\def\bXi{\bar{\Xi}}

\begin{document}

\title{Bosonic Fradkin-Tseytlin equations unfolded.}

\author{O.V.~Shaynkman}

\maketitle
\vspace{-20pt}
\begin{center}
{\small\it I.E.Tamm Theory Department, Lebedev Physical Institute, Leninski
prospect 53,\\ 119991, Moscow, Russia}
\end{center}
\begin{abstract}
We test infinite-dimensional extension of algebra $\asu(k,k)$ proposed by Fradkin and Linetsky as the candidate for conformal higher spin algebra. 
Adjoint and twisted-adjoint
representations of $\asu(k,k)$ on the space of this algebra are carefully explored. For $k=2$ corresponding unfolded system is analyzed and
it is shown to encode
Fradkin-Tseytlin equations for the set of all integer spins $1,2,\ldots$ with infinite multiplicity. 
\end{abstract}

\footnotetext{\scriptsize{\tt e-mail: shayn@lpi.ru}}

\section{Introduction\label{Introduction}}
In this paper we study unfolded formulation of Fradkin-Tseytlin equations \cite{Fradkin_Tseytlin}
\be\label{Fr_Ts_eq}
&\Pi\underbrace{\ptl_\mu\cdots\ptl_\mu}_s\phi_{\nu(s)}=C_{\nu(s),\mu(s)}\,,\\
&\underbrace{\ptl^\mu\cdots\ptl^\mu}_s C_{\nu(s),\mu(s)}=0\,,
\ee
which describe free conformal dynamics of spin $s$ traceless field $\phi_{\nu(s)}$ in 4-dimensional Minkowski space.
Here $C_{\nu(s),\mu(s)}$ is
associated with traceless generalized Weyl tensor separately symmetric with respect to group of indices $\mu$ and $\nu$ and such that symmetrization
with respect to any $s+1$ indices vanishes, $\Pi$ is a projector that carries out necessary symmetrizations and subtracts traces.
Generalized Weyl tensor $C_{\nu(s),\mu(s)}$ is obviously invariant with respect to gauge transformations
\be\label{Fr_Ts_gauge_transform}
\gd \phi_{\nu(s)}=\ptl_\nu\gep_{\nu(s-1)}-\frac{s-1}{2s}\eta_{\nu\nu}\ptl^\mu\gep_{\mu\nu(s-2)}
\ee
with traceless gauge parameter $\gep_{\nu(s-1)}$ and Minkowski metric $\eta_{\nu\nu}$.

If full nonlinear conformal higher spin theory exists, these equations should correspond to its free level.
As nonlinear $AdS$ higher spin theory
teaches us, the main ingredient needed to construct such kind of theories is higher spin algebra that describes gauge symmetries of the theory.
In paper \cite{Fr_Lin_conf_HS_alg} Fradkin and Linetsky proposed a number of candidates for the role of infinite-dimensional $4d$ conformal
higher spin gauge symmetry algebra, which extends ordinary $4d$ conformal algebra\footnote{Strictly speaking in \cite{Fr_Lin_conf_HS_alg}
an extension of superconformal algebra $\asu(2,2|N)$ was considered, but we do not treat super case in the present paper.} $\aso(4,2)\sim\asu(2,2)$.
Their construction is based on the oscillator realization of $\asu(2,2)$
\cite{Gunaydin, Bars_Gunaydin}. Here we give a straightforward generalization of their results for the case of algebra $\asu(k,k)$ with $k\geq 2$ and
briefly discuss the structure of the infinite-dimensional algebras obtained.

Consider star product algebra generated by bosonic oscillators
\begin{equation}\label{oscillators}
    [b_\gb,a^\ga]_*=\gd^\ga_\gb\,, \quad [\bb_\db,\ba^\da]_*=\gd^\da_\db\,,\quad \ga,\da=1,\ldots ,k\,, \quad k\geq 2\,.
\end{equation}
Here $*$ denotes the star product of Weyl ordered symbols of operators $f(a,b,\ba,\bb)$ and $g(a,b,\ba,\bb)$ given by formula
$f*g=f\exp(\overleftrightarrow{\Delta})g\,,$ where
\begin{equation}\label{star_product_Delta}
  \overleftrightarrow{\Delta}=\half\left(\overleftarrow{\dpd{b}}\cdot\overrightarrow{\dpd{a}}-
                                         \overleftarrow{\dpd{a}}\cdot\overrightarrow{\dpd{b}}+
                                         \overleftarrow{\dpd{\bb}}\cdot\overrightarrow{\dpd{\ba}}-
                                         \overleftarrow{\dpd{\ba}}\cdot\overrightarrow{\dpd{\bb}}\right)\,,
\end{equation}
$\overleftarrow{\dpd{}}\,,\overrightarrow{\dpd{}}$ denote correspondingly the left and the right derivatives and
$\cdot$ denotes contractions
\begin{equation}\label{dif_cdot}
  a\cdot b=a^\ga b_\ga\,,\quad\ba\cdot \bb=\ba^\da\bb_\da\,,\quad
  \dpd{a}\cdot\dpd{b}=\dpd{a^\ga}\dpd{b_\ga}\,,\quad\dpd{\ba}\cdot\dpd{\bb}=\dpd{\ba^\da}\dpd{\bb_\da}\,.
\end{equation}

Within this setup bilinears of oscillators \eqref{oscillators} centralized by helicity operator
\begin{equation}\label{particle_number_operator}
  \PNO=\frac{i}{2}(a\cdot b-\ba\cdot\bb)
\end{equation}
furnish algebra $\agl(2k,\oC)$ with respect to commutator
\begin{equation}\label{commutator}
  [f,g]_*=f*g-g*f\,.
\end{equation}

The basis of $\agl(2k,\oC)$ along with $\PNO$ contains elements
\begin{equation}\label{su_symbols}
  \begin{aligned}
    & \uL_\ga{}^\gb=b_\ga a^\gb-\frac{1}{k}\gd_\ga^\gb a\cdot b\,,
    &&\buL_\da{}^\db=\bb_\da \ba^\db-\frac{1}{k}\gd_\da^\db \ba\cdot \bb\,, \\
    & \uP_{\ga\db}=b_\ga\bb_\db\,, && \uK^{\ga\db}=a^\ga\ba^\db\,, \\
    & \uD=\half(a\cdot b+\ba\cdot\bb)\,,
  \end{aligned}
\end{equation}
which satisfy the following commutation relations
\begin{equation}\label{sukk_commut_rel}
  \begin{aligned}
    &[\uL_\ga{}^\gb,\uL_\gga{}^\gd]_*=\gd_\ga^\gd\uL_\gga{}^\gb-\gd_\gga^\gb\uL_\ga{}^\gd\,,
    &&[\buL_\da{}^\db,\buL_\dg{}^\dd]_*=\gd_\da^\dd\buL_\dg{}^\db-\gd_\dg^\db\buL_\da{}^\dd\,,
    \\
    &[\uL_\ga{}^\gb,\uP_{\gga\dd}]_*=-\gd_\gga^\gb\uP_{\ga\dd}+\frac{1}{k}\gd_\ga^\gb\uP_{\gga\dd}\,,
    &&[\buL_\da{}^\db,\uP_{\gga\dd}]_*=-\gd_\dd^\db\uP_{\gga\da}+\frac{1}{k}\gd_\da^\db\uP_{\gga\dd}\,,
    \\
    &[\uL_\ga{}^\gb,\uK^{\gga\dd}]_*=\gd_\ga^\gga\uK^{\gb\dd}-\frac{1}{k}\gd_\ga^\gb\uK^{\gga\dd}\,,
    &&[\buL_\da{}^\db,\uK^{\gga\dd}]_*=\gd_\da^\dd\uK^{\gga\db}-\frac{1}{k}\gd_\da^\db\uK^{\gga\dd}\,,
    \\
    &[\uD,\uP_{\ga\db}]_*=-\uP_{\ga\db}\,,
    &&[\uD,\uK^{\ga\db}]_*=\uK^{\ga\db}\,,
    \\
    &[\uP_{\ga\db},\uK^{\gga\dd}]_*=\gd_\ga^\gga\buL_\db{}^\dd+\gd_\db^\dd\uL_\ga{}^\gga+
        \frac{2}{k}\gd_\ga^\gga\gd_\db^\dd\uD\,.
  \end{aligned}
\end{equation}

Consider the involution $\dag$ of the star product algebra defined by
\be\label{involution}
& (a^\ga)^\dag=i\ba^\da\,,\quad (b_\ga)^\dag=i\bb_\da\,,\quad (\ba^\da)^\dag=ia^\ga\,,\quad (\bb_\da)^\dag=ib_\ga\,,\\
& f^\dag(a,b,\ba,\bb)=\bar{f}(a^\dag,b^\dag,\ba^\dag,\bb^\dag)\,.
\ee
Real form of $\agl(2k,\oC)$ singled out by the requirement
\begin{equation}\label{reality_condition}
\compconj(f)=f\,,\quad\mbox{where $\compconj(f)=-f^\dag$}
\end{equation}
is identified with algebra $\au(k,k)$.
General element of $\au(k,k)$ has, thus, form
\begin{equation}\label{su_gen_el}
  \uX=X^\ga{}_\gb\uL_\ga{}^\gb+\bar{X}^\da{}_\db\buL_\da{}^\db+
  X^{\ga\db}\uP_{\ga\db}+X_{\ga\db}\uK^{\ga\db}+X\uD+X'\PNO\,,
\end{equation}
where $X^\ga{}_\gb$ and $\bar{X}^\da{}_\db$ are mutually complex conjugate $k\times k$-matrices,
$X^{\ga\db}$ and $X_{\ga\db}$ are Hermitian $k\times k$-matrices and $X$, $X'$ are real numbers.
Algebra $\au(k,k)$ decomposes into direct sum
\begin{equation}\label{ukk_decomp}
  \au(k,k)=\asu(k,k)\oplus\au(1)\,,
\end{equation}
where $\au(1)$ is spanned by $\PNO$.

To construct an infinite-dimensional extension of $\au(k,k)$ let us bring all polynomials (not only bilinear) of oscillators \eqref{oscillators}
into the play still requiring them to be centralized by $\PNO$
\begin{align}
  \label{iukk_gen_el_central_rel}
  &[\PNO,f]_*=\frac{i}{2}(\na-\nba-\nb+\nbb)f=0
\end{align}
and to satisfy reality condition \eqref{reality_condition}. 
Corresponding Lie algebra with respect to commutator \eqref{commutator} was called $\aiu(k,k)$
in \cite{Fr_Lin_conf_HS_alg}, where letter i means infinite.

Decomposing general element of $\aiu(k,k)$ into a sum of traceless components multiplied by powers of $a\cdot b$ and $\ba\cdot\bb$
and taking into account that $a\cdot b=\uD-i\PNO$, $\ba\cdot \bb=\uD+i\PNO$
one gets
\begin{equation}\label{iukk_gen_el}
  f=\sum_{u,v=0}^\infty i^{u+v}\PNO^u\uD^vf_{u,v}(a,b,\ba,\bb)\,.
\end{equation}
Here $f_{u,v}$ is traceless with respect to $a\,, b$ and $\ba\,, \bb$, i.e.
\begin{equation}\label{iukk_trless_cond}
  \ddpddot{a}{b}f_{u,v}=0\,,\quad \ddpddot{\ba}{\bb}f_{u,v}=0.
\end{equation}
Also note that due to \eqref{iukk_gen_el_central_rel} $f_{u,v}$ is even function $f_{u,v}(a,b,\ba,\bb)=f_{u,v}(-a,-b,-\ba,-\bb)$.


Algebra $\aiu(k,k)$ contains an infinite chain of ideals $\ideal^m$
\begin{equation}\label{isukk_ideals_embedded}
  \aiu(k,k)\supset\ideal^1\supset\ideal^2\supset\cdots\supset\ideal^m\supset\cdots\,.
\end{equation}
Here ideal $\ideal^m$ is spanned by the elements of the form
\begin{equation}\label{ideal_n}
\ideal^m :\qquad  \underbrace{\PNO*\cdots *\PNO}_u*A=\Big(\PNO-\frac{i}{8}(\ddpd{a\cdot}{b}-\ddpd{\ba\cdot}{\bb})\Big)^uA=
  \PNO^u A+\cdots\,,\quad u\geq m\,,
\end{equation}
where dots on the right-hand side denote the lower power terms.

Let quotient algebras $\aiu(k,k)/\ideal^m$ be denoted as $\aisun{m-1}(k,k)$
\begin{equation}\label{isunkk_embeddings}
  \aisun{0}(k,k)\subset\aisun{1}(k,k)\subset\cdots\subset\aisun{m}(k,k)\subset\cdots\subset\aiu(k,k)\,.
\end{equation}
Algebra $\aisun{0}(k,k)$
is semi-simple, in what follows we omit index 0 and denote it as $\aisu(k,k)$.

Let us note that in paper \cite{Fr_Lin_conf_HS_alg} algebra $\aisu(2,2)$ was denoted as $\ahsc$,
where hsc means higher spin conformal and 4 indicates that it extends 4-dimensional conformal algebra. In the later paper \cite{Vasiliev_conf_HS_sym in _four_dimm}
algebra $\aiu(2,2)$ was denoted as $\acu$ and algebra $\aisu(2,2)$ was denoted as $\ahuzero$,
where 8 indicates the number of oscillators used and pair 1,0 points out that above algebras have
trivial structure in spin 1 Yang-Mills sector.

In the present paper we analyze unfolded system corresponding to algebra $\aiu(2,2)$ and show that it describes a collection of Fradkin-Tseytlin equations
that corresponds to all bosonic spins with infinite degeneracy. Let us note that some other approaches
to Fradkin-Tseytlin equations were suggested in papers \cite{Segal}, \cite{Bekaert_Grigoriev}.

The rest of the paper is organized as follows. In section \ref{section_Unf_formul_preliminary}
we recall some relevant facts about unfolded formulation.
Structure of algebra's $\asu(k,k)$ adjoint representations on the vector space of $\aiu(k,k)$
is discussed in section \ref{section_Adjoint_module}. In section \ref{section_Twist_adjoint_module}
we study twisted-adjoint representation of $\asu(k,k)$. In section \ref{section_Fr_Ts_eq_unfolded} unfolded formulation of conformal higher spin bosonic
equations is analyzed for $k=2$.
Section \ref{section_Conclusion} contains conclusions.
In Appendix \ref{section_slk_plus_slk_irreps} we recall relevant
facts concerning finite-dimensional $\asl(k)\oplus\asl(k)$ irreps. In Appendix \ref{section_Basis_of_direct_decompos} we find
basises where adjoint and twisted-adjoint modules from sections \ref{section_Adjoint_module} and
\ref{section_Twist_adjoint_module} decompose into submodules. In appendix \ref{section_sigma_min_cohomology}
$\adsm$ and $\twsm$-cohomology corresponding to the gauge sector and Weyl sector of unfolded systems under consideration are found.

\section{Unfolded formulation: preliminary remarks\label{section_Unf_formul_preliminary}}
Let $\cM^d$ be some $d$-dimensional manifold with coordinates $x^1,\ldots,x^d$.
Any dynamical system on $\cM^d$ can be reformulated in unfolded form of the first order differential equations 
\cite{Vasiliev_unfolded_repr_for_eq_in_two_plus_one_AdS}
(see \cite{BCIV_review} for a review)
\be\label{UF_gen}
\extdiff W^\Go=F^\Go(W)\,.
\ee
Here $W^\Go(x)$ is a collection of differential forms (numerated by multiindex $\Go$) of ranks $\deg(W^\Go)=\formrank{\Go}$, $\extdiff$ is exterior differential 
and
\be\label{UF_F}
F^\Go(W)=\sum_{n=0}^{\infty}f^\Go_{\Phi_1\ldots\Phi_n}W^{\Phi_1}\cdots W^{\Phi_n}
\ee
is a $\formrank{\Go}+1$ rank form. $F^\Go(W)$ is composed
from elements of $W^\Go(x)$, which are multiplied by virtue of exterior product\footnote{In this paper all products of
differential forms are supposed to be exterior and we omit the designation of exterior product $\wedge$ in formulae.} and are contracted with constant functions
\be\label{UF_gen_struct_const}
f^\Go_{\ldots\Phi\Psi\ldots}=(-1)^{\formrank{\Phi}\formrank{\Psi}}f^\Go_{\ldots\Psi\Phi\ldots}\,.
\ee

Compatibility conditions of \eqref{UF_gen} require $F^\Go(W)$ to satisfy identities
\be\label{UF_compat_cond}
F^\Phi\frac{\gd}{\gd W^\Phi}F^\Go\equiv 0\,,
\ee
where $\frac{\gd}{\gd W^\Phi}$ is the left derivative. In terms of constants $f^\Go_{\Phi_1\ldots\Phi_n}$ conditions \eqref{UF_compat_cond}
have a form of generalized Jacobi identities
\be\label{UF_gen_Jacobi_ident}
\sum_{m=0}^\infty\sum_{n=1}^\infty nf^\Theta_{[\Psi_1\ldots\Psi_m}f^\Go_{\Theta\Phi_1\ldots\Phi_{n-1}\}}\equiv 0\,,
\ee
where the left-hand side of \eqref{UF_gen_Jacobi_ident} is (anti)symmetrised according to \eqref{UF_gen_struct_const}.
Any solution of \eqref{UF_gen_Jacobi_ident} defines a free differential algebra (FDA) \cite{Sullivan_FDA}. In what follows we
assume that \eqref{UF_gen_Jacobi_ident} holds independently of the
value of space-time dimension $d$. In this case FDA defined by \eqref{UF_gen_Jacobi_ident} is called universal.

Unfolded system \eqref{UF_gen} corresponding to universal FDA is invariant with respect to gauge transformations
\be\label{UF_gauge_transform}
\gd W^\Go=\extdiff \gep^\Go+\gep^\Phi\frac{\gd}{\gd W^\Phi}F^\Go(W)\,,
\ee
where $\gep^\Go(x)$ are $\formrank{\Go}-1$-form gauge parameters.

Let us analyze system \eqref{UF_gen} perturbatively assuming that fields of zeroth order form a subclass of 1-forms $W^A(x)\subseteq W^\Go(x)$.
The most general form of $F^A(W)$ in the sector of zero order fields is
\be\label{UF_F_zero_oder}
F^A(W)=-\frac{1}{2}f^A_{BC}W^BW^C\,,
\ee
where constants $f^A_{BC}=-f^A_{CB}$ due to \eqref{UF_gen_Jacobi_ident} are required to satisfy ordinary Jacobi identities.
Therefore $W^A(x)$ can be identified with connection 1-form taking values in some Lie algebra $\agen$ with structure constants $f^A_{BC}$
and system \eqref{UF_gen} reduces to the zero curvature condition
\be\label{UF_zero_curv_cond}
\extdiff W^A+\frac{1}{2}f^A_{BC}W^BW^C=0\,.
\ee
Gauge transformations \eqref{UF_gauge_transform} become usual gauge transformation of a connection 1-form  in this case
\be\label{UF_gauge_transform_connect_one_form}
\gd W^A=\extdiff \gep^A+f^A_{BC}W^B\gep^C\,,
\ee
where $\gep^C(x)$ is 0-form gauge parameter.

Let us treat all other fields from $W^\Go(x)$ as fluctuations of $W^A(x)$. For our purposes it is sufficient to consider
the case when $W^\Go(x)$ consists of 1-forms $\go^a(x)$ and 0-forms $C^i(x)$ only (the general case
is considered in \cite{Vasiliev_on_conf_sl_four_and_sp_eight_symm}). System \eqref{UF_gen} linearized over $W^A(x)$ reduces to
\begin{align}
\label{UF_lin_gauge}
&(D\go)^a=\frac{1}{2}W^AW^BH_{AB}{}^a{}_jC^j\,,&&\mbox{where }(D\go)^a=\extdiff \go^a+W^A(T_A)^a{}_b\go^b\,,\\
\label{UF_lin_Weyl}
&(\tilde{D}C)^i=0\,,&&\mbox{where }(\tilde{D}C)^i=\extdiff C^i+W^A(\tilde{T}_A)^i{}_jC^j.
\end{align}
As one can readily see compatibility conditions \eqref{UF_gen_Jacobi_ident} require matrices $(T_A)^a{}_b$ and $(\tilde{T}_A)^i{}_j$
to form some representations of Lie algebra $\agen$. Let corresponding modules be denoted as $\cM$ and $\tilde{\cM}$\,.
Then $D$ and $\tilde{D}$ from \eqref{UF_lin_gauge} and \eqref{UF_lin_Weyl} define $\cM$ and $\tilde{\cM}$-covariant derivatives respectively.
Both derivatives are nilpotent
\be\label{UF_covar_der_nilpotent}
D^2=0\,,\qquad\tilde{D}^2=0
\ee
as a consequence of zero curvature condition \eqref{UF_zero_curv_cond}.

As it was argued in \cite{Vasiliev_on_conf_sl_four_and_sp_eight_symm} the term
\be\label{UF_Chev_Eilen_term}
\frac{1}{2}W^AW^BH_{AB}{}^a{}_j
\ee
on the right-hand side of \eqref{UF_lin_gauge} should belong to nontrivial class of 2-nd Chevalley-Eilenberg cohomology taking
values in $\agen$-module $\cM\otimes\tilde{\cM}^*$. Indeed, compatibility of \eqref{UF_lin_gauge} in the sector of $C^i(x)$ is equivalent to the closedness of
\eqref{UF_Chev_Eilen_term}
\be\label{UF_Chev_Eilen_cocycle}
(\gd_{ChE})^a{}_b{}^i{}_j \Big(W^AW^BH_{AB}{}^b{}_i\Big)=0
\ee
with respect to Chevalley-Eilenberg differential (see \cite{Chevalley_Eilenberg_cohom_theor_of_Lie_groups_and_Lie_alg})
\be\label{UF_Chev_Eilen_differential}
&(\gd_{ChE})^a{}_b{}^i{}_j=\frac{1}{2}\gd^a_b\gd^i_j f^A_{BC}W^BW^C\frac{\gd}{\gd W^A}-W^A(T\otimes\tilde{T}^*)_A{}^a{}_b{}^i{}_j\,,\\
&\gd_{ChE}{}^2=0\,.
\ee
Here
\be\label{UF_M_times_tildeM_repr}
(T\otimes\tilde{T}^*)_C{}^a{}_b{}^i{}_j=(T_C)^a{}_b\gd^i_j-\gd^a_b(\tilde{T}_C)^i{}_j
\ee
are matrices acting on module $\cM\otimes\tilde{\cM}^*$.
If \eqref{UF_Chev_Eilen_term} is $\gd_{ChE}$-exact
\be\label{UF_Chev_Eilen_coboundery}
W^AW^BH_{AB}{}^a{}_i=(\gd_{ChE})^a{}_b{}^j{}_i\Big(W^A\theta_A{}^b{}_j\Big)\,,
\ee
the right-hand side of \eqref{UF_lin_gauge} can be removed by the field redefinition
\be\label{UF_field_redifinition}
\go^a=\go'^a-\frac{1}{2}W^A\theta_A{}^a{}_iC^i\,.
\ee
And conversely if some field redefinition removing the right-hand side of \eqref{UF_lin_gauge} exists, it should necessarily have form
\eqref{UF_field_redifinition} with $W^A\theta_A{}^a{}_i$ satisfying \eqref{UF_Chev_Eilen_coboundery}.

System \eqref{UF_zero_curv_cond}, \eqref{UF_lin_gauge} and \eqref{UF_lin_Weyl} is locally invariant with respect to gauge
transformation\footnote{There are also gauge transformations with parameters associated to $\go^a$, which are discussed later
(see \eqref{UF_gauge_transform_lin_gauge_vacuum_connection}).}
\eqref{UF_gauge_transform_connect_one_form} of connection 1-form $W^A(x)$ and the following gauge transformations of fields $\go^a(x)$ and $C^i(x)$
\be
\label{UF_gauge_transform_lin_gauge_Weyl}
&\gd\go^a=-\gep^A(T_A)^a{}_b\go^b+\gep^AW^BH_{AB}{}^a{}_iC^i\,,\\
&\gd C^i=-\gep^A(\tilde{T}_A)^i{}_jC^j\,.
\ee
If some solution
\be\label{UF_background_connection}
W^A=W_0^A
\ee
of zero curvature condition \eqref{UF_zero_curv_cond} is fixed, the gauge symmetry above
breaks down to the global symmetry that keeps $W_0^A$ stable. Parameter of this symmetry $\gep_0^A(x)$ should obviously satisfy equation
\be\label{UF_global_symm_parameter}
\gd W_0^A=\extdiff \gep_0^A+f^A_{BC}W_0^B\gep_0^C=0\,,
\ee
which is consistent due to zero curvature condition \eqref{UF_zero_curv_cond}. Equation \eqref{UF_global_symm_parameter} reconstructs $\gep_0^A(x)$
in terms of its value $\gep_0^A(x_0)$ at any given point $x_0$. So $\gep_0^A(x_0)$ plays a role of the moduli space of $W_0^A$ global symmetry algebra,
which therefore can be identified with $\agen$. When substituted to \eqref{UF_lin_gauge}, \eqref{UF_lin_Weyl},
$W_0^A$ plays a role of vacuum connection describing $\agen$-invariant background geometry.
We only require component of $W_0^A$ corresponding to the generator of generalized translation (i.e. generalized coframe) to be
of maximal possible rank.

Let us consider system \eqref{UF_lin_gauge}, \eqref{UF_lin_Weyl} with $W^A=W_0^A$ substituted.
\begin{align}
\label{UF_lin_gauge_vacuum_connection}
&(D_0\go)^a=\frac{1}{2}W_0^AW_0^BH_{AB}{}^a{}_jC^j\,,&&\mbox{where }(D_0\go)^a=\extdiff \go^a+W_0^A(T_A)^a{}_b\go^b\,,\\
\label{UF_lin_Weyl_vacuum_connection}
&(\tilde{D}_0C)^i=0\,,&&\mbox{where }(\tilde{D}_0C)^i=\extdiff C^i+W_0^A(\tilde{T}_A)^i{}_jC^j.
\end{align}
As follows from above consideration it is globally $\agen$-invariant with
respect to transformations \eqref{UF_gauge_transform_lin_gauge_Weyl}
with $W^A=W_0^A$ and $\gep^A=\gep_0^A$ substituted. This system is also gauge invariant with respect to gauge transformations
\be\label{UF_gauge_transform_lin_gauge_vacuum_connection}
\gd\go^a=D_0\gep^a\,,
\ee
where $\gep^a(x)$ is 0-form gauge parameter associated to field $\go^a(x)$.

To analyze dynamical content of system \eqref{UF_lin_gauge_vacuum_connection}, \eqref{UF_lin_Weyl_vacuum_connection}
let us first suppose that the right-hand side of \eqref{UF_lin_gauge_vacuum_connection} is zero.
In this case equations \eqref{UF_lin_gauge_vacuum_connection}, \eqref{UF_lin_Weyl_vacuum_connection}
are independent and both have form of covariant constancy conditions. Suppose that modules $\cM$ and $\tilde{\cM}$ are graded with grading
bounded from below. Decompose covariant derivatives \eqref{UF_lin_gauge_vacuum_connection}, \eqref{UF_lin_Weyl_vacuum_connection}
into the summands with definite gradings.
We assume that each covariant derivative contains a single operator of negative grading
(the case when there are several operators with negative gradings was considered in \cite{Ponomarev_Vasiliev_Unfolded_scalar_supermultiplet})
\be\label{UF_covariant_deriv_decomp}
D_0=\cD_0+\gs_-+\sum_\eta\gs^\eta_+\,,\qquad\tilde{D}_0=\tilde{\cD}_0+\tilde{\gs}_-+\sum_\theta\tilde{\gs}^\theta_+\,.
\ee
Here $\cD_0$, $\tilde{\cD}_0$ denote operators of zero grading which include exterior differential,
$\gs^\eta_+$, $\tilde{\gs}^\theta_+$ denote purely algebraic operators of various positive gradings and $\gs_-$, $\tilde{\gs}_-$ are purely algebraic
operators of negative grading. Operators $\gs_-$ and $\tilde{\gs}_-$ are nilpotent due to the nilpotency of covariant derivatives \eqref{UF_covar_der_nilpotent}.

Let subspace of $\cM$ with fixed grading $n$ be called the $n$-th level of $\cM$.
Analyzing equation \eqref{UF_lin_gauge_vacuum_connection} and its gauge symmetries
\eqref{UF_gauge_transform_lin_gauge_vacuum_connection} level by level starting from the lowest grading one can see
\cite{Shaynkman_Vasiliev_Scalar_field_from_the_HST_perspective} that those fields which are not $\gs_-$-closed (they are called auxiliary fields) are expressed
by \eqref{UF_lin_gauge_vacuum_connection} as a derivatives of the lower level fields. Here space-time indices of derivatives are converted into
algebraic indices by virtue of coframe. $\gs_-$-exact fields can be gauged to zero with the use of
Stueckelberg part of gauge symmetry transformations \eqref{UF_gauge_transform_lin_gauge_vacuum_connection}. Leftover fields (that are called
dynamical fields) belong to $H^1_{\gs_-}$ the 1-st cohomology of $\gs_-$. We also get that differential gauge parameters (i.e. those that do not
correspond to Stueckelberg gauge symmetry) belong to $H^0_{\gs_-}$.

Let $E_n$ denote the left-hand side of \eqref{UF_lin_gauge_vacuum_connection} corresponding to the $n$-th level.
Suppose equation $E_m=0$ is solved up to the $n-1$-st level inclusive, which means that all auxiliary fields up to
the $n$-th level properly expressed in terms of derivatives of dynamical fields.
Bianchi identities
\be\label{UF_Bianchi_identities}
D_0D_0\go\equiv 0
\ee
on the $n-1$-st level require $E_n$ to be $\gs_-$-closed. If $H^2_{\gs_-}$ the 2-nd cohomology of $\gs_-$ is trivial on the $n$-th level, equation
$E_n=0$ can be satisfied by appropriate choice of auxiliary field on the $n+1$-st level.
In other case $E_n=0$ also imposes some differential restriction on dynamical fields requiring that $E_n$ belongs to the trivial cohomology class.
Therefore nontrivial differential equations on dynamical fields are in one-to-one correspondence with $H^2_{\gs_-}$. Moreover,
if $\asubgen\subset\agen$ is a subalgebra of $\agen$ that acts horizontally (i.e. keeps levels invariant), differential equations imposed by
\eqref{UF_lin_gauge_vacuum_connection} and $H^2_{\gs_-}$ are isomorphic as $\asubgen$-modules.

Summarizing, the dynamical content of equation \eqref{UF_lin_gauge_vacuum_connection} with the zero right-hand side is described by
$H^0_{\gs_-}$, $H^1_{\gs_-}$, $H^2_{\gs_-}$ which correspond to differential gauge parameters, dynamical fields and differential equations on the dynamical
fields respectively. Analogously for equation \eqref{UF_lin_Weyl_vacuum_connection} the dynamical fields and differential equations
correspond to $\tilde{H}^0_{\tilde{\gs}_-}$ and $\tilde{H}^1_{\tilde{\gs}_-}$.

To analyze system \eqref{UF_lin_gauge_vacuum_connection}, \eqref{UF_lin_Weyl_vacuum_connection} with nonzero right-hand side let us consider
operator
\be\label{UF_D_hat}
\hat{D}_0=\hat{\cD}_0+\hat{\gs}_-+\hat{\gs}_+\,,
\ee
where
\be\label{UF_D_hat_z_m_p}
&\hat{\cD}_0=\cD_0+\tilde{\cD}_0\,,\\
&\hat{\gs}_-=\gs_-+\tilde{\gs}_-+\gs\,,\qquad (\gs)^a{}_i=-\frac{1}{2}W_0^AW_0^BH_{AB}{}^a{}_j\,,\\
&\hat{\gs}_+=\sum_\eta\gs^\eta_++\sum_\theta\tilde{\gs}^\theta_+\,.
\ee
In these notation system \eqref{UF_lin_gauge_vacuum_connection}, \eqref{UF_lin_Weyl_vacuum_connection} can be rewritten in the following form
\be\label{UF_lin_hat}
\hat{D}_0\hat{\Psi}=0\,,
\ee
where new field $\hat{\Psi}$ is a pair $\hat{\Psi}=(\go, C)$ incorporating 1-forms $\go$ and 0-forms $C$. Here all operators
are extended by zero
on the spaces where they undefined. Operator $\hat{D}_0$ is nilpotent due to compatibility conditions of system
\eqref{UF_lin_gauge_vacuum_connection}, \eqref{UF_lin_Weyl_vacuum_connection}. Gauge transformations \eqref{UF_gauge_transform_lin_gauge_vacuum_connection}
take form
\be\label{UF_D_hat_gauge_transform}
\gd\hat{\Psi}=\hat{D}_0\hat{\Upsilon}\,,
\ee
where $\hat{\Upsilon}=(\gep,0)$.

Let us consider $\hat{\gs}_-$-cocomplex $\hat{\mathfrak{C}}=(\hat{\cS},\hat{\gs}_-)$ with $p$-form element $\hat{\Psi}^p\in\hat{\cS}$ defined as a pair
$\hat{\Psi}^p=(\go^p,C^{p-1})$, where $\go^p$ is $p$-form taking values in module $\cM$ and $C^{p-1}$ is $p-1$-form
taking values in module $\tilde{\cM}$ ($C^{-1}\equiv 0$). Standard definition of $\hat{\gs}_-$-closed $p$-forms subspace
$\hat{\cC}^p=(\go^p_\cC,C^{p-1}_\cC)$ gives in components the following relations
\be\label{UF_gs_hat_closed_space_components}
&\gs_-\go^p_\cC+\gs C^{p-1}_\cC=0\,,\\
&\tilde{\gs}_-C^{p-1}_\cC=0\,.
\ee
Subspace of $\hat{\gs}_-$-exact $p$-forms $\hat{\cE}^p=(\go^p_\cE,C^{p-1}_\cE)$ is defined in components by
\be\label{UF_gs_hat_exact_space_components}
&\go^p_\cE=\gs_-\go^{p-1}+\gs C^{p-2}\,,\\
&C^{p-1}_\cE=\tilde{\gs}_-C^{p-2}
\ee
for some elements $\go^{p-1}$, $C^{p-2}$. Let $p$-th $\hat{\gs}_-$-cohomology be defined as quotient
\be
\label{UF_gs_hat_cohomology_definition}
\hat{H}^p_{\hat{\gs}_-}=\hat{\cC}^p/\hat{\cE}^p\,.
\ee

Since above analysis of equation \eqref{UF_lin_gauge_vacuum_connection} with the zero right-hand side is based on Bianchi identities\footnote{In fact
we also required $D$ to have unique $\gs_-$ and grading to be bounded from below which is obviously also true for \eqref{UF_lin_hat}.}
\eqref{UF_Bianchi_identities} only it is applicable to equation \eqref{UF_lin_hat}.
We therefore obtain that dynamical content of equation \eqref{UF_lin_hat} (or equivalently of system
\eqref{UF_lin_gauge_vacuum_connection}, \eqref{UF_lin_Weyl_vacuum_connection} with nonzero right-hand side) is defined by $\hat{H}_{\hat{\gs}_-}$.
Namely:
\begin{itemize}
  \item differential gauge parameters are given by $\hat{H}^0_{\hat{\gs}_-}$;
  \item dynamical fields are given by $\hat{H}^1_{\hat{\gs}_-}$;
  \item differential equations on dynamical fields are in one-to-one correspondence with
        $\hat{H}^2_{\hat{\gs}_-}$.
\end{itemize}

In section \ref{section_Fr_Ts_eq_unfolded} we use above technic to analyze dynamical content of $\asu(2,2)$-invariant unfolded system
that was originally introduced in \cite{Vasiliev_progr_in_HS_gauge_theory}. Before it we explore structure of underlying $\asu(k,k)$-modules.

\section{Structure of adjoint module\label{section_Adjoint_module}}
Consider adjoint action of algebra $\asu(k,k)$ on the vector space of algebra $\aiu(k,k)$, which is given by commutators
$\adX=[\uX,\cdot]=2\uX\overleftrightarrow{\Delta}$. We have
\begin{equation}\label{adinf_repr_ukk}
  \begin{aligned}
    &\adinfL_\ga{}^\gb=a^\gb\dpd{a^\ga}-b_\ga\dpd{b_\gb}-\frac{1}{k}\gd_\ga^\gb(\na-\nb)\,,\\
    &\adinfbL_\da{}^\db=\ba^\db\dpd{\ba^\da}-\bb_\da\dpd{\bb_\db}-\frac{1}{k}\gd_\da^\db(\nba-\nbb)\,,\\
    &\adinfP_{\ga\db}=b_\ga\dpd{\ba^\db}+\bb_\db\dpd{a^\ga}\,,\quad
    \adinfK^{\ga\db}=-a^\ga\dpd{\bb_\db}-\ba^\db\dpd{b_\ga}\,,\\
    &\adinfD=\frac{1}{2}(\na+\nba-\nb-\nbb)\,,\quad
    \adinfPNO=\frac{i}{2}(\na-\nba-\nb+\nbb)\,,
  \end{aligned}
\end{equation}
where $\na\,, \nb\,, \nba\,, \nbb$ denote Euler operators counting the number of corresponding variables.
Let $\admoduleinf$ denote corresponding $\asu(k,k)$-module.

Obviously, operators
\be\label{sinf_one_sinf_two}
\opsinfo=\na+\nbb+1\,,\qquad \opsinft=\nb+\nba+1
\ee
commute with adjoint action of $\asu(k,k)$ \eqref{adinf_repr_ukk}. Moreover due to centralization requirement
\eqref{iukk_gen_el_central_rel}
\be
\opsinfo f=\opsinft f
\ee
for any $f\in\admoduleinf$.
Therefore module $\admoduleinf$ decomposes into finite-dimensional submodules $\admoduleinfs[s]$
\be\label{admoduleinf_s}
\opsinfo\admoduleinfs[s]=\opsinft\admoduleinfs[s]=s\admoduleinfs[s]\,.
\ee

Modules $\admoduleinf$ and $\admoduleinfs[s]$ are reducible with submodules
\be
&\admoduleinf\supset\ideal^1\supset\cdots\supset\ideal^m\supset\cdots\,,\\
&\admoduleinfs[s]\supset\ideal^1_s\supset\cdots\supset\ideal^{s-1}_s\,,
\ee
where $\ideal^m$ is ideal \eqref{ideal_n} and
\be
\ideal^m_s=\ideal^m\cap\admoduleinfs[s]\,.
\ee
Note that $\ideal^m_s\equiv 0$ for $m\geq s$.

Consider quotient modules
\be\label{adquatient}
&\admodule^0\subset\admodule^1\subset\cdots\subset\admoduleinf\,,\\
&\admodules[s]^0\subset\admodules[s]^1\subset\cdots\subset\admodules[s]^{s-2}\subset\admoduleinfs[s]\,,
\ee
where
\be
\admodule^m=\admoduleinf/\ideal^{m+1}\,,\qquad
\admodules[s]^m=\admoduleinfs[s]/\ideal^{m+1}_s\,,
\qquad m=0,1,\ldots\,.
\ee
Note that $\admodules[s]^m\equiv\admoduleinfs[s]$ for $m\geq s-1$.
In what follows we omit index 0 and denote $\admodule^0$, $\admodules[s]^0$ as $\admodule$, $\admodules[s]$, respectively.

As shown in Appendix \ref{section_admoduleinfs} module $\admoduleinfs[s]$ admits the following decomposition
\be
\label{admodule_inf_s_decomposition}
\admoduleinfs[s]=\oplus_{s'=1}^s\admodules[s']\,.
\ee
And therefore
\be
\label{admodule_n_s_decomposition}
\admodules[s]^m=\oplus_{s'=s-m}^s\admodules[s']\,,\quad m=0,1,\ldots,s-2
\ee
and
\be
\label{admodule_n_decomposition}
\admodule^m=\oplus_{s'=1}^\infty\admodules[s']^m=\oplus_{s'=1}^\infty(m+1)\admodules[s']\,,\quad m=0,1,\ldots,\infty\,,
\ee
where the number $m+1$ on the right-hand side of \eqref{admodule_n_decomposition} indicates multiplicity of modules $\admodules[s']$.

The basis where decomposition \eqref{admodule_inf_s_decomposition} becomes straightforward has form
\be
\label{admodule_inf_s_decomposition_basis}
\adbasis^v_{s,s'}=i^s\PNO^{s-s'}\adbasefunc^v_{s'}(\PNO,\uD)\admonomial_{s'-v}(a,b,\ba,\bb)\,,\qquad s'=1,\ldots,s\,,\quad v=0,\ldots,s'-1\,,
\ee
where subset with the fixed value of $s'$ corresponds to the basis of submodule $\admodules[s']\subset\admoduleinfs[s]$.
Here $\adbasefunc^v_{s'}(\PNO,\uD)$ is homogeneous polynomial of degree $v$ in two variables $\PNO$ and $\uD$,
which particular form is found in Appendix \ref{section_admoduleinfs}. 
It is important to note that in \eqref{admodule_inf_s_decomposition_basis} $\PNO$ and $\uD$ are treated as a new variables independent on oscillators. 
Elements $\admonomial_{s'-v}(a,b,\ba,\bb)$ are traceless (see \eqref{iukk_trless_cond})
eigenvectors of operators
$\opsinfo$ and $\opsinft$ \eqref{sinf_one_sinf_two} corresponding to eigenvalue $s'-v$ 
\be
\label{f_eigenvec}
\opsinfo \admonomial_{s'-v}=\opsinft \admonomial_{s'-v}=(s'-v)\admonomial_{s'-v}\,,
\ee
with some reality conditions discussed later.
In other words $\admonomial_{s'-v}$ are monomials of the form
\be\label{admodule_s_monomial}
\admonomial_{s'-v}(\na,\nb,\nba,\nbb)=x_{\ga(\na)}^{\gb(\nb)}{}_{;\,\da(\nba)}^{;\,\db(\nbb)}a^{\ga(\na)}\ba^{\da(\nba)}b_{\gb(\nb)}\bb_{\db(\nbb)}\,.
\ee
Here $x_{\ga(\na)}^{\gb(\nb)}{}_{;\,\da(\nba)}^{;\,\db(\nbb)}$
are traceless complex tensors symmetric separately with respect to each group of indices
$\ga(\na)\,, \da(\nba)\,, \gb(\nb)\,, \db(\nbb)$, where number
in parentheses indicates the number of indices in the group,
and $a^{\ga(\na)}=a^{\ga_1}\cdots a^{\ga_{\na}}$ denotes $\na$-th power of oscillator $a$ and analogous notation
for oscillators $\ba, b$ and $\bb$. Certainly
values of $\na,\nba,\nb,\nbb$ in \eqref{admodule_s_monomial} should be coordinated with $s'$ and $v$
through formula \eqref{f_eigenvec}.

Due to above arguments $\adbasis^v_{s,s'}$
forms, with respect to generators $\adinfL_\ga{}^\gb$, $\adinfbL_\da{}^\db$,
irreducible $\asl(k)\oplus\asl(k)$-module corresponding to Young tableau
\be\label{admodule_s_monomial_Young_tableau}
\bt{rll}
&\hspace{-2mm} undotted &\hspace{1mm} dotted\\[1mm]
upper\hspace{4mm} & \bt{l}
                    \begin{Young}{3}{1}{\USkip\renewcommand{\ULabelMidVShift}{3}}
                    \Put{\Block{3}{1}}\ULabelMid{\na}
                    \end{Young}
                    \et &
                            \bt{l}
                            \begin{Young}{4}{1}{\USkip\renewcommand{\ULabelMidVShift}{3}}
                            \Put{\Block{4}{1}}\ULabelMid{\nba}
                            \end{Young}
                            \et\\[2mm]
lower\hspace{4mm} & \bt{l}
                    \begin{Young}{5}{1}{\USkip\renewcommand{\ULabelMidVShift}{3}}
                    \Put{\Block{5}{1}}\ULabelMid{\nb}
                    \end{Young}
                    \et &
                            \bt{l}
                            \begin{Young}{6}{1}{\USkip\renewcommand{\ULabelMidVShift}{3}}
                            \Put{\Block{6}{1}}\ULabelMid{\nbb}
                            \end{Young}
                            \et
\et
\ee
(see Appendix \ref{section_slk_plus_slk_irreps} for more details).

In what follows we study the structure of module $\admodules[s']$ and in particular show that it is irreducible.
Elements $\uL_\ga{}^\gb\,,\buL_\da{}^\db\,,\uD\,,\PNO$ of $\au(k,k)$ commute with $\PNO^{s-s'}\adbasefunc^v_{s'}$ and therefore
are represented in $\admodules[s']$ by the same operators as in module $\admoduleinfs[s]$.
As shown in Appendix \ref{section_admoduleinfs} elements $\uP_{\ga\db}$ and $\uK^{\ga\db}$ are represented in $\admodules[s']$ 
by the following operators (up to an overall factor $i^s\PNO^{s-s'}$)
\begin{equation}\label{ad_repr_ukk}
  \begin{aligned}
    &\adP_{\ga\db}\cdot\adbasefunc^v_{s'}\admonomial_{s'-v}=\left[v\adbasefunc^{v-1}_{s'}\trProj b_\ga\bb_\db+
\adbasefunc^v_{s'}\Big(v \func(n)+1\Big)\trProj\bb_\db\dpd{a^\ga}+\adbasefunc^v_{s'}\Big(v \func(\bn)+1\Big)\trProj b_\ga\dpd{\ba^\db}+\right.\\
    &\quad\qquad\qquad\qquad\qquad\left.{}+\Big(2s'+2k-v-4\Big)\adbasefunc^{v+1}_{s'}
\func(n)\func(\bn)\ddpd{a^\ga}{\ba^\db}\right]\admonomial_{s'-v}\,,\\
    &\adK^{\ga\db}\cdot\adbasefunc^v_{s'}\admonomial_{s'-v}=-\left[v\adbasefunc^{v-1}_{s'}\trProj a^\ga\ba^\db+
\adbasefunc^{v}_{s'}\Big(v\func(n)+1\Big)\trProj\ba^\db\dpd{b_\ga}+\adbasefunc^{v}_{s'}\Big(v\func(\bn)+1\Big)\trProj a^\ga\dpd{\bb_\db}+\right.\\
    &\qquad\qquad\qquad\qquad\qquad\left.+\Big(2s'+2k-v-4\Big)\adbasefunc^{v+1}_{s'}
\func(n)\func(\bn)\ddpd{b_\ga}{\bb_\db}\right]\admonomial_{s'-v}\,,
  \end{aligned}
\end{equation}
where $n=\na+\nb$, $\bn=\nba+\nbb$, $\func(n)=1/(n+k)$ and
$\trProj=(\trProj)^2$ is projector to the traceless component \eqref{iukk_trless_cond}
\be\label{trProj_formulas}
&\trProj a^\ga = a^\ga-\,a\cdot b\:\func(n)\dpd{b_\ga}\,,&&
\trProj b_\ga = b_\ga-\,a\cdot b\:\func(n)\dpd{a^\ga}\,,\\
&\trProj \ba^\db = \ba^\db-\,\ba\cdot \bb\:\func(\bn)\dpd{\bb_\db}\,,&&
\trProj \bb_\db = \bb_\db-\,\ba\cdot \bb\:\func(\bn)\dpd{\ba^\db}\,,\\
&\trProj a^\ga\ba^\db = \trProj a^\ga \trProj \ba^\db\,,&&
\trProj b_\ga\bb_\db = \trProj b_\ga \trProj \bb_\db\,.
\ee

Every element $\adbasis^v_{s,s'}$ has a definite conformal weight
\be\label{admodule_conf_weight}
\adD\adbasis^v_{s,s'}=\frac{1}{2}(\na+\nba-\nb-\nbb)\adbasis^v_{s,s'}=\confweight\adbasis^v_{s,s'}\,,
\ee
which due to \eqref{f_eigenvec} ranges for fixed value of $v=0,1,\ldots, s'-1$ from
\be\label{adlowest_conf_w}
\confweight_{\rm min}(v)=-s'+v+1 \mbox{ for }\adbasis^v_{s,s'}{}_{\rm min}=i^s\PNO^{s-s'}\adbasefunc^v_{s'}
x^{\gb(s'-v-1);\db(s'-v-1)}b_{\gb(s'-v-1)}\bb_{\db(s'-v-1)}
\ee
to
\be\label{adhighest_conf_w}
\confweight_{\rm max}(v)=s'-v-1 \mbox{ for }\adbasis^v_{s,s'}{}_{\rm max}=i^s\PNO^{s-s'}\adbasefunc^v_{s'}
x_{\ga(s'-v-1);\da(s'-v-1)}a^{\ga(s'-v-1)}\ba^{\da(s'-v-1)}\,.
\ee
Elements \eqref{adlowest_conf_w} and \eqref{adhighest_conf_w}
with $v=0$ have the lowest conformal weight $-s'+1$ and
the highest conformal weight $s'-1$ correspondingly.

All the elements $\adbasis^v_{s,s'}$ that form basis of $\admodules[s']$ (i.e. with fixed $s'=1,\ldots s$) can be arranged on the following diagram
\be\label{admodule_s_diagram}
\begin{picture}(100,130)%
\newsavebox{\adLevelLine}\sbox{\adLevelLine}{\qbezier[70](0,0)(160,0)(320,0)}%
\put(-40,65)%
{%
\put(0,0){\adModuleStruct{4}}\put(100,0){\adModuleStruct{3}}\put(170,0){\ensuremath{\cdots}}%
\put(230,0){\adModuleStruct{1}}\put(260,0){\adModuleStruct{0}}%
\put(-8,-65){$\scriptstyle{v=0}$}\put(92,-65){$\scriptstyle{v=1}$}\put(170,-65){$\cdots$}%
\put(210,-65){$\scriptstyle{v=s'-2}$}\put(252,-65){$\scriptstyle{v=s'-1}$}%
\put(-60,-55){\vector(0,1){120}}\put(-62,60){\llap{$\scriptstyle{\confweight}$}}%
\put(-62,0){\llap{$\scriptstyle{0}$}\line(1,0){4}}%
\put(-62,12){\llap{$\scriptstyle{1}$}\line(1,0){4}}%
\put(-62,-12){\llap{$\scriptstyle{-1}$}\line(1,0){4}}%
\put(-62,48){\llap{$\scriptstyle{s'-1}$}\line(1,0){4}}%
\put(-62,-48){\llap{$\scriptstyle{-s'+1}$}\line(1,0){4}}%
\put(-62,27){\llap{$\vdots$}}%
\put(-62,-30){\llap{$\vdots$}}%
}%
\put(-40,65)%
{%
\put(-62,0){\usebox{\adLevelLine}}%
\put(-62,12){\usebox{\adLevelLine}}%
\put(-62,-12){\usebox{\adLevelLine}}%
\put(-62,48){\usebox{\adLevelLine}}%
\put(-62,-48){\usebox{\adLevelLine}}%
}%
\end{picture}
\ee
Here every dot ($\bullet$) indicates some $\adbasis^v_{s,s'}$. All dots in the same row correspond to $\adbasis^v_{s,s'}$-s with the same
conformal weight indicated on the left axis. Dots compose the collection of rhombuses,
which are distinguished by the value of $v=0,1,\ldots,s'-1$ indicated on the bottom of the diagram.
The lowest (the highest) dot in each rhombus corresponds to $\adbasis^v_{s,s'}{}_{\min}$ ($\adbasis^v_{s,s'}{}_{\max}$)
(see \eqref{adlowest_conf_w}, \eqref{adhighest_conf_w})
of the lowest (the highest) conformal weight for given $v$. The arrows indicate
transformations which change orders of $\adbasis^v_{s,s'}$-s with respect to oscillators $a, b, \ba, \bb$
in such a way that $s'$ is kept constant and $\confweight$ increases by 1. Namely
\begin{equation}
\begin{picture}(10,10)%
\put(0,0){\vector(1,1){10}}%
\end{picture}:
\quad\left\{
\begin{aligned}
\na\rightarrow\na+1\,,\\
\nbb\rightarrow\nbb-1\,.
\end{aligned}
\right.\qquad
\begin{picture}(10,10)%
\put(10,0){\vector(-1,1){10}}%
\end{picture}:
\quad\left\{
\begin{aligned}
\nb\rightarrow\nb-1\,,\\
\nba\rightarrow\nba+1\,,
\end{aligned}
\right.
\end{equation}

It is convenient to introduce independent\footnote{Note that variables $\na$, $\nb$, $\nba$ and $\nbb$ are not independent on $\admodules[s']$
since they obey relations \eqref{f_eigenvec}.} "coordinates" on diagram \eqref{admodule_s_diagram}
\be\label{admodule_s_coordinates}
(v,q,t)\,,\qquad v=0,\ldots,s'-1\,,\quad q,t=0,\ldots,s'-v-1\,.
\ee
Here $v$ numerates the rhombus in \eqref{admodule_s_diagram} and $q$ ($t$) indicates the number of upper-right (upper-left)
arrows one should pass from the very bottom dot to get to the indicated dot. For instance coordinates
$(v,0,0)$, $(v,s'-v-1,0)$, $(v,0,s'-v-1)$ and $(v,s'-v-1,s'-v-1)$ indicate on the bottom, the right, the left, and the upper corners 
of rhombus $v$ correspondingly. In these terms all other variables are expressed as
\be\label{ad_variab_through_coord}
&\na=q\,,&&\nb=s'-v-t-1\,,\\
&\nba=t\,,&&\nbb=s'-v-q-1\,,\\
&\confweight=-s'+v+q+t-1\,.
\ee

Let us note that complex conjugation \eqref{reality_condition}
transforms $\adbasis^v_{s,s'}(v,q,t)$ corresponding to the dot $(v,q,t)$ to 
$\bar{\adbasis}^v_{s,s'}(v,t,q)$ corresponding to the dot $(v,t,q)$ symmetric
with respect to reflection in a line connecting the top
and the bottom of rhombus $v$. Therefore due to reality condition \eqref{reality_condition} coordinate-tensors (i.e. tensors like $x$
in \eqref{admodule_s_monomial}) of $\adbasis^v_{s,s'}(v,q,t)$ and $\adbasis^v_{s,s'}(v,t,q)$
are mutually complex conjugate if $q\neq t$ and
coordinate-tensor of $\adbasis^v_{s,s'}(v,q,q)$ is self Hermitian conjugate.

Finally, let us show that $\admodules[s']$ is irreducible. Suppose $\admodules[s']$ is reducible then it decomposes into
direct sum
\be\label{admodule_s_if_reducible}
\admodules[s']=\admodules[s']'\oplus\admodules[s']''\,.
\ee
Each module in \eqref{admodule_s_if_reducible} has a lowest conformal weight subspace $\lcws_{s'}'\subset\admodules[s']'$ and
$\lcws_{s'}''\subset\admodules[s']''$ annihilated by $\adP_{\ga\db}$. Both $\lcws_{s'}'$ and $\lcws_{s'}''$ form $\asl(k)\oplus\asl(k)$-modules.
Let $m'$ and $m''$ denote some $\asl(k)\oplus\asl(k)$-irreducible submodules of $\lcws_{s'}'$ and $\lcws_{s'}''$ correspondingly.
Since all $\adbasis^v_{s,s'}$-s in \eqref{admodule_s_diagram} with fixed conformal weight (i.e. those contained in fixed row of \eqref{admodule_s_diagram})
have different $\asl(k)\oplus\asl(k)$-structure, one necessarily concludes that $m'$ and $m''$ coincide with some $\adbasis^v_{s,s'}$ 
from \eqref{admodule_s_diagram}. On the other hand as one can easily see from \eqref{ad_repr_ukk} the only $\adbasis^v_{s,s'}$ 
annihilated by $\adP_{\ga\db}$ is $\adbasis^0_{s,s'}{}_{\rm min}$ given by formula \eqref{adlowest_conf_w} with $v=0$. We, thus, conclude that module 
$\admodules[s']$ is irreducible.

From \eqref{ad_repr_ukk} one finds that quadratic Casimir operator
\be\label{Casimir_gen}
\cC^2_{\au(k,k)}=\uL_\ga{}^\gb\uL_\gb{}^\ga+\buL_\da{}^\db\buL_\db{}^\da+\frac{2}{k}(\uD^2+\PNO^2)-\{\uP_{\ga\db},\uK^{\ga\db}\}
\ee
of algebra $\au(k,k)$ takes in module $\admodules[s']$ the following value
\be\label{Casimir}
\cC^2_{\au(k,k)}=2(s'-1)(s'+2k-2)\,.
\ee

\section{Structure of twist-adjoint module\label{section_Twist_adjoint_module}}
Let us now consider twisted-adjoint $\asu(k,k)$-module $\twmoduleinf$. It is spanned by oscillators $a^\ga\,, b_\ga\,, \ba^\da$ and $\tbb^\da$,
where oscillator $\tbb^\da$ is obtained from oscillator $\bb_\da$ by twist transformation
\be\label{twist_b}
\bb_\da&\rightarrow\dpd{\tbb^\da}\,,\\
\dpd{\bb_\da}&\rightarrow-\tbb^\da
\ee
such that commutator
\be\label{twist_commut_conserv}
[\dpd{\bb_\da},\bb_\db]=[-\tbb^\da,\dpd{\tbb^\db}]
\ee
conserves.
Due to conservation of commutator \eqref{twist_commut_conserv} operators that represent $\au(k,k)$ on $\twmoduleinf$ can be obtained from that of $\admoduleinf$
by simple replacement \eqref{twist_b}. We have (cf. \eqref{adinf_repr_ukk})
\begin{equation}\label{twinf_repr_ukk}
  \begin{aligned}
    &\twinfL_\ga{}^\gb=\adinfL_\ga{}^\gb\,,\\
    &\twinfbL_\da{}^\db=\ba^\db\dpd{\ba^\da}+\tbb^\db\dpd{\tbb^\da}-\frac{1}{k}\gd_\da^\db(\nba+\ntbb)\,,\\
    &\twinfP_{\ga\db}=b_\ga\dpd{\ba^\db}+\ddpd{a^\ga}{\tbb^\db}\,,\quad
    \twinfK^{\ga\db}=a^\ga\tbb^\db-\ba^\db\dpd{b_\ga}\,,\\
    &\twinfD=\frac{1}{2}(\na+\nba-\nb+\ntbb+k)\,,\quad
    \twinfPNO=\frac{i}{2}(\na-\nba-\nb-\ntbb-k)\,,
  \end{aligned}
\end{equation}
where $\ntbb$ is Euler operator for oscillator $\tbb$. We require $\twmoduleinf$ to be annihilated by $\twinfPNO$
\be\label{tPNO_annihil}
(\na-\nba-\nb-\ntbb-k)\twmoduleinf=0\,.
\ee

Analogously to $\admoduleinf$ module $\twmoduleinf$ can be decomposed into direct sum of submodules $\twmoduleinfs[s]$ picked out by requirement
\be\label{twmoduleinf_s}
\optsinfo\twmoduleinfs[s]=\optsinft\twmoduleinfs[s]=s\twmoduleinfs[s]\,,
\ee
where
\be\label{optsinf_o_t}
\optsinfo=\na-\ntbb-k+1\,,\qquad\optsinft=\nb+\nba+1
\ee
are operators that commute with \eqref{twinf_repr_ukk}. Note that due to
\eqref{tPNO_annihil} $\optsinfo f=\optsinft f$ for any element $f\in\twmoduleinf$.

Twist transformation \eqref{twist_b} applied to the basis elements of $\admoduleinfs[s]$
\eqref{admodule_inf_s_decomposition_basis} gives rise to the following elements of $\twmoduleinfs[s]$
\be
\label{twmodule_inf_s_decomposition_basis}
\twbasis^v_{s,s'}=i^s\tPNO^{s-s'}\twbasefunc^v_{s'}(\tPNO,\tD)\twmonomial_{s'-v}(a,b,\ba,\tbb)\,,\qquad s'=1,\ldots,s\,,\quad v=0,\ldots,s'-1\,.
\ee
Here
\be\label{tPNO}
\tPNO=\frac{i}{2}(a\cdot b-\ba\cdot\dpd{\tbb})
\ee
is annihilated by twist-adjoint action of $\asu(k,k)$ \eqref{twinf_repr_ukk},
\be
\label{tD}
\tD=\half(a\cdot b+\ba\cdot\dpd{\tbb})
\ee
and $\twbasefunc^v_{s'}(\tPNO,\tD)$ is twist transformed $\adbasefunc^v_{s'}(\PNO,\uD)$.
Finally, $\twmonomial_{s'-v}(a,b,\ba,\tbb)$ are eigenvectors of operators $\optsinfo\,,\optsinft$ \eqref{optsinf_o_t} that correspond to eigenvalue $s'-v$
\be
\label{tf_eigenvec}
\optsinfo \twmonomial_{s'-v}=\optsinft \twmonomial_{s'-v}=(s'-v)\twmonomial_{s'-v}
\ee
and satisfy twisted tracelessness relations (cf. \eqref{iukk_trless_cond})
\be
\label{twtracelessness}
&\tbb\cdot\dpd{\ba}\twmonomial_{s'-v}=0\,,\\
&\ddpddot{a}{b}\twmonomial_{s'-v}=0\,.
\ee
In other words $\twmonomial_{s'-v}$ are monomials
\be\label{twmodule_s_monomial}
\twmonomial_{s'-v}(\na,\nba,\nb,\ntbb)=
\tilde{x}^{\gb(\nb)}_{\ga(\na)}{}^;_{;\,\db(\ntbb),\da(\nba)}a^{\ga(\na)}b_{\gb(\nb)}\tbb^{\db(\ntbb)}\ba^{\da(\nba)}\,,
\ee
where $\tilde{x}^{\gb(\nb)}_{\ga(\na)}{}^;_{;\,\db(\ntbb),\da(\nba)}$
are complex traceless tensors separately symmetric with respect to upper and lower group of undotted indices and
of the symmetry type described by two-row Young tableau with first(second) row
of length $\ntbb$($\nba$)  with respect to dotted indices. Certainly values of $\na\,,\nb\,,\nba\,,\ntbb$ in \eqref{twmodule_s_monomial}
should be coordinated with $s'$ and $v$ through formula \eqref{tf_eigenvec}.

Due to above arguments $\twbasis^v_{s,s'}$
forms, with respect to generators $\twinfL_\ga{}^\gb$, $\twinfbL_\da{}^\db$,
irreducible $\asl(k)\oplus\asl(k)$-module corresponding to Young tableau
\be\label{twmodule_s_monomial_Young_tableau}
\bt{rll}
&\hspace{9mm} undotted &\hspace{7mm} dotted\\[1mm]
upper\hspace{0mm} &
                    \bt{l}
                    \begin{Young}{12}{1}{\USkip\DSkip\renewcommand{\ULabelMidVShift}{4}}
                    \Put{\Block{12}{1}}\ULabelMid{\na}
                    \end{Young}
                    \et
                  &\hspace{5mm}
                    \bt{l}
                    \begin{Young}{3}{1}{\USkip\DSkip\renewcommand{\ULabelMidVShift}{4}}
                    \Put{\Block{3}{1}}\ULabelMid{\ntbb}
                    \PutToDown{\Block{2}{1}}\DLabelMid{\nba}
                    \end{Young}
                    \et
                  \\[2mm]
lower\hspace{0mm} &
                    \bt{l}
                    \begin{Young}{5}{1}{\USkip\renewcommand{\ULabelMidVShift}{4}}
                    \Put{\Block{5}{1}}\ULabelMid{\nb}
                    \end{Young}
                    \et
\et
\ee
(see Appendix \ref{section_slk_plus_slk_irreps} for more details).

Analogously to adjoint case let $\twsubmodulen[m]$ denote submodule of $\twmoduleinf$ spanned by elements \eqref{twmodule_inf_s_decomposition_basis}
with the power of $\tPNO$ grater or equal to $m$
and let $\twsubmodulens[m][s]=\twsubmodulen[m]\cap\twmodules[s]$ denote corresponding submodule of $\twmodules[s]$.
Note that $\twsubmodulens[m][s]\equiv 0$ for $m\geq s$. Let us define quotient modules
\be
&\twmodulen[0]\subset\twmodulen[1]\subset\cdots\subset\twmoduleinf\,,\\
&\twmodulens[0][s]\subset\twmodulens[1][s]\subset\cdots\subset\twmodulens[s-2][s]\subset\twmoduleinfs[s]\,,
\ee
where
\be
\twmodulen[m]=\twmoduleinf/\twsubmodulen[m+1]\,,\quad
\twmodulens[m][s]=\twmoduleinfs[s]/\twsubmodulens[m+1][s]\,.
\ee
Note that $\twmodulens[m][s]\equiv \twmoduleinfs[s]$ for $m\geq s-1$.
In what follows we omit index 0 and denote $\twmodulen[0]$ and $\twmodulens[0][s]$ as $\twmodule$ and $\twmodules[s]$, respectively.

To find how $\uP_{\ga\db}$ and $\uK^{\ga\db}$ are represented in basis \eqref{twmodule_inf_s_decomposition_basis} one should apply
twist transformation to \eqref{ad_repr_ukk}, \eqref{trProj_formulas}. We have
\begin{equation}\label{tw_repr_ukk}
  \begin{aligned}
    &\twP_{\ga\db}\cdot\twbasefunc^v_{s'}\twmonomial_{s'-v}=\left[v\twbasefunc^{v-1}_{s'}\ttrProj b_\ga\dpd{\tbb^\db}+
    \twbasefunc^v_{s'}\Big(v \func(n)+1\Big)\ttrProj\ddpd{a^\ga}{\tbb^\db}+\twbasefunc^v_{s'}\Big(v \func(\tbn)+1\Big)\ttrProj b_\ga\dpd{\ba^\db}+\right.\\
    &\quad\qquad\qquad\qquad\qquad\left.{}+\Big(2s'+2k-v-4\Big)
    \twbasefunc^{v+1}_{s'}\func(n)\func(\tbn)\ddpd{a^\ga}{\ba^\db}\right]\twmonomial_{s'-v}\,,\\
    &\twK^{\ga\db}\cdot\twbasefunc^v_{s'}\twmonomial_{s'-v}=-\left[v\twbasefunc^{v-1}_{s'}\ttrProj a^\ga\ba^\db+
    \twbasefunc^{v}_{s'}\Big(v\func(n)+1\Big)\ttrProj\ba^\db\dpd{b_\ga}-\twbasefunc^{v}_{s'}\Big(v\func(\tbn)+1\Big)\ttrProj a^\ga\tbb^\db-\right.\\
    &\qquad\qquad\qquad\qquad\qquad\left.-\Big(2s'+2k-v-4\Big)
    \twbasefunc^{v+1}_{s'}\func(n)\func(\tbn)\tbb^\db\dpd{b_\ga}\right]\twmonomial_{s'-v}\,,
  \end{aligned}
\end{equation}
where $\tbn=\nba-\ntbb-k$ and
$\ttrProj$ is projector to component satisfying \eqref{twtracelessness}
\be\label{ttrProj_formulas}
&\ttrProj a^\ga = a^\ga-\func(n-2)\,a\cdot b\:\dpd{b_\ga}\,,&&
\ttrProj b_\ga = b_\ga-\func(n-2)\,a\cdot b\:\dpd{a^\ga}\,,\\
&\ttrProj \ba^\db = \ba^\db+\func(\tbn-2)(\ba^\db+\tbb^\db\: \ba\cdot \dpd{\tbb})\,,&&
\ttrProj \dpd{\tbb^\db} = \dpd{\tbb^\db}-\func(\tbn-2)\,\ba\cdot \dpd{\tbb}\:\dpd{\ba^\db}\,,\\
&\ttrProj a^\ga\ba^\db = \ttrProj a^\ga \ttrProj \ba^\db\,,&&
\ttrProj b_\ga\dpd{\tbb^\db} = \ttrProj b_\ga \ttrProj \dpd{\tbb^\db}\,.
\ee

Although all the above formulae were obtained by application of twist transformation \eqref{twist_b} (which conserves commutators \eqref{twist_commut_conserv})
to the analogous formulae corresponding to the adjoint modules
the structure of the twist-adjoint modules and their analysis have some important nuances in comparison with the adjoint case.

Firstly, the twist-adjoint modules are infinite-dimensional. This is because operator $\optsinfo$ contains the difference of $\na$ and $\ntbb$ and,
thus, requirement \eqref{tf_eigenvec} does not bound the order of $\twmonomial_{s'-v}$ with respect to $a$ and $\tbb$.

Secondly, contrary to the adjoint case elements $\twbasis^v_{s,s'}$ of the twist-adjoint module are not linearly independent.
Indeed, as was discussed above  $\twmonomial_{s'-v}$ forms with respect to the dotted indices $\asl(k)$-module corresponding to the
two-row Young tableaux with the first row of length $\ntbb$ and the second row of length $\nba$ (see \eqref{twmodule_s_monomial_Young_tableau} and
Appendix \ref{section_slk_plus_slk_irreps}).
Therefore
\be\label{twmodule_specific}
(\ba\cdot\dpd{\tbb})^u\twmonomial_{s'-v}=(\tD+i\tPNO)^u\twmonomial_{s'-v}=0\quad\mbox{for $u>v_{\max}$,}
\ee
where
\be
\label{v_max}
v_{\max}\twmonomial_{s'-v}=(\ntbb-\nba)\twmonomial_{s'-v}\,,
\ee
i.e.
\be
\label{tD_u_trough_tPNO}
\tD^u\twmonomial_{s'-v}=-\sum_{j=1}^u {u\choose j}(i\tPNO)^j\tD^{n-j}\twmonomial_{s'-v}\quad\mbox{for $u>v_{\max}$}\,.
\ee
From \eqref{tD_u_trough_tPNO} one gets in particular that
\be
\label{twbasis_leniar_dependence}
\twbasis^{v_{\max}+1}_{s,s'}=\sum_{j=1}^{\min(v_{\max}+1,s'-1)}(\cdots)\twbasis^{v_{\max}-j+1}_{s,s'-j}\,,
\ee
where $(\cdots)$ are some coefficients.

In what follows let us reduce set \eqref{twmodule_inf_s_decomposition_basis} to linearly independent subset 
\be\label{twmodule_inf_s_decomposition_basis_lin_indep}
\twbasis^v_{s,s'}\,,\qquad s'=1,\ldots,s\,,\quad v=0,\ldots,\min(v_{\max},s'-1)\,.
\ee
Let the elements $\twbasis^v_{s,s'}$ and corresponding monomials $\twmonomial_{s'-v}$ with
$v=v_{\max}$ be called terminal and let denote them as $\twbasisterm^v_{s,s'}$ and $\twmonomialterm_{s'-v}$.
Finally let us note that after we have reduced to linearly independent subset \eqref{twmodule_inf_s_decomposition_basis_lin_indep}
it is possible to treat $\tPNO$ and $\tD$ as independent of oscillators variables analogously to adjoint case. 

For fixed $s'$ all linearly independent $\twbasis^v_{s,s'}$
can be arranged in the following diagram
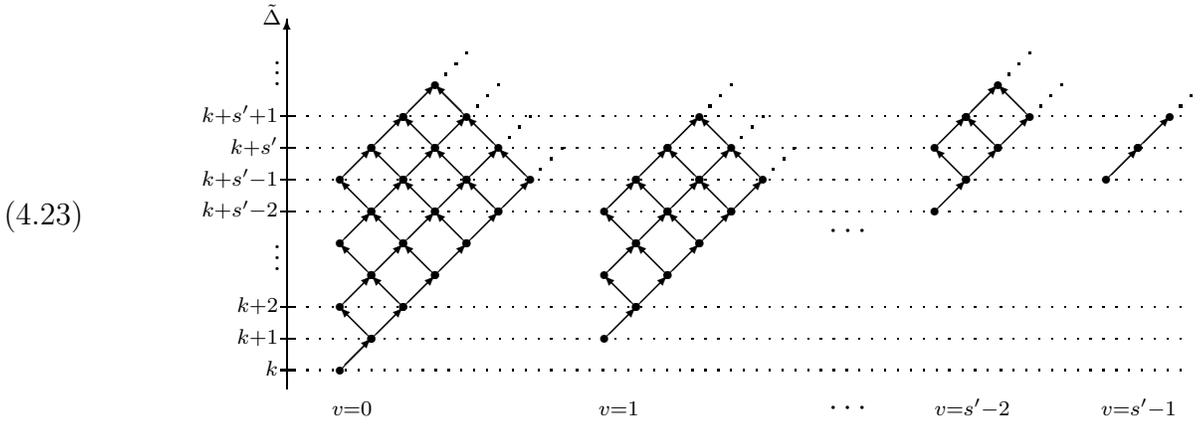
\be\label{twmodule_s_diagram}
\begin{picture}(100,155)%
\newsavebox{\twLevelLine}\sbox{\twLevelLine}{\qbezier[70](0,0)(170,0)(340,0)}%
\put(-40,65)%
{%
\put(-40,-48){\put(0,0){\twModuleStruct{6}{3}}\put(100,12){\twModuleStruct{5}{2}}\put(185,50){\ensuremath{\cdots}}%
\put(225,60){\twModuleStruct{3}{1}}\put(290,72){\twModuleStruct{2}{0}}}%
\put(-43,-65){$\scriptstyle{v=0}$}\put(58,-65){$\scriptstyle{v=1}$}%
\put(145,-65){$\cdots$}\put(185,-65){$\scriptstyle{v=s'-2}$}\put(248,-65){$\scriptstyle{v=s'-1}$}%
\put(-60,-55){\vector(0,1){140}}\put(-62,83){\llap{$\scriptstyle{\tilde{\confweight}}$}}%
\put(-62,-48){\put(0,-2){\llap{$\scriptstyle{k}\,$}}\line(1,0){4}}%
\put(-62,-36){\put(0,-2){\llap{$\scriptstyle{k+1}\,$}}\line(1,0){4}}%
\put(-62,-24){\put(0,-2){\llap{$\scriptstyle{k+2}\,$}}\line(1,0){4}}%
\put(-62,-10){\llap{$\vdots$}}%
\put(-62,12){\put(0,-2){\llap{$\scriptstyle{k+s'-2}\,$}}\line(1,0){4}}%
\put(-62,24){\put(0,-2){\llap{$\scriptstyle{k+s'-1}\,$}}\line(1,0){4}}%
\put(-62,36){\put(0,-2){\llap{$\scriptstyle{k+s'}\,$}}\line(1,0){4}}%
\put(-62,48){\put(0,-2){\llap{$\scriptstyle{k+s'+1}\,$}}\line(1,0){4}}%
\put(-62,60){\llap{$\vdots$}}%
%
\put(-62,-48){\usebox{\twLevelLine}}%
\put(-62,-36){\usebox{\twLevelLine}}%
\put(-62,-24){\usebox{\twLevelLine}}%
\put(-62,12){\usebox{\twLevelLine}}%
\put(-62,24){\usebox{\twLevelLine}}%
\put(-62,36){\usebox{\twLevelLine}}%
\put(-62,48){\usebox{\twLevelLine}}%
}%
\end{picture}
\ee
Here every dot ($\bullet$) indicates some $\twbasis^v_{s,s'}$. All dots in the same row correspond to different $\twbasis^v_{s,s'}$ with the same
conformal weight (indicated on the left axis)
\be\label{twmodule_conf_weight}
\twD\twbasis^v_{s,s'}=\frac{1}{2}(\na+\nba-\nb+\ntbb+k)\twbasis^v_{s,s'}=\tilde{\confweight}\twbasis^v_{s,s'}\,,
\ee
which due to \eqref{v_max}, \eqref{optsinf_o_t} range for fixed value of $v=0,1,\ldots, s-1$ from
\be\label{twlowest_conf_w}
\tilde{\confweight}_{\rm min}(v)=v+k \mbox{ for }
\twbasis^v_{s,s'}{}_{\rm min}=i^s\tPNO^{s-s'}\twbasefunc^v_{s'}\,\tilde{x}^{\gb(s'-v-1)}_{\ga(s'+k-1)}{}^;_{;\,\db(v)}a^{\ga(s'+k-1)}b_{\gb(s'-v-1)}\tbb^{\db(v)}
\ee
to infinity.

Dots compose the collection of strips of the width (number of dots) $s'-v$ and of the infinite length,
which are distinguished by the value of $v=0,1,\ldots,s'-1$ indicated on the bottom of the diagram.
The lowest dot in each strip corresponds to $\twbasis^v_{s,s'}{}_{\rm min}$ \eqref{twlowest_conf_w}
of the lowest conformal weight for given $v$. The arrows indicate
transformations which change orders of $\twbasis^v_{s,s'}$ with respect to oscillators $a, b, \ba, \tbb$
in such a way that $s'$ is kept constant and $\tilde{\confweight}$ increases by 1. Namely
\begin{equation}
\label{tw_diagramm_oscil_order_change}
\begin{picture}(10,10)%
\put(0,0){\vector(1,1){10}}%
\end{picture}:
\quad\left\{
\begin{aligned}
\na\rightarrow\na+1\,,\\
\ntbb\rightarrow\ntbb+1\,,
\end{aligned}
\right.\qquad
\begin{picture}(10,10)%
\put(10,0){\vector(-1,1){10}}%
\end{picture}:
\quad\left\{
\begin{aligned}
\nb\rightarrow\nb-1\,,\\
\nba\rightarrow\nba+1\,.
\end{aligned}
\right.
\end{equation}

Introduce independent coordinates on \eqref{twmodule_s_diagram}
\be\label{twcoord}
\twcoord{v,q,t}\,,\qquad v=0,\ldots,s'-1\,,\quad q=0,\ldots,\infty\,,\quad t=0,\ldots,\min(q,s'-v-1)\,,
\ee
where $v$ numerates the stripe and $q$ ($t$) indicates the number of upper-right (upper-left)
arrows one should pass from the very bottom dot to get to the indicated dot.
In these terms all other variables are expressed as
\be\label{tw_variab_through_coord}
&\na=s'+k+q-1\,,&&\nb=s'-v-t-1\,,\\
&\nba=t\,,&&\ntbb=v+q\,,\\
&\tilde{\confweight}=k+v+q+t\,,&&v_{\max}=v+q-t\,.
\ee
From the expression for $v_{\max}$ one finds that the terminal terms, which are defined by requirement $v_{\max}=v$, correspond to the dots with
coordinates $\twcoord{v,t,t}$, $t=0,\ldots,s'-v-1$, i.e. the most left dots of each stripe.

Now we are going to show that elements listed in diagram \eqref{twmodule_s_diagram} form the basis of $\asu(k,k)$-submodule
$\twmodules[s']\subset\twmoduleinfs[s]$.
Operators $\twL_\ga{}^\gb\,,\twbL_\da{}^\db\,,\twD\,,\twPNO$ \eqref{twinf_repr_ukk}
commute with $i^s\tPNO^{s-s'}\twbasefunc^v_{s'}$ and therefore
conserve set \eqref{twmodule_s_diagram}.
Looking at operators $\twP_{\ga\db}$ and $\twK^{\ga\db}$ \eqref{tw_repr_ukk} one sees that their 1-st, 3-d and 4-s terms also conserve \eqref{twmodule_s_diagram},
but 2-nd terms once act at the terminal element $\twbasisterm^v_{s,s'}$ decreases $v_{\max}$ by 1, but keeps $v$ the same and, thus,  maps it to
$\twbasis^{v_{\max}+1}_{s,s'}$, which due to \eqref{twbasis_leniar_dependence}
is equivalent to the sum of terms corresponding to $s'-1\,,s'-2\,,\ldots$ So it looks like that $\twmodules[s']$ is not $\asu(k,k)$-invariant.
In Appendix \ref{section_twmoduleinfs}\label{problem_with_terminal_terms} it is shown that this is not the case,
since these problem terms zero out.

We, thus, shown that analogously to the adjoint case modules $\twmoduleinfs[s]$, $\twmodules[s]^m$ and $\twmodule^m$ admit decompositions
\be
\label{twmodule_decomposition}
&\twmoduleinfs[s]=\oplus_{s'=1}^s\twmodules[s']\,,\\
&\twmodules[s]^m=\oplus_{s'=s-m}^s\twmodules[s']\,,\quad m=0,1,\ldots,s-2\,,\\
&\twmodule^m=\oplus_{s'=1}^\infty\twmodules[s']^m=\oplus_{s'=1}^\infty(m+1)\twmodules[s']\,,\quad m=0,1,\ldots,\infty\,.
\ee
The basis of $\twmoduleinfs[s]$ coordinated with this decomposition has form
\be
\label{tw_module_inf_basis}
&\twbasis^v_{s,s'}=i^s\tPNO^{s-s'}\twbasefunc^v_{s'}\twmonomial_{s'-v}(v,q,t)\,,\\
&s'=1,\ldots s\,,\quad v=0,\ldots s'-1\,,\quad q=0,\ldots,\infty\,,\quad t=0,\ldots \min(q,s'-v-1)\,,
\ee
where elements with fixed $s'$, which are listed in diagram \eqref{twmodule_s_diagram}, form basis of $\twmodules[s']$.

In the same manner as for module $\admodules[s']$ one can show that $\twmodules[s']$ is irreducible with quadratic Casimir operator given by formula
\eqref{Casimir}.

In what follows we also need complex conjugate modules $\btwmodulens[m][s]$, $m=0,1,\ldots,\infty$, which are given by
\be
\btwmodulens[m][s]=-\twmodulens[m][s]{}^\dag\,.
\ee
Here involution $\dag$ is defined in \eqref{involution} and $(\tbb^\da)^\dag=i\tb^\ga$, $(\tb^\ga)^\dag=i\tbb^\da$, 
where oscillator $\tb^\ga$ is the twist transformed oscillator $b_\ga$.

\section{Unfolded formulation of Fradkin-Tseytlin equations\label{section_Fr_Ts_eq_unfolded}}
In this section we set the number of oscillators $k=2$. According to procedure described in section \ref{section_Unf_formul_preliminary} consider
zero curvature equation \eqref{UF_zero_curv_cond} for 4-d connection 1-form $W(a,b,\ba,\bb)=\xi^\mu W_\mu$ taking values in algebra
$\aiu(2,2)$. Where $W$ is defined on 4-d Minkowski space-time with coordinates $x^\mu$ and basic 1-forms $\xi^\mu$,
$\mu=0,\ldots,4$.

Let us fix
vacuum solution
\be\label{vacuum_connect_one_form}
W_0=\xi^{\ga\db}b_\ga\bb_\db\,,
\ee
where
\be\label{xi_diffenition}
\xi^{\ga\db}=\xi^\mu\gs_\mu{}^{\ga\db}\,,\qquad \compconj(\xi^{\ga\db})=\xi^{\gb\da}
\ee
and $\gs_\mu{}^{\ga\db}$ are 4 Pauli matrices. It obviously satisfies \eqref{UF_zero_curv_cond} and corresponds to Cartesian coordinate choice
in 4-d Minkowski space-time.

Let $\GaugeFinf(a,b,\ba,\bb)$ denote 1-form gauge fields taking values in $\admoduleinf$.
2-form curvatures $\Curvinf(a,b,\ba,\bb)$ of algebra $\aiu(2,2)$ linearized with respect to the vacuum $W_0$ are given by
\be\label{linear_curv_inf}
\Curvinf=(\extdiff +\adsminf)\GaugeFinf\,,
\ee
where
\be
\label{ext_diff_k_two}
\extdiff=\xi^\mu\dpd{x^\mu}=\xi^{\ga\db}\dpd{x^{\ga\db}}
\ee
and
\be\label{ad_sm_inf}
\adsminf=\xi^{\ga\db}\adinfP_{\ga\db}=\xi^{\ga\db}\Big(b_\ga\dpd{\ba^\db}+\bb_\db\dpd{a^\ga}\Big)\,.
\ee

Consider 0-form fields $\WeylFinf(a,b,\ba,\tbb)$ and $\bWeylFinf(a,\tb,\ba,\bb)$ taking values in twist-adjoint $\asu(2,2)$-modules
$\twmoduleinf$ and $\btwmoduleinf$ correspondingly.
Unfolded system of $\asu(2,2)\sim\aso(4,2)$-invariant equations (Fradkin-Tseytlin equations)
was formulated in paper \cite{Vasiliev_progr_in_HS_gauge_theory}. In our notation it has the following form
\be
\label{Fr_Ts_unf_inf_Curv_with Weyl_glued}
&\Curvinf=\ChEsminf\WeylFinf+\bChEsminf\bWeylFinf\,,\\
&(\extdiff+\twsminf)\WeylFinf=0\,,\\
&(\extdiff+\btwsminf)\bWeylFinf=0\,,
\ee
where
\be
\label{tw_sm_inf}
&\twsminf=\xi^{\ga\db}\twP_{\ga\db}=\xi^{\ga\db}\Big(b_\ga\dpd{\ba^\db}+\ddpd{a^\ga}{\tbb^\db}\Big)\,,\\
%
&\btwsminf=\xi^{\ga\db}\btwP_{\ga\db}=\xi^{\ga\db}\Big(\bb_\db\dpd{a^\ga}+\ddpd{\ba^\db}{\tb^\ga}\Big)
\ee
and
\be\label{tw_sm_inf_}
&\ChEsminf=\Xi^{\ga\ga}\ddpd{a^\ga}{a^\ga}\Big|_{\tbb=0}\,,\qquad
%
&&\bChEsminf=-\bXi^{\da\da}\ddpd{\ba^\da}{\ba^\da}\Big|_{\tb=0}\,,\\[3mm]
&\Xi^{\ga\ga}=\xi^{\ga\db}\xi^{\ga\dg}\gep_{\db\dg}\,,&&\compconj(\Xi^{\ga\ga})=\bXi^{\da\da}=\xi^{\gb\da}\xi^{\gga\da}\gep_{\gb\gga}\,.
\ee
Here $\gep^{\ga\gb}$, $\gep_{\ga\gb}$ and $\gep^{\da\db}$, $\gep_{\da\db}$ are totally antisymmetric rank 2 tensors
fixed by relation $\gep^{12}=\gep_{12}=1$. They are used to rise and lower indices $x^\ga\gep_{\ga\gb}=x_\gb$, $x_\gb\gep^{\ga\gb}=x^\ga$
and analogous formulae for dotted indices.

Compatibility conditions of system \eqref{linear_curv_inf}, \eqref{Fr_Ts_unf_inf_Curv_with Weyl_glued} require
\be\label{comp_cond}
&\adsminf{}^2=\twsminf{}^2=\{\extdiff,\adsminf\}=\{\extdiff,\twsminf\}=0\,,\\
&\adsminf\ChEsminf=\ChEsminf\twsminf
\ee
and analogous relations for $\btwsminf$ and $\bChEsminf$. They can be easily checked by direct computation.

As was shown above operators $\adsminf$ and $\twsminf$ ($\btwsminf$) commute with $\opsinfo$, $\opsinft$ and $\optsinfo$, $\optsinft$
($\opbtsinfo$, $\opbtsinft$) correspondingly.
One can also see that operator $\ChEsminf+\bChEsminf$ maps differential forms taking values in submodule $\twmoduleinfs[s]\oplus\btwmoduleinfs[s]$ into
these taking values\footnote{Really operator $\ChEsminf$ decreases $\na$ by 2 and sets $\ntbb=0$. So it gives non-zero result for $\twmoduleinfs[s]$
elements independent of $\tbb$ only. Therefore the value $\optsinfo=\na-\ntbb-k+1$ of the element before transformation is equal to
the value $\opsinfo=(\na-2)+\nbb+1$ of the element after transformation for $k=2$. The same is true for $\bChEsminf$.} in $\admoduleinfs[s]$.
Therefore system \eqref{linear_curv_inf}, \eqref{Fr_Ts_unf_inf_Curv_with Weyl_glued}
decomposes into infinite number of subsystems on
fields $\GaugeFinf_s$, $\WeylFinf_s$ and $\bWeylFinf_s$, $s=1,2,\ldots$ taking values in submodules
$\admoduleinfs[s]$, $\twmoduleinfs[s]$, $\btwmoduleinfs[s]$.

Moreover, we argue that above system decomposes into the collection
of subsystems on fields $\GaugeF_{s'}$, $\WeylF_{s'}$ and $\bWeylF_{s'}$, corresponding to submodules
$\admodules[s']$, $\twmodules[s']$, $\btwmodules[s']$, $s'=1,\ldots, s$.
Indeed, since $\adsminf$, $\twsminf$ and $\btwsminf$ are constructed in terms of $\asu(2,2)$-generators (namely generator of translation in adjoint and
twist-adjoint representation) that admit decompositions \eqref{admodule_inf_s_decomposition}, \eqref{twmodule_decomposition}
into these submodules the same decomposition is valid for operators $\adsminf$, $\twsminf$ and $\btwsminf$.

Let us now study operator $\ChEsminf$. In order to above decomposition take place one needs to show that operator $\ChEsminf$
some how transforms $\tPNO^{s-s'}$ in front of $\twmodules[s']$ basis (see \eqref{tw_module_inf_basis}) into $\PNO^{s-s'}$ in front of $\admodules[s']$ basis
(see \eqref{admodule_inf_s_decomposition_basis}) and analogously for operator $\bChEsminf$ and $\bar{\tPNO}^{s-s'}$.
First consider $\ChEsminf\tPNO\tilde{h}$, where $\tilde{h}$ is an arbitrary function of $a^\ga,b_\ga,\ba^\da,\tbb^\da$ and $\xi^{\ga\db}$.
One has
\be\label{qwerty1}
\ChEsminf\tPNO\tilde{h}=\PNO\ChEsminf\tilde{h}+
\left. i\xi^{\ga\db}\xi^{\ga\dg}\gep_{\db\dg}b_\ga\dpd{a^\ga}\right|_{\tbb=0}\tilde{h}+
\left. \frac{i}{2}\xi^{\ga\db}\xi^{\ga\dg}\ba^{\dd}\ddpd{a^\ga}{a^\ga}\right|_{\tbb=0}\Big(\gep_{\db\dg}\bb_{\dd}-
\gep_{\db\dg}\dpd{\tbb^{\dd}}\Big)\tilde{h}\,.
\ee
Taking into account that the 3-rd term of \eqref{qwerty1} zero out if dotted indices in parentheses are antisymmetrized one has
\be
\label{ChEsminf_acts_twmoduleinf_basis_Z_one}
\ChEsminf\tPNO\tilde{h}=\PNO\ChEsminf\tilde{h}+\adsminf\oppsi^1\tilde{h}+\oppsi^1\twsminf\tilde{h}\,,
\ee
where
\be
\label{oppsi_one}
\oppsi^1=-i\xi^{\ga\db}\epsilon_{\db\dg}\left.\ba^{\dg}\dpd{a^{\ga}}\right|_{\tbb=0}\,.
\ee
From \eqref{ChEsminf_acts_twmoduleinf_basis_Z_one} one gets that for general power of $\tPNO$
\be
\label{ChEsminf_acts_twmoduleinf_basis_Z_general}
\ChEsminf\tPNO^{q}\tilde{h}=\PNO^{q}\ChEsminf\tilde{h}+\adsminf\oppsi^{q}\tilde{h}+\oppsi^{q}\twsminf\tilde{h}\,,
\ee
where
\be
\label{oppsi_s_s_prime}
\oppsi^{q}=\sum_{j=0}^{q-1}\PNO^j\oppsi^1\tPNO^{q-j}
\ee
and analogous formulae for operator $\bChEsminf$.

Therefore after field redefinition
\be
\label{field_redifinition}
\GaugeFinf_{s}\rightarrow\GaugeFinf_{s}-\sum_{j=1}^{s-1}\oppsi^j\WeylFinf_{s}-\sum_{j=1}^{s-1}\boppsi^j\bWeylFinf_{s}
\ee
system \eqref{linear_curv_inf}, \eqref{Fr_Ts_unf_inf_Curv_with Weyl_glued} with $s$ fixed split into subsystems corresponding
to modules $\admodules[s']$, $\twmodules[s']$, $\btwmodules[s']$ for $s'=1,\ldots, s$
\be
\label{Fr_Ts_unf_s_prime}
&\Curv_{s'}=(\extdiff +\adsm)\GaugeF_{s'}=\ChEsm\WeylF_{s'}+\bChEsm\bWeylF_{s'}\,,\\
&(\extdiff+\twsm)\WeylF_{s'}=0\,,\\
&(\extdiff+\btwsm)\bWeylF_{s'}=0\,.
\ee
Here $\GaugeF_{s'}$, $\WeylF_{s'}$, $\bWeylF_{s'}$ are 1 and 0-forms taking values in corresponding irreducible modules and
\be
\label{sigma_minus_operators}
\adsm=\xi^{\ga\db}\adP_{\ga\db}\,,\quad\twsm=\xi^{\ga\db}\twP_{\ga\db}\,,\quad\btwsm=\xi^{\ga\db}\btwP_{\ga\db}\,.
\ee
Operator $\ChEsm$ ($\bChEsm$) in \eqref{Fr_Ts_unf_s_prime} is a restriction of $\ChEsminf$ ($\bChEsminf$) to submodule $\twmodules[s']$ ($\btwmodules[s']$).
Operator $\ChEsm$ acts nontrivially on $\tbb$ independent $p$-forms only, i.e. those taking values in the elements of module $\twmodules[s']$ 
with coordinates $(v,0,0)\,,v=0,\ldots,s'-1$, and transforms them into $p+2$-forms
taking values in the element of module $\admodules[s']$ with coordinates $(0,s-1,v)$. Analogously operator $\bChEsm$ acts nontrivially on $p$-forms taking
values in the elements of module $\btwmodule_{s'}$ with coordinates $(v,0,0)\,,v=0,\ldots,s'-1$, and transforms them into $p+2$-forms
taking values in the complex conjugated elements of module $\admodules[s']$ with coordinates $(0,v,s-1)$.

%

As was discussed in section \ref{section_Unf_formul_preliminary} the dynamical content of system \eqref{Fr_Ts_unf_s_prime} is
encoded by cohomology of operator $\hat{\gs}_-=\adsm+\twsm+\btwsm+\ChEsm+\bChEsm$. Let $\cohom^p_{s';\Gd}$, $\tcohom^p_{s';\Gd}$ and $\btcohom^p_{s';\Gd}$
denote $p$-th cohomology with conformal weight $\Gd$ of operators $\adsm$, $\twsm$ and $\btwsm$ correspondingly.
These cohomology are found in Appendix \ref{section_sigma_min_cohomology}. For cohomology of $\adsm$ we have (up to an overall factor $i^{s-s'}\PNO^{s-s'}$)
\be
\label{adsm_cohomology_k_two}
&\cohom^0_{s';-s'+1}=i^{s'}\gaugepar^{\gb(s'-1)\,;\,\db(s'-1)}\,b_{\gb(s'-1)}\bb_{\db(s'-1)}\,,\\
&\cohom^1_{s';-s'+1}=i^{s'}\xi^{\gga\dd}\,\physf^{\gb(s'-1)\,;\,\db(s'-1)}_{\gga\,;\,\dd}\,b_{\gb(s'-1)}\bb_{\db(s'-1)}\,,\\
&\cohom^2_{s';0}=i^{s'}\Big(\Xi^{\gga\gga}\,\adeq_{\ga(s'-1)\gga(2)}^{\gb(s'-1)}\,a^{\ga(s'-1)}b_{\gb(s'-1)}+
\Xi^{\dg\dg}\,\badeq_{\da(s'-1)\dg(2)}^{\db(s'-1)}\,\ba^{\da(s'-1)}\bb_{\db(s'-1)}\Big)\,,\\
&\cohom^3_{s';s'-1}=i^{s'}\Xi^{\gga\dd}\,\adsyzo_{\ga(s'-1)\gga\,;\,\da(s'-1)\dd}\,a^{\ga(s'-1)}\ba^{\da(s'-1)}\,,\\
&\cohom^4_{s';s'-1}=i^{s'}\Xi\,\adsyzt_{\ga(s'-1)\,;\,\da(s'-1)}\,a^{\ga(s'-1)}\ba^{\da(s'-1)}\,,
\ee
where in addition to $\Xi^{\ga\ga}$ and $\bXi^{\da\da}$ (see \eqref{tw_sm_inf_}) we define
\be
\label{Xi_in_ass}
&\Xi^{\ga\db}=i\xi^{\ga\dg}\xi^{\rho\dd}\xi^{\gt\db}\gep_{\rho\gt}\gep_{\dg\dd}\,,&&\compconj(\Xi^{\ga\db})=\Xi^{\gb\da}\,,\\
&\Xi=i\xi^{\ga\da}\xi^{\gb\db}\xi^{\gga\dg}\xi^{\gd\dd}\gep_{\ga\gb}\gep_{\gga\gd}\gep_{\da\dg}\gep_{\db\dd}\,,&&\compconj(\Xi)=\Xi
\ee
and tensors $\gaugepar\,,\physf\,,\adeq$, $\adsyzt$ are traceless.

For cohomology of $\twsm$ we have (up to an overall factor $i^{s-s'}\tPNO^{s-s'}$)
\be
\label{twsm_cohomology_k_two}
&\tcohom^0_{s';2}=i^{s'}\Weyltens_{\ga(s'+1)}^{\gb(s'-1)}a^{\ga(s'+1)}b_{\gb(s'-1)}\,,\\
&\tcohom^1_{s';s'+1}=i^{s'}\xi^{\gga\dd}\,\twbasefunc^{s'-1}_{s'}\,\tweq_{\ga(s'+1),\gga\,;\,\db(s'-1)\dd}\,a^{\ga(s'+1)}\tbb^{\db(s'-1)}\,.
%
\ee
where tensors $\Weyltens\,,\tweq\,,\twsyz$ are traceless and symmetry type of $\tweq$ ($\twsyz$) with respect to undotted indices corresponds
to two row Young tableaux with the first row of length $s'-1$ and the second row of length 1 (2)
(see Appendix \ref{section_slk_plus_slk_irreps} for more details). Cohomology of $\btwsm$ are complex conjugate to \eqref{twsm_cohomology_k_two}.

As one can easily see operator $\ChEsm+\bChEsm$ maps $\tcohom^0_{s';2}+\btcohom^0_{s';2}$ to $\cohom^2_{s';0}$. So to speak
cohomology $\cohom^2_{s';0}$ is "glued up" by $\ChEsm+\bChEsm$. In other words 0-form
$\tcohom^0_{s';2}+\btcohom^0_{s';2}$ is not closed with respect to operator $\hat{\gs}$ 
and $\cohom^2_{s';0}$ is $\hat{\gs}$-exact. We, thus, have that 0-th and 1-st cohomology of $\hat{\gs}$
(see \eqref{UF_gs_hat_closed_space_components}-\eqref{UF_gs_hat_cohomology_definition}) are
\be
\label{hatsm_cohomology_k_two}
&\hcohom^0_{s';-s'+1}=(\cohom^0_{s';-s'+1},0)\,,\\
&\hcohom^1_{s';-s'+1}=(\cohom^1_{s';-s'+1},0)\,,\\
&\hcohom^2_{s';s'+1}=(0,\tcohom^1_{s';s'+1}+\btcohom^1_{s';s'+1})\,.
\ee

From \eqref{hatsm_cohomology_k_two} one gets that $\physf^{\gb(s'-1)\,;\,\db(s'-1)}_{\gga\,;\,\dd}$ is the dynamical field of system \eqref{Fr_Ts_unf_s_prime},
$\gaugepar^{\gb(s'-1)\,;\,\db(s'-1)}$ is the parameter of gauge transformations obeyed by $\physf$,
$\Weyltens_{\ga(s'+1)}^{\gb(s'-1)}+\bar{\Weyltens}_{\da(s'+1)}^{\db(s'-1)}$
is the generalized Weyl tensor, which is expressed in terms of $\physf$, finally
$\tweq_{\ga(s'+1),\gga\,;\,\db(s'-1)\dd}=0$, $\bar{\tweq}_{\da(s'+1),\dg\,;\,\gb(s'-1)\gd}=0$ are differential
equations imposed on $\Weyltens$ and $\bar{\Weyltens}$.
Direct form of these equations can be also easily obtained. We have
\be
\label{spin_s_prime_diff_eq}
&\Weyltens^{\ga(2s')}=\underbrace{\dpd{x_{\ga}{}^{\da}}\cdots\dpd{x_{\ga}{}^{\da}}}_{s'}\physf^{\ga(s')\,;\,\da(s')}\,,
&&\bar{\Weyltens}^{\da(2s')}=\underbrace{\dpd{x^{\ga}{}_{\da}}\cdots\dpd{x^{\ga}{}_{\da}}}_{s'}\physf^{\ga(s')\,;\,\da(s')}\,,\\
&\underbrace{\dpd{x^{\ga}{}_{\da}}\cdots\dpd{x^{\ga}{}_{\da}}}_{s'}\Weyltens^{\ga(2s')}=0\,,
&&\underbrace{\dpd{x_{\ga}{}^{\da}}\cdots\dpd{x_{\ga}{}^{\da}}}_{s'}\bar{\Weyltens}^{\da(2s')}=0\,,
\ee
where $\physf$ obeys the gauge transformations
\be
\label{spin_s_prime_gauge_trans}
\gd\physf^{\ga(s')\,;\,\db(s')}=\dpd{x_{\ga\db}}\gaugepar^{\ga(s'-1)\,;\,\db(s'-1)}\,.
\ee
Here symmetrization over the indices denoted by the same latter is implied and
to avoid projectors to the traceless and/or Young symmetry components we rose and lowered indices by means of $\gep^{\ga\gb}$, $\gep_{\ga\gb}$,
$\gep^{\da\db}$, $\gep_{\da\db}$.

If transformed from spinor indices $\ga,\da$ to vector indices $\mu$ (by means of Pauli matrices) equations \eqref{spin_s_prime_diff_eq},
\eqref{spin_s_prime_gauge_trans} coincide with equations
\eqref{Fr_Ts_eq}, \eqref{Fr_Ts_gauge_transform} for spin $s'$ field. Here $\Weyltens^{\ga(2s')}$ and $\bar{\Weyltens}^{\da(2s')}$
correspond to selfdual and antiselfdual parts of $C_{\nu(s'),\mu(s')}$. We, thus, showed that system
\eqref{Fr_Ts_unf_s_prime} realizes unfolded formulation of spin $s'$ Fradkin-Tseytlin equations.

According to above consideration, unfolded systems corresponding to
modules $\admodule^m_s$, $\twmodule^m_s$, $\btwmodule^m_s$ with fixed $s=1,2,\ldots$ and $m=0,\ldots, s-2, \infty$ or
to modules $\admodule^m$, $\twmodule^m$, $\btwmodule^m$ with fixed $m=0,1,\ldots\infty$ decompose into independent subsystems \eqref{Fr_Ts_unf_s_prime}.
Therefore such systems describes the collection of Fradkin-Tseytlin equations for the  spins given in decompositions
\eqref{admodule_inf_s_decomposition}-\eqref{admodule_n_decomposition}, \eqref{twmodule_decomposition}. In particular unfolded system proposed in
\cite{Vasiliev_progr_in_HS_gauge_theory} (see \eqref{Fr_Ts_unf_inf_Curv_with Weyl_glued}), which corresponds to modules 
$\admoduleinf$, $\twmoduleinf$, $\btwmoduleinf$, describes all integer spin fields
$s'=1,2,\ldots$ with infinite multiplicity.

\section{Conclusion\label{section_Conclusion}}
We have proposed unfolded system \eqref{Fr_Ts_unf_s_prime} that describes linear conformal dynamics of spin $s'$ gauge field
(spin $s'$ Fradkin-Tseytlin equations).
We also have shown that any unfolded system based on $\asu(2,2)$ adjoint and twisted-adjoint modules $\admodule^m$ and $\twmodule^m$, $\btwmodule^m$, 
$m=0,1,\ldots,\infty$ can be decomposed into independent subsystems of form \eqref{Fr_Ts_unf_s_prime}
by means of appropriate field redefinition and found spectrum of spins for any $m$.
In particular we have shown that system of equations proposed in \cite{Vasiliev_progr_in_HS_gauge_theory}
\eqref{Fr_Ts_unf_inf_Curv_with Weyl_glued} describes conformal fields
of all integer spins greater or equal than 1 with infinite multiplicity.

This work can be considered as a first modest step towards construction of the full nonlinear conformal theory of higher spins. One of the main
ingredients of the higher spin theories is a higher spin algebra. Our results pretend to be a probe of different candidates to this role. We see
that $\asu(2,2)$ modules $\admodule^m$, $\twmodule^m$, $\btwmodule^m$, $m=1,2\ldots$ mediate between those with $m=0$ and $m=\infty$.
One can speculate that the same is true for algebras $\aisu^m(2,2)$, which mediate between $\aisu(2,2)$ and $\aiu(2,2)$.

Having in mind that conformal higher spin theory has to be somehow related to $AdS$ higher spin theory one can suppose that
algebra $\aisu(2,2)$ is more preferable since its spectrum just literally coincides with the spectrum of some $AdS$ higher spin theory. On the other hand
equations proposed in \cite{Vasiliev_progr_in_HS_gauge_theory}, which correspond to $\aiu(2,2)$, are considerably simpler than
\eqref{Fr_Ts_unf_s_prime}. Therefore an interesting question arises wether it is possible to simplify \eqref{Fr_Ts_unf_s_prime}
maybe by mixing again gauge and Weyl sectors of the theory. Another important area of investigation is to consider
super extensions of $\aisu^m(2,2)$ and, thus, bring fermions into the play.

\section*{Acknowledgement}
I am grateful to Mikhail Vasiliev for extremely useful discussions at all stages of the work.
I would like to thank Vyacheslav Didenko for helpful comments on the manuscript.
The work is supported in part by RFBR grant No 14-02-01172.

\appendix
\section{Finite-dimensional $\asl(k)\oplus\asl(k)$ irreps\label{section_slk_plus_slk_irreps}}
Algebra $\asl(k)\oplus\asl(k)\subset\au(k,k)$
is generated by $\uL_\ga{}^\gb$ and $\buL_\da{}^\db$
(see (\ref{su_symbols})). Let the first summand of $\asl(k)\oplus\asl(k)$ that is generated by
$\uL_\ga{}^\gb$ be referred to as undotted $\asl(k)$ and the second summand that is generated by $\buL_\da{}^\db$
be referred to as dotted $\asl(k)$.
All finite-dimensional irreps of $\asl(k)\oplus\asl(k)$ are given by tensor products of finite-dimensional
irreps of undotted and dotted $\asl(k)$. In what follows we recall some well-known facts about $\asl(k)$-irreps
taking as an example undotted $\asl(k)$. Needless to say that the same arguments work for dotted $\asl(k)$
once undotted indices are replaced by  dotted indices.

Finite-dimensional irreps of $\asl(k)$ are given by $\asl(k)$-tensors
\be\label{symm_basis}
&T_{\gb_1(\lurl_1),\ldots,\gb_k(\lurl_k)}^{\ga_1(\uurl_1),\ldots,\ga_k(\uurl_k)}\,,\\
\uurl_1\geq\cdots\geq\uurl_k & \in\oZ_+\,,\quad \lurl_1\geq\cdots\geq\lurl_k\in\oZ_+\,,
\ee
written in symmetric basis or equivalently by $\asl(k)$-tensors
\be\label{antisymm_basis}
&T_{\gb_1[\lucl_1],\ldots,\gb_{\lurl_1}[\lucl_{\lurl_1}]}^{\ga_1[\uucl_1],\ldots,\ga_{\uurl_1}[\uucl_{\uurl_1}]}\,,\\
k\geq\uucl_1\geq\cdots\geq\uucl_{\uurl_1} & \in\oZ_+\,,\quad k\geq\lucl_1\geq\cdots\geq\lucl_{\lurl_1}\in\oZ_+\,,
\ee
written in antisymmetric basis.
Here $\ga_i(\uurl_i)$ denotes $\uurl_i$ totally symmetrized indices $\ga^1_i,\ldots,\ga^{\uurl_i}_i$ and
$\ga_i[\uucl_i]$ denotes $\uucl_i$ totally antisymmetrized indices $\ga^1_i,\ldots,\ga^{\uucl_i}_i$. Irreducibility conditions
require tensor $T$
\begin{enumerate}
\item
to be traceless, i.e. such that contraction of any pair of indices $\ga_i$ and $\gb_j$ gives zero;
\item
to satisfy Young symmetry conditions implying for upper indices in (anti)symmetric basis that total (anti)symmetrization
of all indices from the
set $\ga_i(\uurl_i)$ ($\ga_i[\uucl_i]$) with some index from the set $\ga_j(\uurl_j)$ ($\ga_j[\uucl_j]$)
gives zero for $j>i$ and analogous conditions for lower indices.
\end{enumerate}
It is often very useful to visualize the structure of tensor $T$ with the help of the pair of Young
tableaux
\be\label{slk_irrep}
\bt{rl}
upper\hspace{3mm} & \bt{l}\UUCase\LUEnumMDotsLabelBlockD{7}{2}{5}{1}{2}{1}{1}{1}{k}\et\\[10mm]
lower\hspace{3mm} & \bt{l}\LUCase\LUEnumMDotsLabelBlockD{6}{1}{4}{2}{3}{1}{2}{1}{k}\et
\et\,,
\ee
where $i$-th row of the upper(lower) tableau corresponds to $i$-th upper(lower) group of totally symmetric indices of tensor
$T$ in symmetric basis (\ref{symm_basis}) or equivalently
$i$-th column of the upper(lower) tableau corresponds to $i$-th upper(lower) group of totally antisymmetric indices of tensor
$T$ in antisymmetric basis (\ref{antisymm_basis}).
Let Young tableau with rows of lengths $\gl_1\geq\cdots\geq\gl_k$ be denoted as $\Ydiag(\gl_1,\ldots,\gl_k)$ and
Young tableau with columns of heights $\mu_1\geq\cdots\geq\mu_\mcol$\footnote{Note that the number of columns
does not limited by $k$ contrarily to the number of rows.} be denoted as  $\Ydiag[\mu_1,\ldots,\mu_\mcol]$.

Using totally antisymmetric tensors $\gep^{\ga[k]}$, $\gep_{\ga[k]}$ one can rise and lower indices of~$T$
\begin{align}
T{}_{\ldots,\ga_i[\lucl_i],\ldots}^{\ldots}\mapsto\gep^{\gb[k-\lucl_i]\ga_i[\lucl_i]}
T{}_{\ldots,\ga_i[\lucl_i],\ldots}^{\ldots}\,,\label{riseind}\\
T{}^{\ldots,\ga_i[\uucl_i],\ldots}_{\ldots}\mapsto\gep_{\ga_i[\uucl_i]\gb[k-\uucl_i]}
T{}^{\ldots,\ga_i[\uucl_i],\ldots}_{\ldots}\,.\label{lowerind}
\end{align}
Let us consider composition of (\ref{riseind}) and (\ref{lowerind})
\be\label{riselowerind}
T{}_{\ldots,\gb_j[\lucl_j],\ldots}^{\ldots,\ga_i[\uucl_i],\ldots}\mapsto
\gep^{\gd[k-\lucl_j]\gb_j[\lucl_j]}\gep_{\ga_i[\uucl_i]\gga[k-\uucl_i]}
T{}_{\ldots,\gb_j[\lucl_j],\ldots}^{\ldots,\ga_i[\uucl_i],\ldots}\,.
\ee
Taking into account that tensor product of two $\gep$-tensors is equal to alternative sum
of Kronecker deltas product
\be\label{two_ep}
\gep^{\ga^1\ldots\ga^k}\gep_{\gb^1\ldots\gb^k}=\sum_{\gs\in S_k}(-1)^{\pi(\gs)}\gd^{\ga^1}_{\gb^{\gs(1)}}
\cdots\gd^{\ga^k}_{\gb^{\gs(k)}}\,,
\ee
where sum is taken over all permutations of $(1,\ldots, k)$ and $\pi(\gs)$ is the oddness of permutation $\gs$,
one can readily see that (\ref{riselowerind}) vanishes if $\uucl_i+\lucl_j>k$\label{big_Young_tab_vanish}.
Indeed, in this case at least one of Kronecker deltas in every summand of
(\ref{two_ep}) contracts with $T$ that is considered to be traceless. Therefore, the
irreducibility conditions above are consistent only if $\uucl_i+\lucl_j\leq k$ for any $i,j$.
Due to the same arguments operations of rising and of lowering indices (\ref{riseind}), (\ref{lowerind})
result in a tensor satisfying irreducibility conditions only when applied to
the highest (i.e. the first) column of $T$.

Let us define rising and lowering Hodge conjugations by formulas
\begin{align}
{}^{*_{\rm r}}(T{}_{\ga_1[\lucl_1],\ldots}^{\ldots})=\frac{1}{\lucl_1!}
\gep^{\gb[k-\lucl_1]\ga_1[\lucl_1]}T{}_{\ga_1[\lucl_1],\ldots}^{\ldots}\,,\label{Hodge_r}\\
{}^{*_{\rm l}}(T{}^{\ga_1[\uucl_1],\ldots}_{\ldots})=\frac{1}{\uucl_1!}
\gep_{\ga_1[\uucl_1]\gb[k-\uucl_1]}T{}^{\ga_1[\uucl_1],\ldots}_{\ldots}\,.\label{Hodge_l}
\end{align}
They map $\asl(k)$-irrep described by Young tableau (\ref{slk_irrep}) into the equivalent $\asl(k)$-irrep
described by different Young tableau
\be\label{Hdiag}
\bt{rllr}
\raisebox{2mm}{upper}\hspace{3mm} &
\bt{l}\UUCase\renewcommand{\UEnumNumb}{3}\raisebox{3mm}{\UEnumDotsBlockD{7}{2}{5}{1}{2}{1}{1}{1}}\et &&
\bt{r}
\UUCase\begin{Young}{7}{7}{\renewcommand{\USkipValue}{20}}
\Put{\Block{1}{7}}\PutToRight{\BlockIV{7}{2}{5}{1}{2}{1}{1}{1}}
\newcounter{VectorX}\newcounter{VectorY}
\SetToCol{1}
\Add{VectorX}{\value{CurrentColX}}{5}\Add{VectorY}{\value{ULYoungY}}{10}
\put(\value{VectorX},\value{VectorY}){\vector(0,-1){10}}
\addtocounter{VectorX}{-7}
\put(\value{VectorX},\value{VectorY}){$\LabelStyle{}k-\lucl_1$}
\SetToCol{2}\UEnum{1}{3}\UDots
\end{Young}
\et
\\[-8mm]
&&$\rule{0mm}{10mm}^{\displaystyle\stackrel{*_{\rm r}}{\longrightarrow}}_{\displaystyle\stackrel{*_{\rm l}}{\longleftarrow}}$&
\\[-8mm]
lower\hspace{3mm} &
\bt{l}\LUCase\renewcommand{\UEnumNumb}{3}\UEnumDotsBlockD{6}{1}{4}{2}{3}{1}{2}{1}\et &&
\bt{r}
\LUCase\raisebox{-15mm}{\begin{Young}{5}{5}{\USkip}
\Put{\BlockIV{5}{1}{3}{2}{2}{1}{1}{1}}
\SetToCol{1}\UEnum{2}{2}\UDots
\end{Young}}
\et
\et\,.
\ee
Here ${}^{*_{\rm r}}$ kills the first column of the lower Young tableau and adds the column of height $k-\lucl_1$ to the
left-hand side of the upper Young tableau (since $\uucl_1+\lucl_1\leq k$ we, thus, get proper Young tableau) and
${}^{*_{\rm l}}$ acts in the opposite way. Young tableaux that result in consequent application
of transformations (\ref{Hdiag}) describe one and the same $\asl(k)$-module.
Coefficients in (\ref{Hodge_r}) and (\ref{Hodge_l}) are chosen such that
\be
{}^{*_{\rm r}}{}^{*_{\rm l}}T={}^{*_{\rm l}}{}^{*_{\rm r}}T=T\,.
\ee

Quadratic Casimir operator of algebra $\asl(k)$ is given by formula
\be\label{slk_Casimir}
\Casimir_{\asl(k)}=\uL_\ga{}^\gb\uL_\gb{}^\ga\,.
\ee
For the $\asl(k)$-irrep described by Young tableaux (\ref{slk_irrep})
$\Casimir_{\asl(k)}$ is equal to
\be\label{slk_Casimir_sym}
\Casimir_{\asl(k)}=(k+1)(\UUTotal+\LUTotal)-\frac{1}{k}(\UUTotal-\LUTotal)^2+\sum_i\uurl_i(\uurl_i-2i)+
\sum_i\lurl_i(\lurl_i-2i)\,,
\ee
where $\UUTotal$ is the total number of upper indices (total number of cells in the upper Young tableau) and
$\LUTotal$ is that of the lower indices. $\Casimir_{\asl(k)}$ can be also expressed through the heights of the Young
tableaux (\ref{slk_irrep})
\be\label{slk_Casimir_antisym}
\Casimir_{\asl(k)}=(k-1)(\UUTotal+\LUTotal)-\frac{1}{k}(\UUTotal-\LUTotal)^2-\sum_i\uucl_i(\uucl_i-2i)
-\sum_i\lucl_i(\lucl_i-2i)\,.
\ee

Irreducible representations of $\asl(k)\oplus\asl(k)$ are given by tensor product of two $\asl(k)$-irreps
\begin{align}
T_{\ga_1(\lurl_1),\ldots,\ga_k(\lurl_k)}^{\gb_1(\uurl_1),\ldots,\gb_k(\uurl_k)}\otimes
T_{\da_1(\ldrl_1),\ldots,\da_k(\ldrl_k)}^{\db_1(\udrl_1),\ldots,\db_k(\udrl_k)}=
T_{\ga_1(\lurl_1),\ldots,\ga_k(\lurl_k);\da_1(\ldrl_1),\ldots,\da_k(\ldrl_k)}^{
\gb_1(\uurl_1),\ldots,\gb_k(\uurl_k);\db_1(\udrl_1),\ldots,\db_k(\udrl_k)}\,,\\[3mm]
T_{\ga_1[\lucl_1],\ldots,\ga_{\lurl_1}[\lucl_{\lurl_1}]}^{\gb_1[\uucl_1],\ldots,\gb_{\uurl_1}[\uucl_{\uurl_1}]}\otimes
T_{\da_1[\ldcl_1],\ldots,\da_{\ldrl_1}[\ldcl_{\ldrl_1}]}^{\db_1[\udcl_1],\ldots,\db_{\udrl_1}[\udcl_{\udrl_1}]}=
T_{\ga_1[\lucl_1],\ldots,\ga_{\lurl_1}[\lucl_{\lurl_1}];\da_1[\ldcl_1],\ldots,\da_{\ldrl_1}[\ldcl_{\ldrl_1}]}^{
\gb_1[\uucl_1],\ldots,\gb_{\uurl_1}[\uucl_{\uurl_1}];\db_1[\udcl_1],\ldots,\db_{\udrl_1}[\udcl_{\udrl_1}]}\,,
\end{align}
where first $\asl(k)$ acts on undotted indices and second $\asl(k)$ acts on dotted indices. Tensors $T$ are supposed to
satisfy irreducibility conditions above separately within undotted and dotted indices. This representation
can be described by four Young tableaux
\be\label{slk_slk_irrep}
\bt{rll}
&\hspace{8mm} undotted &\hspace{10mm} dotted\\[3mm]
upper\hspace{3mm} & \bt{l}\UUCase\LUEnumMDotsLabelBlockD{7}{2}{5}{1}{2}{1}{1}{1}{k}\et &
\bt{l}\UDCase\LUEnumMDotsLabelBlockD{8}{1}{6}{2}{3}{1}{1}{1}{k}\et\\[10mm]
lower\hspace{3mm} & \bt{l}\LUCase\LUEnumMDotsLabelBlockD{6}{1}{4}{2}{3}{1}{2}{1}{k}\et &
\bt{l}\LDCase\LUEnumMDotsLabelBlockC{5}{2}{2}{2}{1}{1}{k}\et
\et\,,
\ee
Quadratic Casimir operator
\be\label{slkslk_Casimir}
\Casimir_{\asl(k)\oplus\asl(k)}=\uL^\ga{}_\gb\uL^\gb{}_\ga+\buL^\da{}_\db\buL^\db{}_\da
\ee
of $\asl(k)\oplus\asl(k)$-irrep described by
(\ref{slk_slk_irrep}) is given by the sum of quadratic $\asl(k)$-Casimirs for undotted and dotted parts of
(\ref{slk_slk_irrep}) (see (\ref{slk_Casimir_sym}) and/or (\ref{slk_Casimir_antisym})).

\section{The basis of $\admoduleinfs[s]$ and $\twmodule^{\infty}_{\tilde{s}}$
direct decomposition\label{section_Basis_of_direct_decompos}}
\subsection{Module $\admoduleinfs[s]$\label{section_admoduleinfs}}

Let us find the basis of module $\admoduleinfs[s]$ in which it decomposes into submodules $\admodules[s']$ \eqref{admodule_inf_s_decomposition}.
Since elements of $\asu(k,k)$ commute with helicity operator $\PNO$ and also
\be
\label{sinf_one_sinf_two_PNO_D_commutator}
[\opsinf_{1,2},\PNO]=\PNO\,,\qquad [\opsinf_{1,2},\uD]=\uD\,,
\ee
submodule $\admodules[s']$ enters to $\admoduleinfs[s]$
with the factor $\PNO^{s-s'}$. Taking it into account the natural ansatz for the basis under consideration is
\be
\label{admodule_inf_s_decomposition_basis_ansatz}
i^s\PNO^{s-s'}\adbasefunc^v_{s'}(\PNO,\uD)f_{s'-v}\,,\qquad s'=1,\ldots,s\,,\quad v=0,\ldots,s'-1\,.
\ee
Here $f_{s'-v}(a,b,\ba,\bb)$ are traceless (i.e. satisfying \eqref{iukk_trless_cond})
eigenvectors of operators $\opsinfo$ and $\opsinft$ corresponding to eigenvalue $s'-v$.
And
\be
\label{g_v_ansatz}
\adbasefunc^v_{s'}(\PNO,\uD)=\sum_{j=0}^v\PNO^{v-j}\uD^j\adbasisanscoef^v_{s';j}
\ee
is homogeneous polynomial of degree $v$ in two variables $\PNO$ and $\uD$, with coefficients $\adbasisanscoef^v_{s';j}(\na,\nb,\nba,\nbb)$
to be found from the requirement
that elements of \eqref{admodule_inf_s_decomposition_basis_ansatz} with fixed $s'$ span invariant subspace of module $\admoduleinfs[s]$, which is
submodule $\admodules[s']$ in decomposition \eqref{admodule_inf_s_decomposition}.

Operators $\adinfL_\ga{}^\gb\,,\adinfbL_{\da}{}^{\db}\,,\adinfD\,,\adinfPNO$ (see \eqref{adinf_repr_ukk}) obviously have the same form in new basis and conserve
elements of \eqref{admodule_inf_s_decomposition_basis_ansatz}, thus, keeping $\admodules[s']$ invariant. Suppose that operator corresponding to $\uP_{\ga\db}$
also keeps $\admodules[s']$ invariant for some particular choice of $\adbasefunc^v_{s'}$. Then as one can easily see
it has in basis \eqref{admodule_inf_s_decomposition_basis_ansatz} the following form
\be
\label{amodule_s_P_ansatz}
&\adP_{\ga\db}\,\,\,i^s\PNO^{s-s'}\adbasefunc^v_{s'}f_{s'-v}=i^s\PNO^{s-s'}\left(\adbasefunc^{v-1}_{s'}P^{v-1}_{s';-}\trProj b_\ga\bb_\db+
\adbasefunc^{v}_{s'}P^{v}_{s';0}\trProj \bb_\db\dpd{a^\ga}+\right.\\
&\qquad\qquad\qquad\qquad\qquad\qquad\qquad\quad\left.{}+\adbasefunc^{v}_{s'}P^{v}_{s';\bar{0}}\trProj b_\ga\dpd{\ba^\db}+
\adbasefunc^{v+1}_{s'}P^{v+1}_{s';+}\ddpd{a^\ga}{\ba^\db}\right)f_{s'-v}\,,
\ee
where $P^v_{s';-}\,,P^v_{s';0}\,,P^v_{s';\bar{0}}\,,P^v_{s';+}$ are some unknown coefficients and $\trProj$ is projector to the
traceless component \eqref{trProj_formulas}.

The requirement that
$\adP_{\ga\db}$ keeps $\admodules[s']$ invariant reduces to the following system of recurrence equations
\be
\label{admodule_inf_s_decomposition_recurence_eq}
&(j\!+\!1)\adbasisanscoef^v_{s';j+1}(\na,\nb,\nba,\nbb)=P^{v-1}_{s';-}(\na,\nb\!+\!1,\nba,\nbb\!+\!1)\adbasisanscoef^{v-1}_{s';j}(\na,\nb\!+\!1,\nba,\nbb\!+\!1)\,,\\[3mm]
&(j\func(n\!-\!1)+1)\adbasisanscoef^v_{s';j}(\na,\nb,\nba,\nbb)-i(j\!+\!1)\func(n\!-\!1)\adbasisanscoef^v_{s';j+1}(\na,\nb,\nba,\nbb)=\\
&\qquad\qquad\qquad\qquad\qquad\qquad\qquad{}=P^{v}_{s';0}(\na\!-\!1,\nb,\nba,\nbb\!+\!1)\adbasisanscoef^{v}_{s';j}(\na\!-\!1,\nb,\nba,\nbb\!+\!1)\,,\\[3mm]
&(j\func(\bn\!-\!1)+1)\adbasisanscoef^v_{s';j}(\na,\nb,\nba,\nbb)+i(j\!+\!1)\func(\bn\!-\!1)\adbasisanscoef^v_{s';j+1}(\na,\nb,\nba,\nbb)=\\
&\qquad\qquad\qquad\qquad\qquad\qquad\qquad{}=P^{v}_{s';\bar{0}}(\na,\nb\!+\!1,\nba\!-\!1,\nbb)\adbasisanscoef^{v}_{s';j}(\na,\nb\!+\!1,\nba\!-\!1,\nbb)\,,\\[3mm]
&(2s'+2k-2v+j-5)\func(n\!-\!1)\func(\bn\!-\!1)\adbasisanscoef^v_{s';j-1}(\na,\nb,\nba,\nbb)+\\
&{}\!+\!(j\!+\!1)\func(n\!-\!1)\func(\bn\!-\!1)\adbasisanscoef^v_{s';j+1}(\na,\nb,\nba,\nbb)+
i(n-\bn)\func(n\!-\!1)\func(\bn\!-\!1)\adbasisanscoef^v_{s';j}(\na,\nb,\nba,\nbb)=\\
&\qquad\qquad\qquad\qquad\qquad\qquad\qquad{}=P^{v+1}_{s';+}(\na\!-\!1,\nb,\nba\!-\!1,\nbb)\adbasisanscoef^{v+1}_{s';j}(\na\!-\!1,\nb,\nba\!-\!1,\nbb)
\ee
with boundary condition $\adbasisanscoef^v_{s';v}\equiv 1$.
Here $n=\na+\nb\,,\quad\bn=\nba+\nbb$ and $\func(n)=1/(n+k)$.

One can show that solution of system \eqref{admodule_inf_s_decomposition_recurence_eq} is
\be
\label{admodule_inf_s_decomposition_recurence_solution}
&P^{v-1}_{s';-}=v\,,\qquad P^{v}_{s';0}=v\func(n)+1\,,\qquad P^{v}_{s';\bar{0}}=v\func(\bn)+1\,,\\
&P^{v+1}_{s';+}=(2s'+2k-v-4)\func(n)\func(\bn)\,,\\
&\adbasisanscoef^v_{s';j}={v\choose j}\frac{\gd_{s';v-j}}{\prod_{h=j}^{v-1}(2s'+2k-v-3+h)}\,,
\ee
where $\gd_{s';m}$ satisfies the following recurrence equation
\be
\label{admodule_inf_s_decomposition_recurence_eq_for_gd}
\gd_{s';m}=i(n-\bn)\gd_{s';m-1}+(m-1)(2s'+2k-m-2)\gd_{s';m-2}\,,\qquad j=0,\ldots, v-1\,,
\ee
with boundary conditions $\gd_{s';0}\equiv 1\,, \quad \gd_{s';m<0}\equiv 0$.

Consider involution of Heisenberg algebra
\be
\label{symbols_oscillator_involution}
\tau:\quad\left\{
\begin{aligned}
&a^\ga\leftrightarrow \dpd{a^\ga}\,,&&\ba^\da\leftrightarrow \dpd{\ba^\da}\,,\\
&b_\ga\leftrightarrow \dpd{b_\ga}\,,&&\bb_\da\leftrightarrow \dpd{\bb_\da}\,.
\end{aligned}\right.
\ee
It induces involution of algebra $\asu(k,k)$
\be
\label{sukk_involution}
\tau:\quad\left\{
\begin{aligned}
&\uL_\ga{}^\gb\leftrightarrow \uL_\gb{}^{\ga}\,,&&\uP_{\ga\db}\leftrightarrow -\uK^{\ga\db}\,,&&\uD\leftrightarrow \uD\,,\\
&\buL_\da{}^\db\leftrightarrow \buL_\db{}^{\da}\,,&&&&\PNO\leftrightarrow -\PNO\,,
\end{aligned}\right.
\ee

As follows from \eqref{admodule_inf_s_decomposition_recurence_solution}, \eqref{admodule_inf_s_decomposition_recurence_eq_for_gd} Euler operators
$\na\,,\nb\,,\nba\,,\nbb$ contribute to coefficients $\adbasisanscoef^v_{s';j}$ through the combination $n-\bn$ only. Therefore elements of
\eqref{admodule_inf_s_decomposition_basis_ansatz} are invariant up to the factor -1 with respect to $\tau$.
Since $\uP_{\ga\db}$ and $\uK^{\ga\db}$ are $\tau$-conjugated one concludes that $\uK^{\ga\db}$ also keeps $\admodules[s']$ invariant.

The elements of \eqref{admodule_inf_s_decomposition_basis_ansatz} are obviously linearly independent and span the whole $\admoduleinfs[s]$. Therefore
they form the basis of $\admoduleinfs[s]$ under consideration.
Substituting values of $P^v_{s';-}$, $P^v_{s';0}$, $P^v_{s';\bar{0}}$, $P^v_{s';+}$ found in \eqref{admodule_inf_s_decomposition_recurence_solution}
to \eqref{amodule_s_P_ansatz} one gets representation of $\asu(k,k)$ on $\admodules[s']$ (see \eqref{ad_repr_ukk} for exact formulae).


Few lower examples of $\adbasefunc^v_{s'}$ are the following
\be
\label{lower_adbasefunc}
&\adbasefunc^0_{s'}=1\,,\\
&\adbasefunc^1_{s'}=\PNO\frac{i(n-\bn)}{2s'+2k-4}+\uD\,,\\
&\adbasefunc^2_{s'}=\PNO^2\frac{2s'+2k-4-(n-\bn)^2}{(2s'+2k-4)(2s'+2k-5)}+2\PNO\uD\frac{i(n-\bn)}{2s'+2k-4}+\uD^2\,,\\
&\adbasefunc^3_{s'}=\PNO^3i(n-\bn)\frac{6s'+6k-14-(n-\bn)^2}{(2s'+2k-4)\cdots(2s'+2k-6)}+3\PNO^2\uD\frac{2s'+2k-4-(n-\bn)^2}{(2s'+2k-4)(2s'+2k-5)}+\\[-1mm]
&\qquad\qquad\qquad\qquad\qquad\qquad\qquad\qquad\qquad\qquad\qquad\qquad\qquad{}+3\PNO\uD^2\frac{i(n-\bn)}{2s'+2k-4}+\uD^3\,.
\ee

\subsection{Module $\twmoduleinfs[s]$\label{section_twmoduleinfs}} Let us now consider module $\twmoduleinfs[s]$, which is obtained from
$\admoduleinfs[s]$ by twist transformation \eqref{twist_b}. Twist transformation of basis elements \eqref{admodule_inf_s_decomposition_basis_ansatz}
gives
\be
\label{twmodule_inf_s_decomposition_basis_ansatz}
i^s\tPNO^{s-s'}\twbasefunc^v_{s'}(\tPNO,\tD)\tf_{s'-v}\,,\qquad s'=1,\ldots,s\,,\quad v=0,\ldots,s'-1\,,
\ee
where $\tPNO$, $\tD$ are given by formulae \eqref{tPNO}, \eqref{tD} correspondingly,
$\tf_{s'-v}$ satisfies to \eqref{tf_eigenvec}, \eqref{twtracelessness} and
$\twbasefunc^v_{s'}(\tPNO,\tD)$ is twist transformed $\adbasefunc^v_{s'}(\PNO,\uD)$.

Now according to arguments given in page \pageref{problem_with_terminal_terms} we need to show that the 2-nd terms of the right-hand sides of \eqref{tw_repr_ukk}
vanish when acting on terminal element $\twbasisterm^{v}_{s'}$, i.e.
\be
\label{to_show}
&\twbasefunc^{v}_{s'}\Big(v\func(n)+1\Big)\ttrProj\ddpd{a^\ga}{\tbb^\db}\twmonomialterm_{s'-v}=0\,,\\
&\twbasefunc^{v}_{s'}\Big(v\func(n)+1\Big)\ttrProj\ba^\db\dpd{b_\ga}\twmonomialterm_{s'-v}=0\,.
\ee
Suppose first that $v_{\max}=v=0$, i.e. consider terminal terms $\twbasisterm^{0}_{s'}(0,q,q)$, $q=0,\ldots,s-1$ (see diagram
\eqref{twmodule_s_diagram}). The structure of dotted indices of all such terms is described by
the two-row Young tableau with the rows of equal length and, thus, terms \eqref{to_show} project it to the two-row Young tableau with the first row less 
than the second, which is zero.

Now lets $v_{\max}=v>0$, i.e. consider terminal terms $\twbasisterm^{v}_{s'}(v,q,q)$, $q=0,\ldots,s-v-1$ in diagram
\eqref{twmodule_s_diagram}. As one can easily see from \eqref{twlowest_conf_w}, \eqref{tw_diagramm_oscil_order_change}
monomials $\twmonomialterm_{s'-v}(v,q,q)$ corresponding to such terms have value
\be
(n-\tbn)\twmonomialterm_{s'-v}=2(s'+k-1)\twmonomialterm_{s'-v}
\ee
and operators $\ddpd{a^\ga}{\tbb^\db}$, $\ba^\db\dpd{b_\ga}$ from \eqref{to_show} decrease it by 2. Let us find the form of function $\twbasefunc^v_{s'}$
in \eqref{to_show}. Substituting the value of $n-\tbn$ to twist transformed equation
\eqref{admodule_inf_s_decomposition_recurence_eq_for_gd} one finds that
\be
\label{tgd_to_show_solution}
\tilde{\gd}_{s';m}=i^m\prod_{h=0}^{m-1}(2s'+2k-4-h)
\ee
and, thus,
\be
\label{twbasefunc_to_show}
\twbasisanscoef^v_{s';j}={v\choose j}i^{v-j}\,,\qquad\twbasefunc^v_{s'}=(\tD+i\tPNO)^v\,.
\ee
Taking into account that operators $\ddpd{a^\ga}{\tbb^\db}$, $\ba^\db\dpd{b_\ga}$ from \eqref{to_show} decrease the value of $v_{\max}$ by 1
one gets that due to \eqref{twmodule_specific} elements of \eqref{to_show} vanish.

%

\section{$\gs_-$-cohomology\label{section_sigma_min_cohomology}}
Let $\comp=\{\csp,\diffg\}$ be some co-chain complex. Here
$\csp=\oplus_{p=0}\csp^p$ is a graded space and
\be
\diffg:\quad\csp^p\mapsto\csp^{p+1}\,,\quad \diffg^2=0
\ee
is a differential.
The powerful tool to calculate cohomology of $\comp$ consists in consideration of homotopy operator
$\diffg^*$
\be
\diffg^*:\quad \csp^{p+1}\mapsto\csp^p\,,\quad(\diffg^*)^2=0\,.
\ee
According to the standard result of the cohomological algebra (see e.g. \cite{Homotopy_standard})
every element of $\comp$-cohomology group $\cohom$ has representative belonging to kernel of the
anticommutator $\anticomm=\{\diffg,\diffg^*\}$ provided that $\anticomm$ is diagonalizable on $\csp$.

Indeed, $\anticomm$ obviously commutes with $\diffg$ and, thus, both operators have common set of eigenvectors. Suppose $\psi\in\csp$ is
$\diffg$-closed vector from this set, such that $\anticomm\psi=q\psi$, $q\neq 0$. Then acting on $\psi$ by operator $\frac{1}{q}\anticomm$
one gets that $\psi=\frac{1}{q}\diffg\diffg^*\psi$ is $\diffg$-exact.

In paper \cite{Vas_conf_cohom} it was observed that if along with the above assumptions operator $\anticomm$ is
positive or negative semi-definite, representatives of $\cohom$ are in one-to-one correspondence with the elements from the
kernel of $\anticomm$.

Let us recall the arguments of \cite{Vas_conf_cohom}. Suppose for definiteness that $\anticomm$ is nonnegative. Then one can define 
positive scalar product $\langle\;|\;\rangle$ on $\csp$ such that operators $\diffg$ and $\diffg^*$ are mutually Hermitian conjugate $\diffg^\dag=\diffg^*$
with respect to it.
One then has that:
\begin{enumerate}
\item\label{item_one} $\langle\psi|\anticomm\psi\rangle=\langle\psi\diffg |\diffg \psi\rangle+\langle\psi\diffg^* |\diffg^* \psi\rangle$ and, thus,
elements from the kernel of $\anticomm$ are necessarily $\diffg$ and $\diffg^*$-closed;
\item those $\diffg$-closed $\psi$ that are not annulated by $\anticomm$ are $\diffg$-exact due to above arguments;
\item oppositely, if $\psi=\diffg\chi\neq 0$ is $\diffg$-exact, $\langle\psi|\psi\rangle=\langle\chi|\diffg^*\psi\rangle\neq 0$, i.e.
$\diffg^*\psi\neq 0$ and, thus, due to item \ref{item_one}, $\anticomm\psi\neq 0$.
\end{enumerate}

Let differential $\diffg$ and homotopy $\diffg^*$ are given by
\be\label{diffg_and_homotopy}
\diffg=\xi^{\ga\db}\gP_{\ga\db}\,,\qquad\diffg^*=\dpd{\xi^{\ga\db}}\gK^{\ga\db}\,,
\ee
where $\gP_{\ga\db}$ and $\gK^{\ga\db}$ are $\asu(k,k)$ generators of translations and special conformal transformations
represented by operators acting in some $\asu(k,k)$-module $\gModule$.
Let $\csp$ be space of differential forms (graded by the rank of the form)
taking values in $\gModule$. In other words $\csp=\Lambda\otimes\gModule$, where
$\Lambda$ denotes $k^2$-dimensional external algebra generated by $\xi$.

Consider subspace $\Lambda^p\subset\Lambda$ of the $p$-th order monomials in $\xi$.  Algebra $\asl(k)\oplus\asl(k)$ is represented on $\Lambda^p$
by operators
\be\label{slkslk_xi}
\xiuL_\ga{}^\gb=\xi^{\gb\dg}\dpd{\xi^{\ga\dg}}-\frac{1}{k}\gd_\ga^\gb n_\xi\,,\quad
\xibuL_\da{}^\db=\xi^{\gga\db}\dpd{\xi^{\gga\da}}-\frac{1}{k}\gd_\da^\db n_\xi
\ee
with quadratic Casimir operator equal to
\be\label{C_xi}
\xiCasimir_{\asl(k)\oplus\asl(k)}=\xiuL_\ga{}^\gb\xiuL_\gb{}^\ga+\xibuL_\da{}^\db\xibuL_\db{}^\da=
2kn_\xi-\frac{2}{k}n_\xi^2=2kp-\frac{2}{k}p^2\,.
\ee

In these notation operator $\anticomm$ has the following form
\be\label{deltag}
\anticomm=\{\diffg,\diffg^*\}=-\frac{1}{2}\Casimir_{\au(k,k)}+\frac{1}{2}\Casimir_{\asl(k)\oplus\asl(k)}+
\frac{1}{k}(\gD+p)(\gD+p-k^2)+\frac{1}{k}\gPNO^2\,,
\ee
where $p$ is the rank of differential form, $\gD$ and $\gPNO$ are dilatation and helicity operators represented in $\gModule$,
$\Casimir_{\au(k,k)}$ is quadratic $\au(k,k)$-Casimir operator \eqref{Casimir_gen}
calculated on module $\gModule$ and $\Casimir_{\asl(k)\oplus\asl(k)}$
is quadratic $\asl(k)\oplus\asl(k)$-Casimir operator \eqref{slkslk_Casimir}
calculated on module $\csp=\Lambda\otimes\gModule$, i.e.
\be
\Casimir_{\asl(k)\oplus\asl(k)}=(\adL_\ga{}^\gb+\xiuL_\ga{}^{\gb})(\adL_\gb{}^\ga+\xiuL_\gb{}^{\ga})+
(\adbL_\da{}^\db+\xibuL_\da{}^{\db})(\adbL_\db{}^\da+\xibuL_\db{}^{\da})\,.
\ee
Note that all the ingredients of (\ref{deltag}) commute and their common eigenvectors form the basis diagonalizing $\anticomm$.

For the subsequent analysis we also need to study $\asl(k)\oplus\asl(k)$-tensorial structure of $\Lambda^p$. To this end consider the basis element of $\Lambda^p$
\be\label{p_xis}
\Xi(p)^{\ga_1\da_1\cdots\ga_p\da_p}=\xi^{\ga_1\da_1}\cdots\xi^{\ga_p\da_p}\,.
\ee
Since $\xi$-s anticommute one can easily see that symmetrization of any group of undotted indices
of $\Xi(p)$ imply the antisymmetrization of the corresponding group of dotted indices and conversely.
Therefore, if undotted indices are projected to obey some symmetry conditions\footnote{For instance that
can be done by contracting undotted indices of $\Xi(p)$ with $\asl(k)$-tensor
of the symmetry type under consideration.}
corresponding to Young tableau $\Ydiag$ with the rows of lengths $\gl_1,\ldots,\gl_k$,
$\gl_1+\cdots+\gl_k=p$, dotted indices are automatically projected to obey symmetry conditions
corresponding to Young tableau $\Ydiag^\transpose$ with the columns of heights $\gl_1,\ldots,\gl_k$. Note
that all the rows of $\Ydiag$ are, thus, required to be not greater than $k$ since in opposite case antisymmetrization
of more than $k$ dotted indices implied. On the other hand projection of undotted indices of $\Xi(p)$
to symmetry conditions corresponding to any
Young tableau not longer than $k$ (i.e. such that any of its rows are not longer than $k$) leads to nonzero result.

Let us define an operation of transposition $\transpose$ that maps Young tableau $\Ydiag$ with rows $\gl_1,\ldots,\gl_k$ to
Young tableau $\Ydiag^\transpose$ with columns $\gl_1,\ldots,\gl_k$ and let
$\Ydiag^\transpose$ be called the transpose of $\Ydiag$.
In these notation decomposition of $\Xi(p)$ into $\asl(k)\oplus\asl(k)$-irreducible components
is the following
\be\label{p_xis_irr}
\Xi(p):\quad\bigoplus_{\Ydiag^{p,k}}
\raisebox{3mm}{\bt{rcc}
&undotted&dotted\\
upper&$\Ydiag^{p,k}$&$(\Ydiag^{p,k})^\transpose$\\
lower
\et}\,,
\ee
where $\Ydiag^{p,k}$ denote any Young tableau of the length not longer than $k$ and with the total number of cells equal to $p$.

Using formulas (\ref{slk_Casimir_sym}) and (\ref{slk_Casimir_antisym}) one can calculate Casimir operator
$\xiCasimir_{\asl(k)\oplus\asl(k)}$ of the $\asl(k)\oplus\asl(k)$-representations corresponding to
Young tableaux listed in
decomposition (\ref{p_xis_irr}). It can be easily seen that the terms of (\ref{slk_Casimir_sym}) depending on the lengths of the rows of
$\Ydiag^{p,k}$ are cancelled out by the terms of (\ref{slk_Casimir_antisym}) depending on the heights of the columns of
$(\Ydiag^{p,k})^\transpose$ and one finally
gets the same result as in (\ref{C_xi}).

\subsection{Gauge sector\label{section_Gauge_sector}}
Consider co-chain complex $\comp_s=(\csp_s,\adsm)$, where $\csp_s=\Lambda\otimes\admodules[s]$ is the space
of differential forms taking values in $\asu(k,k)$-adjoint module $\admodules[s]$ and
operator $\adsm=\xi^{\ga\db}\adP_{\ga\db}$ is differential.
$\comp_s$ is obtained from above consideration if one sets $\gModule$, $\gP_{\ga\db}$ and $\gK^{\ga\db}$ equal to
$\admodules[s]$, $\adP_{\ga\db}$ and $\adK^{\ga\db}$ correspondingly.
To coordinate notation let homotopy $\diffg^*=\dpd{\xi^{\ga\db}}\adK^{\ga\db}$ be denoted as $\adsm^*$ in this case.

Note that we should also require reality of $\csp_s$, i.e.
\be\label{csp_s_real_cond}
\compconj(\omega_s)=\omega_s\,,
\ee
where $\omega_s\in\csp_s$ and $\compconj$ is given by \eqref{reality_condition} and \eqref{xi_diffenition}.
However one can ignore \eqref{csp_s_real_cond} until all $\adsm$-cohomology are found. Indeed, suppose \eqref{csp_s_real_cond}
is disregarded. If $\cohom_s^p$ is some $\adsm$-cohomology,
$\compconj(\cohom_s^p)$ is also $\adsm$-cohomology since differential $\adsm$ is real. So the combinations $\cohom_s^p+\compconj(\cohom_s^p)$ give
us all real $\adsm$-cohomology.

Let $\csp_s^p$ denote subspace of $p$-forms in $\csp_s$.
Consider such a scalar product on $\csp_s$ with respect to which involution \eqref{symbols_oscillator_involution}
defined on $\xi^{\ga\db}$ by
\be\label{tau_additional}
\tau:\;\xi^{\ga\db}\leftrightarrow\dpd{\xi^{\ga\db}}
\ee
plays a role of Hermitian conjugation. Such scalar product is obviously positive definite\footnote{Note that it is not $\asl(k)\oplus\asl(k)$-invariant.
However it does not effect our consideration.}.

Due to \eqref{sukk_involution} and \eqref{tau_additional} operators $\adsm$ and $-\adsm^*$ are mutually
$\tau$-conjugate and, thus, $\anticomm=\{\adsm,\adsm^*\}$ is negative semidefinite. Therefore $\adsm$-cohomology $\cohom_s$ coincide with
the kernel of operator $\anticomm$, which can be found by analyzing those elements of $\csp_s$ that correspond to the maximal eigenvalues of $\anticomm$.

Substituting the value of $\Casimir_{\asu(k,k)}$ (see \eqref{Casimir}) to \eqref{deltag} one gets
\be\label{Delta_s}
\anticomm=\frac{1}{2}\Casimir_{\asl(k)\oplus\asl(k)}-(s-1)(s+2k-2)+
\frac{1}{k}(\confweight+p)(\confweight+p-k^2)\,.
\ee
If conformal weight $\confweight$ and differential form rank $p$ are fixed the maximal value of $\anticomm$ corresponds to the maximum of
$\Casimir_{\asl(k)\oplus\asl(k)}$.

The general element of $\csp_s^p$ has the following form
\be\label{csp_s_p_gen_el}
\go^p_s=\go^p_{s\,}{}_{\gga_1\ldots\gga_p;\dg_1\ldots\dg_p;\,}{}_{\ga(\na);\da(\nba)}^{\gb(\nb);\db(\nbb)}
\xi^{\gga_1\dg_1}\cdots\xi^{\gga_p\dg_p}a^{\ga(\na)}\ba^{\da(\nba)}b_{\gb(\nb)}\bb_{\db(\nbb)}\,.
\ee
Its $\asl(k)\oplus\asl(k)$-tensorial structure is described by Young tableaux found in tensor product of those describing
$\asl(k)\oplus\asl(k)$-structure of $\Xi(p)$ (see \eqref{slkslk_xi}) and of $\admodules[s]$
(see \eqref{admodule_s_monomial_Young_tableau}), i.e.
\be\label{cspirr}
\omega^p_s:\;\; \bigoplus_{\Ydiag^{p,k}}
\raisebox{3mm}{\bt{rll}
&\hspace{4mm}undotted&\hspace{15mm}dotted\\[1mm]
upper&$\Ydiag^{p,k}$\hspace{1mm}\raisebox{-5mm}{$\otimes$}\hspace{1mm}\symm{$\na$}&
\hspace{3mm}$(\Ydiag^{p,k})^\transpose$\hspace{1mm}\raisebox{-5mm}{$\otimes$}\hspace{1mm}\symm{$\nba$}\\
lower&\phantom{$\Ydiag^{p,k}$\hspace{1mm}$\otimes$\hspace{1mm}}\symm{$\nb$}&
\phantom{\hspace{3mm}$(\Ydiag^{p,k})^\transpose$\hspace{1mm}\raisebox{-5mm}{$\otimes$}\hspace{1mm}}\symm{$\nbb$}
\et}\,,
\ee
where $\symm{n}$ denotes one-row Young tableau of length $n$.

So now we need to find such an irreducible component from tensor product \eqref{cspirr} that maximizes $\Casimir_{\asl(k)\oplus\asl(k)}$.
Let us first consider undotted part of \eqref{cspirr}.
Suppose $\Ydiag^{p,k}=(\gl_1,\ldots,\gl_h)$, i.e. consists of $h\leq k$ rows of lengths $k\geq\gl_1\geq\cdots\geq\gl_h>0$, $\gl_1+\cdots+\gl_h=p$.
As one can readily see from formula \eqref{slk_Casimir_sym} for the value of $\asl(k)$-Casimir operator depending of the rows of Young tableau,
the maximal value of $\asl(k)$-Casimir corresponds to the undotted component of \eqref{cspirr} with the upper row $\symm{\na}$ symmetrized to the first row
of $\Ydiag^{p,k}$ (i.e. located in the mostly upper manner) and without any contractions done between $\Ydiag^{p,k}$ and the lower row $\symm{\nb}$.
The same arguments true for the dotted part of \eqref{cspirr} also.

Let such component be denoted as
\be\label{comp_notation_def}
\Big\{\compxi{\Ydiag^{p,k}}\Utens\compa{\na}\Ttens\compb{\nb}\,,\compxi{(\Ydiag^{p,k})^\transpose}\Utens\compba{\nba}\Ttens\compbb{\nbb}\Big\}\,,
\ee
where the left-hand side (before the comma) corresponds to the undotted part and the right-hand side (after the comma) to the dotted part of Young tableau,
$\Utens$ denotes the mostly upper component in tensor product, $\Ttens$ denotes traceless component (i.e. without any contractions done) in tensor product.
The letters on the top denote from what constituents (either some oscillators or basis 1-forms $\xi$) corresponding Young tableau is composed.
For instance $\compa{\na}$ denotes the row of length $\na$ composed from oscillators $a$.

To calculate the value of $\Casimir_{\asl(k)\oplus\asl(k)}$ for component \eqref{comp_notation_def} it is convenient to use row-formula
\eqref{slk_Casimir_sym} for its undotted part and column-forma \eqref{slk_Casimir_antisym} for its dotted part.  One gets
\be\label{adCasimir_slkslk}
\Casimir_{\asl(k)\oplus\asl(k)}=&\na(\na+2\gl_1-1)+\nba(\nba+2h-1)+\nb(\nb-1)+\nbb(\nbb-1)+\\
&{}+2k(p+s-v-1)-\frac{1}{k}\Big[(p+\na-\nb)^2+(p+\nba-\nbb)^2\Big]\,.
\ee
As seeing from \eqref{adCasimir_slkslk} $\Casimir_{\asl(k)\oplus\asl(k)}$ does not depend on the shape of
Young tableau $\Ydiag^{p,k}$ except the length of the first row $\gl_1$, the number of rows $h$ and the total number of cells $p$.

Substituting \eqref{adCasimir_slkslk} to \eqref{Delta_s} and expressing all the variables in terms of independent coordinates on $\admodules[s]$
\eqref{admodule_s_coordinates}, \eqref{ad_variab_through_coord} one gets
\be\label{Delta_s_next}
\anticomm=q(q+\gl_1-k-s+1)+t(t+h-k-s+1)+v(v+q+t-2s-2k+3)\,.
\ee
Due to inequalities $q,t\leq s-v-1$ and $k\geq 2$ one can see that the last term in \eqref{Delta_s_next} is negative for $v>0$ and, thus, maximization of
$\anticomm$ requires $v=0$. So we finally arrive at
\be\label{Delta_s_last}
\anticomm=q(q+\gl_1-k-s+1)+t(t+h-k-s+1)\,.
\ee

Let $\cohom^p_{s;\confweight}$ denote $p$-th $\adsm$-cohomology corresponding to the module $\admodules[s]$ with conformal weight $\confweight$.
First consider degenerate case $s=1$. Since $\admodules[1]$ is trivial and $\adsm\equiv 0$ one
gets that cohomology $\cohom_{1,0}^p$ are all real $p$-forms
\be\label{adcohom_s_one}
\cohom_{1;0}^p=\{\compxi{\Ydiag^{p,k}},\compxi{(\Ydiag^{p,k})^\transpose}\}+\mbox{comp. conj.}\,, \quad p=0,\ldots, k^2\,,
\ee
where complex conjugated of $\{\compxi{\Ydiag^{p,k}},\compxi{(\Ydiag^{p,k})^\transpose}\}$ is $\{\compxi{(\Ydiag^{p,k})^\transpose},\compxi{\Ydiag^{p,k}}\}$.

Suppose now that $s>1$. All zeros of \eqref{Delta_s_last} and corresponding $\adsm$-cohomology are listed below.
\begin{enumerate}
\item\label{adcohom_item_one} $v=q=t=0$, i.e. $\na=\nba=0$, $\nb=\nbb=s-1$ and $\confweight=-s+1$.
Formally \eqref{Delta_s_last} does not impose any additional limitations on $\gl_1$ and $h$
(recall that we always require $\gl_1\,,h\leq k$), but according to argument given in Appendix \ref{section_slk_plus_slk_irreps}
(see page \pageref{big_Young_tab_vanish}) traceless tensor identically vanishes if corresponds to Young tableau with total hight of some
upper and lower columns greater than $k$. Since due to \eqref{comp_notation_def} cohomology in this case has form
$\{\compxi{\Ydiag^{p,k}}\Ttens\compb{\nb},\compxi{(\Ydiag^{p,k})^\transpose}\Ttens\compbb{\nbb}\}$ and
$\nb\,,\nbb>0$ for $s>1$ one should require $\gl_1\,,h\leq k-1$. So we have
\be\label{adcohom_s_gen_item_one}
\cohom_{s;-s+1}^p=\{\compxi{\Ydiag^{p,k}}\limone{\gl_1,h\leq k-1}\Ttens\compb{s\!-\!1},\compxi{(\Ydiag^{p,k}}\limone{\gl_1,h\leq k-1})^\transpose
\Ttens\compbb{s\!-\!1}\}+\mbox{c.c.}\,, \quad p=0,\ldots, (k-1)^2\,.
\ee
\item $v=t=0$, $q=s-1$, $\gl_1=k$, i.e. $\na=\nb=0$, $\nba=\nbb=s-1$, $\confweight=0$ and complex conjugated\\
$v=q=0$, $t=s-1$, $h=k$, i.e. $\nba=\nbb=0$, $\na=\nb=s-1$, $\confweight=0$. Analogously to item \ref{adcohom_item_one} to get nonzero result
we additionally require $h\leq k-1$ in first case and $\gl_1\leq k-1$ in complex conjugated case. We, thus, have
\be\label{adcohom_s_gen_item_two}
\cohom_{s;0}^p=&\{\compxi{\Ydiag^{p,k}}\limtwo{\gl_1=k}{h\leq k-1},\compxi{(\Ydiag^{p,k}}\limtwo{\gl_1=k}{h\leq k-1})^\transpose
\Utens\compba{s\!-\!1}\Ttens\compbb{s\!-\!1}\}+\mbox{c.c.}\,, \quad p=k,\ldots, k(k-1)\,.
\ee
\item $v=0$, $q=t=s-1$, $\gl_1=h=k$, i.e. $\na=\nba=s-1$, $\nb=\nbb=0$, $\confweight=s-1$. We have
\be\label{adcohom_s_gen_item_three}
\cohom_{s;s-1}^p=\{\compxi{\Ydiag^{p,k}}\limone{\gl_1=h=k}\Utens\compa{s\!-\!1},\compxi{(\Ydiag^{p,k}}\limone{\gl_1=h=k})^\transpose
\Utens\compba{s\!-\!1}\}+\mbox{c.c.}\,, \quad p=2k-1,\ldots, k^2\,.
\ee
\end{enumerate}
Substituting $k=2$ to above formulae one gets \eqref{adsm_cohomology_k_two}.

\subsection{Weyl sector\label{section_Weyl_sector}}
Consider co-chain complex $\twcomp_s=(\twcsp_s,\twsm)$, where $\twcsp_s=\Lambda\otimes\twmodules[s]$ is the space of differential forms
taking values in $\asu(k,k)$ twist-adjoint module $\twmodules[s]$ and $\twsm=\xi^{\ga\db}\twP_{\ga\db}$. Unfortunately the powerful homotopy
technic described at the beginning of this section is not applicable in the case under consideration. This is because of the sign change in twist transformation
\eqref{twist_b}, which breaks mutual Hermitian conjugacy of $\twP_{\ga\db}$ and $\twK^{\ga\db}$ with respect to any positive definite scalar product.
Therefore anticommutator of $\twsm$ with homotopy $\twsm^*=\dpd{\xi^\ga\db}\twK^{\ga\db}$ is indefinite\footnote{Note that anticommutator
of $\twsm$ and $\dpd{\xi^{\ga\db}}\twP^{*\ga\db}$, where $\twP^{*\ga\db}$ is Hermitian conjugate to $\twP_{\ga\db}$, is semi-definite but nondiagonalizable.}.

For the purposes of the present paper one need to know 0-th and 1-st $\twsm$-cohomology only and also can fix $k=2$.
Let us focus on this case leaving general situation for the future investigation.

Zeroth $\twsm$-cohomology $\tcohom^0_{s;\twconfweight}$ coincide with the elements of $\twmodules[s]$ annihilated by $\twP_{\ga\db}$. Since module $\twmodules[s]$
is irreducible there is a single element of such kind the one with the lowest conformal weight \eqref{twlowest_conf_w}
\be
\tcohom^0_{s;2}=\Weyltens_{\ga(s+1)}^{\gb(s-1)}a^{\ga(s+1)}b_{\gb(s-1)}\,.
\ee

Recall that (up to an overall factor) the basis in $\twmodules[s]$ is $\twbasefunc^v_s\twmonomial_{s-v}(\na,\nba,\nb,\ntbb)$, $v=0,\ldots,s-1$, where
$\twmonomial_{s-v}$ are monomials of form \eqref{twmodule_s_monomial} that can be fixed by independent coordinates $\twcoord{v,q,t}$
\eqref{twcoord}. Since $k=2$ one can rewrite $\twmonomial_{s-v}$ as follows
\be\label{twmodule_s_monomial_k_two}
\twmonomial_{s'-v}(\na,\nba,\nb,\ntbb)=
\tilde{x}_{\ga(\na+\nb)\,;\,\db(\ntbb-\nba)}\underbrace{\gep_{\da\db}\cdots\gep_{\da\db}}_{\nba}
a^{\ga(\na)}b^{\ga(\nb)}\tbb^{\db(\ntbb)}\ba^{\da(\nba)}\,,
\ee
where $b^\ga=b_\gb\gep^{\ga\gb}$ and $\gep^{\ga\gb}$ is totally antisymmetric tensor.
Such monomials form $\asl(k)\oplus\asl(k)$ irrep corresponding to Young tableau\footnote{This Young tableau is obtained from
\eqref{twmodule_s_monomial_Young_tableau} by Hodge conjugation \eqref{Hdiag}.}
\be\label{twmodule_s_monomial_Young_tableau_k_two}
\bt{ll}
\hspace{5mm} undotted &\hspace{10mm} dotted\\[1mm]
                    \bt{l}
                    \begin{Young}{8}{1}{\USkip\DSkip\renewcommand{\ULabelMidVShift}{4}}
                    \Put{\Block{8}{1}}\ULabelMid{\na+\nb}
                    \end{Young}
                    \et
                  &\hspace{5mm}
                    \bt{l}
                    \begin{Young}{5}{1}{\USkip\DSkip\renewcommand{\ULabelMidVShift}{4}}
                    \Put{\Block{5}{1}}\ULabelMid{\ntbb-\nba}
                    \end{Young}
                    \et
\et
\ee
In what follows we denote diagrams like \eqref{twmodule_s_monomial_Young_tableau_k_two} as $(l_1,l_2)$, where $l_1$ and $l_2$ are the numbers of cells of
undotted and dotted rows correspondingly.

Decompose operator $\twsm$ into the sum of three operators in accordance with their action on the basis elements of  $\twmodules[s]$
\be
\twsm=\twsm^-+\twsm^0+\twsm^+\,.
\ee
Namely operator $\twsm^-$ ($\twsm^+$) decreases (increases) $v$ by 1 and $\twsm^0$ does not change it
\be\label{twsm_m_n_p}
&\twsm^-(\twbasefunc_s^v\twmonomial_{s-v})=\twbasefunc^{v-1}_{s}v\xi^{\ga\db}\gep_{\gga\ga}\ttrProj b^\gga\dpd{\tbb^\db}\twmonomial_{s-v}\,,\\
&\twsm^0(\twbasefunc_s^v\twmonomial_{s-v})=\twbasefunc^{v}_{s}\xi^{\ga\db}\ttrProj\Big((v \func(n)+1)\ddpd{a^\ga}{\tbb^\db}
+(v \func(\tbn)+1)b^\gga\gep_{\gga\ga}\dpd{\ba^\db}\Big)\twmonomial_{s-v}\,,\\
&\twsm^+(\twbasefunc_s^v\twmonomial_{s-v})=\twbasefunc^{v+1}_{s}(2s-v)\func(n)\func(\tbn)\xi^{\ga\db}\ddpd{a^\ga}{\ba^\db}\twmonomial_{s-v}\,,
\ee
where $\func(n)=1/(n+2)$ and $\ttrProj$ is projector\footnote{Projector $\ttrProj\, b^\ga\twmonomial_{s-v}$ which acts on the oscillator $b^\ga$ with
risen index carries out symmetrization of $b^\ga$ with all $a$-s and $b$-s in $\twmonomial_{s-v}$
\be\label{ttrProj_formula_k_two}\nn
\ttrProj b^\ga=\frac{1}{n}(\nb b^\ga+a^\ga b^\gb\dpd{a^\gb})\,.
\ee}
given by \eqref{ttrProj_formulas}.
From the nilpotency of $\twsm$ it follows that
\be\label{twsmpmz_relations}
(\twsm^-)^2=(\twsm^+)^2=0\,,\qquad (\twsm^0)^2+\{\twsm^-,\twsm^+\}=0\,.
\ee

Representative of $\twsm$-cohomology can always be chosen to have definite conformal weight and definite irreducible $\asl(k)\oplus\asl(k)$-structure.
General element of $\Lambda^p\otimes \twmodule_{s;\twconfweight}$ with fixed conformal weight $\twconfweight$ can be decomposed as
\be\label{twsm_cohomology_decomposition}
\tgen^p_{s;\twconfweight}=\sum_{v=v_{\min}}^{v=v_{\max}}\tgenc^{p;v}_{s;\twconfweight}\,,
\ee
where summand $\tgenc^{p;v}_{s;\twconfweight}=\Lambda^p\otimes \twbasefunc^v_s (\twmonomial_{s-v}+
\twmonomial^\prime_{s-v}+\cdots)$ is a linear combination of basis elements with $v$ and $\twconfweight$ fixed tensored by $\Lambda^p$.

Within this decomposition $\twsm$-closedness condition for $\tgen^p_{s;\twconfweight}$ split into the system
\begin{align}\label{twsm_closedness_system}
\left\{
\begin{aligned}
&\twsm^-\tgenc^{p;v_{\min}}_{s;\twconfweight}=0\,,\\
&\twsm^-\tgenc^{p;v_{\min}+1}_{s;\twconfweight}+\twsm^0\tgenc^{p;v_{\min}}_{s;\twconfweight}=0\,,\\
&\hspace{15mm}\cdots\\
&\twsm^-\tgenc^{p;v+1}_{s;\twconfweight}+\twsm^0\tgenc^{p;v}_{s;\twconfweight}+
\twsm^+\tgenc^{p;v-1}_{s;\twconfweight}=0\,,\qquad v=v_{\min}+1\,,\ldots\,,v_{\max}-1\,,\\
&\hspace{15mm}\cdots\\
&\twsm^0\tgenc^{p;v_{\max}}_{s;\twconfweight}+\twsm^+\tgenc^{p;v_{\max}-1}_{s;\twconfweight}=0\,,\\
&\twsm^+\tgenc^{p;v_{\max}}_{s;\twconfweight}=0\,.
\end{aligned}
\right.
\end{align}
According to the first equation of \eqref{twsm_closedness_system} $\tgenc^{p;v_{\min}}_{s;\twconfweight}$ is required to be
$\twsm^-$-closed. Suppose it is $\twsm^-$-exact, i.e. such $\varepsilon^{p-1,v_{\min+1}}_{s;\twconfweight+1}$ exists that
$\tgenc^{p;v_{\min}}_{s;\twconfweight}=\twsm^-\varepsilon^{p-1,v_{\min+1}}_{s;\twconfweight+1}$. Then $\twsm$-exact shift
$\tgen^p_{s;\twconfweight}-\twsm\varepsilon^{p-1,v_{\min+1}}_{s;\twconfweight+1}$ zeros out term $\tgenc^{p;v_{\min}}_{s;\twconfweight}$.
So in order to $\tgen^p_{s;\twconfweight}$ be $p$-th $\twsm$-cohomology one can require its term with the lowest value of $v$
to be $p$-th $\twsm^-$-cohomology.

Let us find 1-st $\twsm^-$-cohomology $\tmcohomo$. Note that unlike the whole operator $\twsm$
operators $\twsm^{\pm,0}$ acting separately map monomials into monomials and, thus, one can look for cohomology $\tmcohomo$ among irreducible components of
tensor product $\xi\otimes\twmonomial_{s'-v}$. These components are described by Young tableaux obtained in tensor product of
\eqref{twmodule_s_monomial_Young_tableau_k_two} with one undotted and one dotted cells.
Closedness and exactness of each component can be easily checked by direct computation. In \eqref{tmcohomo_comp} the results are collected
\be\label{tmcohomo_comp}
&(\na\!+\!\nb\!+\!1,\ntbb\!-\!\nba\!+\!1)\,,\!\!\!&&(\na\!+\!\nb\!-\!1,\ntbb\!-\!\nba\!-\!1)\,,\!\!\!&&(\na\!+\!\nb\!+\!1,\ntbb\!-\!\nba\!-\!1)\,,
\!\!\!&&(\na\!+\!\nb\!-\!1,\ntbb\!-\!\nba\!+\!1)\,,\\
&\na\!+\!\nb\geq 0\,,&&\na\!+\!\nb\geq 1\,,&&\na\!+\!\nb\geq 0\,,&&\na\!+\!\nb\geq 1\,,\\
&\ntbb\!-\!\nba\geq 0\,,&&\ntbb\!-\!\nba\geq 1\,,&&\ntbb\!-\!\nba\geq 1\,,&&\ntbb\!-\!\nba\geq 0\,,\\
&\mbox{closed if $\nba=\ntbb=0$} &&\mbox{nonclosed}\,,&&\mbox{nonclosed}\,,&&\mbox{closed}\,,\\
&\mbox{nonexact}&&&&&&\mbox{exact if $\nb>0$}\,.
\ee

We thus have two series of $\tgenc^{1;v_{\min}}_{s;\twconfweight}$ that pretend to contribute to $\tcohom^p_{s;\twconfweight}$.
Let us consider both series separately.

\noindent\textbf{(1)} Component $(\na+\nb+1,1)$ with $\nba=\ntbb=0$. As one can see from diagram \eqref{twmodule_s_diagram} the only element of $\twmodules[s]$ with $\nba=\ntbb=0$
is that with coordinates $\twcoord{0,0,0}$. It has $\na=s+1$, $\nb=s-1$ and the lowest conformal weight $\twconfweight=2$.
One can also see that there are no more elements in $\twmodules[s]$ with
the same conformal weight. So decomposition \eqref{twsm_cohomology_decomposition} reduces to one term
\be\label{candidate_one}
\tgen^1_{s;2}=\xi^{\ga\db}\tilde{x}_{\ga(2s+1)\,;\,\db}a^{\ga(s+1)}b^{\ga(s-1)}\,.
\ee
One can readily see that $\tgen^1_{s;2}$ is $\twsm$-exact
\be\label{candidate_one_exact}
\tgen^1_{s;2}=\twsm\varepsilon^0_{s,3}\,,\qquad\varepsilon^0_{s,3}=\tilde{x}_{\ga(2s+1)\,;\,\db}a^{\ga(s+2)}b^{\ga(s-1)}\tbb^{\db}\,.
\ee
Indeed, $\varepsilon^0_{s,3}$ is $\twsm^{\pm}$-closed and is mapped by $\twsm^0$ to $\tgen^1_{s;2}$.

\noindent\textbf{(2)} Component $(\na-1,\ntbb-\nba+1)$ with $\nb=0$. In this case coordinates of corresponding $\twmodules[s]$ elements $\twmonomial_{\min}$ are
\be\label{candidate_two_coord}
\twcoord{v_{\min},q_{\min}=s\!-\!v_{\min}\!+\!m\!-\!1,t_{\min}=s\!-\!v_{\min}\!-\!1},
\mbox{for some fixed $v_{\min}=0,\ldots,s-1$, and $m=0,1,\ldots$}
\ee
which indicate on some dot at the north-west boundary of the stripe $v_{\min}$ (see diagram \eqref{twmodule_s_diagram}).
Conformal weight and orders with respect to oscillators of $\twmonomial_{\min}$ are
\be\label{candidate_two_variab}
&\na=2s-v_{\min}+m\,, &&\nba=s-v_{\min}-1\,,\\
&\nb=0\,,&&\ntbb=s+m-1\,,\\
&\twconfweight=2s-v_{\min}+m\,,&&v_{\max}=\min(v_{\min}+m,s-1)\,.
\ee
Therefore $\tgenc^{1;v_{\min}}_{s;\twconfweight}$ 
has the following $\asl(k)\oplus\asl(k)$-structure
\be\label{candidate_two_slk_slk_stracture}
(2s-v_{\min}+m-1,v_{\min}+m+1)\,.
\ee

Suppose first that $v_{\min}<s-1$ and consider if there are any other terms in decomposition \eqref{twsm_cohomology_decomposition}.
To construct such terms one should find such elements of $\twmodules[s]$ that
\begin{enumerate}
\item have the same conformal weight;
\item their coordinate $v$ is greater than $v_{\min}$ but less than $v_{\max}$;
\item contribute to component \eqref{candidate_two_slk_slk_stracture} when tensored by $\xi$.
\end{enumerate}

Suppose the element $\monp$ we are looking for has coordinates $\twcoord{\vp,\qp,\tp}$. Let the orders of $\monp$ with respect to
oscillators (which are expressed via coordinates through the formula \eqref{tw_variab_through_coord}) be denoted as $\nap\,,\nbp\,,\nbap\,,\ntbbp$.
Since tensoring by $\xi$ either adds or subtracts one cell to/form Young tableau we require
\be\label{candidate_two_rquirement}
(\nap+\nbp,\ntbbp-\nbap)=(\na\hspace{1mm}({}-2),\ntbb-\nba\hspace{1mm}[{}+2])\,,
\ee
where $\na\,,\nba\,,\ntbb$ are given in \eqref{candidate_two_variab} and numbers in parenthesis (brackets) could be either skipped or taken into account.
Condition \eqref{candidate_two_rquirement} guaranties that $\xi\otimes\monp$ contains component \eqref{candidate_two_slk_slk_stracture}.
In terms of coordinates the requirements above give the following system
\begin{align}\label{candidate_two_syst}
\left\{
\begin{aligned}
&\vp+\qp+\tp+2=2s-v_{\min}+m\,,\\
&\vp-\qp+\tp=v_{\min}-m\hspace{1mm}({}+2)\,,\\
&\vp+\qp-\tp=v_{\min}+m\hspace{1mm}[{}+2]\,.
\end{aligned}
\right.
\end{align}
The solution of \eqref{candidate_two_syst} is
\begin{align}\label{candidate_two_solution}
\left\{
\begin{aligned}
&\vp=v_{\min}\hspace{1mm}({}+1)\hspace{1mm}[{}+1]\,,\\
&\qp=s-v_{\min}+m-1\hspace{1mm}({}-1)=q_{\min}\hspace{1mm}({}-1)\,,\\
&\tp=s-v_{\min}-1\hspace{1mm}[{}-1]=t_{\min}\hspace{1mm}[{}-1]\,,
\end{aligned}
\right.
\end{align}
where $q_{\min}=s-v_{\min}+m-1$ and $t_{\min}=s-v_{\min}-1$ are coordinates of $\twmonomial_{\min}$.
%
Taking into account that $(\vp,\qp,\tp)$ should satisfy \eqref{twcoord} one gets the only solution
\be\label{candidate_two_possib}
&(\vp,\qp,\tp)=(v_{\min}+1,q_{\min},t_{\min}-1)\,,\qquad v_{\min}<s-1\,,\\
&\nap=\na\,,\qquad \nbp=0\,,\qquad \nbap=\nba-1\,,\qquad \ntbbp=\ntbb+1\,.
\ee
In this case
\be\label{candidate_two_form}
\tgen^1_{s;\twconfweight}=&\twbasefunc^{v_{\min}}_s\xi^{\gga\db}
\tilde{x}_{\ga(\na-1)\,;\,\db(\ntbb-\nba+1)}\gep_{\ga\gga}\underbrace{\gep_{\da\db}\cdots\gep_{\da\db}}_{\nba}
a^{\ga(\na)}\tbb^{\db(\ntbb)}\ba^{\da(\nba)}+\\
{}+&\twbasefunc^{v_{\min}+1}_s\xi^{\gga\dd}\tilde{y}_{\ga(\na-1)\,;\,\db(\ntbb-\nba+1)}\gep_{\ga\gga}\gep_{\db\dd}
\underbrace{\gep_{\da\db}\cdots\gep_{\da\db}}_{\nba-1}a^{\ga(\na)}\tbb^{\db(\ntbb+1)}\ba^{\da(\nba-1)}\,,
\ee
where $\na\,,\nba\,,\ntbb$ are fixed by \eqref{candidate_two_variab}. As one can see directly such $\tgen^1_{s;\twconfweight}$
is not $\twsm$-closed.

So we are finally left with the case $v_{\min}=s-1$, when
\be\label{candidate_three_variab}
&\na=s+m+1\,,\qquad&&\nba=0\,,\qquad&&m=0,1,\ldots,\infty\\
&\nb=0\,,&&\ntbb=s+m-1\,,\\
&\confweight=s+m+1
\ee
and
\be\label{candidate_three_form}
\tgen^1_{s;s+m+1}=\twbasefunc^{s-1}_s\xi^{\gga\db}
\tilde{x}_{\ga(s+m)\,;\,\db(s+m)}\gep_{\ga\gga}a^{\ga(s+m+1)}\tbb^{\db(s+m-1)}\,.
\ee
$\tgen^1_{s;s+m+1}$ is obviously not $\twsm^0$-closed for $m>0$. For $m=0$ this is the case since $\tgen^1_{s;s+1}$ is composed of terminal monomial
which is zeroed out by $\twsm^0$ due to the arguments explained in Appendix \ref{section_twmoduleinfs} (see formula \eqref{to_show}).

By means of analogous analysis checking $\varepsilon^0_{s;s+2}$-s that could have contributed to $\tgen^1_{s;s+1}$ one can show that
$\tgen^1_{s;s+1}$ is not $\twsm$-exact. Therefore
\be\label{twcohom_one}
\tcohom^1_{s;s+1}=\twbasefunc^{s-1}_s\xi^{\gga\db}
\tilde{x}_{\ga(s)\,;\,\db(s)}\gep_{\ga\gga}a^{\ga(s+1)}\tbb^{\db(s-1)}\,.
\ee


\begin{thebibliography}{33}
\bibitem{Fradkin_Tseytlin}  E.S. Fradkin and A.A. Tseytlin, {\it Phys.Rept.} {\bf 119} (1985) 233.
\bibitem{Fr_Lin_conf_HS_alg}  E.S. Fradkin and V.Ya. Linetsky, {\it Ann.Phys.} {\bf 198} (1990) 252.
\bibitem{Gunaydin} M. G\"{u}naydin, \emph{Lect.Notes in Phys.}, {\bf 180} (1983) 192.
\bibitem{Bars_Gunaydin} I. Bars and M. G\"{u}naydin, \emph{Commun.Math.Phys.}, {\bf 91} (1983) 31.
\bibitem{Vasiliev_conf_HS_sym in _four_dimm} M.A. Vasiliev, \emph{Phys.Rev.} {\bf D66} 066006 (2002).
\bibitem{Segal} A.Y. Segal, {\it Nucl.Phys.} {\bf B664} (2003) 59-130.
\bibitem{Bekaert_Grigoriev}  X.~Bekaert and M.~Grigoriev, {\it Nucl.Phys.}  {\bf B876} (2013) 667.
\bibitem{Vasiliev_progr_in_HS_gauge_theory} M.A. Vasiliev, "Progress in Higher Spin Gauge Theories", arXiv:hep-th/0104246.
\bibitem{Vasiliev_unfolded_repr_for_eq_in_two_plus_one_AdS} M.A. Vasiliev, \emph{Class.Quant.Grav.}, {\bf 11} (1994) 649.
\bibitem{BCIV_review} X. Bekaert, S. Cnockaert, C. Iazeolla and M.A. Vasiliev, "Nonliniar Higher Spin Theories in Various Dimensions",
                      Proceedings of the First Solvay Workshop on Higher Spin Gauge Theories (Brussels, May 2004), arXiv:hep-th/0503128.
\bibitem{Sullivan_FDA} D. Sullivan, \emph{Publ.Math. IH\'{E}S} {\bf 47} (1977) 269.
\bibitem{Vasiliev_on_conf_sl_four_and_sp_eight_symm} M.A. Vasiliev, \emph{Nucl.Phys.} {\bf B793} (2008) 469.
\bibitem{Chevalley_Eilenberg_cohom_theor_of_Lie_groups_and_Lie_alg} C. Chevalley and S. Eilenberg, \emph{Trans.Amer.Math.Soc.} {\bf 63} (1948) 85.
\bibitem{Ponomarev_Vasiliev_Unfolded_scalar_supermultiplet} D.S. Ponomarev, M.A. Vasiliev, \emph{JHEP} {\bf 1201} (2012) 152.
\bibitem{Shaynkman_Vasiliev_Scalar_field_from_the_HST_perspective} O.V. Shaynkman, M.A. Vasiliev, \emph{Theor.Math.Phys.} {\bf 123} (2000) 683
         [\emph{Teor.Mat.Fiz.} {\bf 123} (2000) 323].
\bibitem{Homotopy_standard}  M. Henneaux and C. Teitelboim, “Quantization of gauge systems”, Prinston University
         Press, Princeton, New Jersey, 1992.
\bibitem{Vas_conf_cohom} M.A. Vasiliev, \emph{Nucl.Phys.} {\bf B829} (2010) 176-224.

\end{thebibliography}
\end{document}